\documentclass[a4paper,fleqn,usenatbib]{mnras}

\usepackage[T1]{fontenc}
\usepackage{ae,aecompl}
\usepackage{graphicx}	
\usepackage{amsmath}	
\usepackage{amssymb}	
\usepackage{bm} 

\newcommand{\beq}{\begin{equation}}
\newcommand{\eeq}{\end{equation}}
\newcommand{\beqy}{\begin{eqnarray}}
\newcommand{\eeqy}{\end{eqnarray}}
\newcommand{\bmlet}{\begin{subequations}}
\newcommand{\emlet}{\end{subequations}}

\title[Unified equations of state for neutron stars]{Unified equations of state for
cold non-accreting neutron stars with Brussels-Montreal functionals. 
I. Role of symmetry energy}

\author[J.~M.~Pearson et al.]{
J. M. Pearson,$^{1}$\thanks{E-mail: pearson@LPS.UMontreal.CA}
N. Chamel,$^{2}$
A. Y. Potekhin,$^{3}$
A. F. Fantina,$^{4,2}$
\newauthor
C. Ducoin,$^{5}$
A. K. Dutta,$^{1,2,6}$
and S. Goriely$^{2}$
\\
$^{1}$D\'ept. de Physique, Universit\'e de Montr\'eal, Montr\'eal
(Qu\'ebec), H3C 3J7 Canada
\\
$^{2}$Institut d'Astronomie et d'Astrophysique, CP-226, Universit\'e
Libre de Bruxelles, 1050 Brussels, Belgium
\\
$^{3}$Ioffe Institute, Politekhnicheskaya 26,
194021 St. Petersburg, Russia
\\
$^{4}$Grand Acc\'el\'erateur National d'Ions Lourds (GANIL), CEA/DRF -
 CNRS/IN2P3, Boulevard Henri Becquerel, 14076 Caen, France
\\
$^{5}$Institut de Physique Nucl\'eaire de Lyon, CNRS/IN2P3, Universit\'e
 Claude Bernard Lyon 1, Villeurbanne, France
\\
$^{6}$School of Physics, Devi Ahilya University, Indore, 452001 India
}

\date{Originally published in \textit{MNRAS} \textbf{481}, 
2994--3026 (2018),
corrected according to Erratum (2019, accepted).}

\begin{document}
\label{firstpage}
\pagerange{\pageref{firstpage}--\pageref{lastpage}}
\maketitle

\begin{abstract}
The theory of the nuclear energy-density functional is used to provide a 
unified and thermodynamically consistent treatment of all regions of 
cold non-accreting neutron stars. In order to assess the impact of our lack of complete knowledge of the density 
dependence of the symmetry energy on the constitution and the global structure 
of neutron stars, we employ four different functionals. All of them were 
precision fitted to essentially all the nuclear-mass data with the 
Hartree-Fock-Bogoliubov method and two different neutron-matter equations of 
state based on realistic nuclear forces. For each functional, we calculate the 
composition, the pressure-density relation, and the chemical potentials 
throughout the star. We show that uncertainties in the symmetry energy can 
significantly affect the theoretical results for the composition and global 
structure of neutron stars. To facilitate astrophysical applications, we 
construct analytic fits to our numerical results.
\end{abstract}



\section{Introduction}
\label{intro}

Three distinct regions can be recognized in a neutron star below its thin 
atmosphere: a locally homogeneous core and two concentric shells characterized 
by different inhomogeneous phases \citep*[e.g.,][]{hae07,lrr}. The outermost 
of the shells, the `outer crust', consists of a lattice of nuclei and electrons
that globally is electrically neutral. In the absence of accreted material, the
surface of this crust is expected to be made of $^{56}$Fe.
However, on moving towards the interior, more and more neutron 
rich nuclei appear \citep*{ruester06,roca2008,pgc11,hemp2013,wolf13,chamel2015c,utama2016,chamel2017}, 
until at a mean baryon number density $\bar{n}$ of around $2.6 \times 10^{-4}$ 
nucleons fm$^{-3}$ (a mass-energy density $\rho$ of around $4.3 \times 10^{11}$~g~cm$^{-3})$ 
neutron drip sets in (see, e.g., \citealt{cfzh15} for a recent discussion). This marks the 
transition to the `inner crust', an inhomogeneous assembly of neutron-proton 
clusters and unbound neutrons, neutralized by electrons (proton drip can also 
set in at higher densities). By the point where 
$\bar{n}$ has risen to about half the equilibrium density $n_0$ of 
infinite (homogeneous) nuclear matter (INM) that is charge-symmetric (we denote
by SNM this special case of INM), the inhomogeneities have been smoothed out 
and we enter the core of the star. The homogeneous medium of which the core is 
comprised is known as `neutron-star matter' (N*M), and is made up primarily
of neutrons, with an admixture of protons neutralized by electrons and,
at densities above $n\simeq 0.12$ fm$^{-3}$, by muons. At higher 
densities, other particles such as hyperons might 
appear~\citep{hae07,weber2007}, but we do not consider this possibility here.

In this paper we continue our project of developing a unified treatment of 
neutron stars within the framework of the picture of `cold catalysed matter'
(CCM), by which we mean that thermal, nuclear and beta equilibrium prevail at a 
temperature $T$ low enough that thermal effects are negligible for the 
composition and pressure. The equilibrium conditions can 
reasonably be expected to be valid in any neutron star that is not accreting
from a neighbour, but will otherwise fail because of the relative slowness with
which accreted matter acquires nuclear equilibrium. For densities greater than 
around $10^{-10}$ nucleons fm$^{-3}$ (10$^{6}$ g cm$^{-3})$ all the atoms of 
the outer crust are completely ionized and the electrons form a
degenerate gas. Since nuclear degeneracy likewise holds everywhere it follows 
that the CCM picture is valid throughout the star except in a thin layer of 
$\rho\lesssim10^6$ g cm$^{-3}$, where the atomic ionization 
and electron degeneracy can be incomplete. The unified treatment 
of all three regions of neutron stars that we present in this paper therefore 
does not include these outermost parts, which do not in any case involve any 
new nuclear physics, and have been extensively discussed by \citet{hae07}.

Our calculations of the degenerate equation of state (EoS) and the
composition of the three regions are microscopic, and the unifying feature
to which we have alluded lies in the fact that in each region we use 
the energy-density functional theory with the same functional. 
For this we shall take one or other of the 
functionals that we have developed in the last few years not only for the study
of neutron-star structure but also for the 
general purpose of providing a unified treatment of a wide variety of 
phenomena associated with the birth and death of neutron stars, such as 
supernova-core collapse and neutron-star mergers, along with the r-process of 
nucleosynthesis (both in the neutrino-driven wind and via the decompression of 
N*M). These functionals are based on generalized Skyrme-type 
forces and contact pairing forces, the formalism for which is presented in the 
Appendix of \citet*{cgp09}. The first set of such functionals \citep*{gcp10} 
that we devised, labelled BSk19, BSk20 and BSk21, has already been applied 
to a unified treatment of neutron-star structure by 
\citet*{pgc11,cfpg11,pcgd12,pfncpg12,fcpg13}. 

The parameters of these functionals were determined primarily by fitting to 
essentially all the data of the 2003 Atomic Mass Evaluation, 
AME 2003~\citep*{ame03}, with the nuclear masses being
calculated using the Hartree-Fock-Bogoliubov (HFB) method. To do this it was 
necessary to add to the HFB energy phenomenological Wigner terms and correction
terms for the spurious collective energy. Note that our HFB code takes account of axial deformations~\citep{sam02}. For each functional complete
HFB nuclear-mass tables, running from one drip line to the other, and labelled 
HFB-19, HFB-20, and HFB-21, respectively, were constructed \citep{gcp10}. The 
supplementary terms that we had to add to the HFB energy to calculate nuclear masses do
not enter into our calculation of the inner crust and core, the functional 
alone being sufficient, while for the outer crust the entire nuclear dependence
is subsumed into the nuclear mass. Thus this latter region is calculated 
directly from the appropriate mass table (or from experimental mass data, where
available).

Since all the astrophysical applications that we envisage involve a long-range 
extrapolation from experimentally accessible environments to highly 
neutron-rich environments our functionals incorporate as much well established 
theoretical knowledge of neutron-rich systems as possible. The most significant
such constraint that we impose is to require that our nuclear-mass fits be 
consistent, up to the densities prevailing in neutron-star cores, with the 
EoS of homogeneous pure neutron matter (NeuM), as calculated by many-body 
theory from realistic two- and three-nucleon forces. Several such EoSs have
have been published, differing considerably in their stiffness at 
supersaturation densities. Functional BSk19 was fitted to the EoS of
\citet{fp81} (FP), BSk20 to the somewhat stiffer EoS of
\citet*{apr98}, which they label as `A18 + $\delta\,v$ + UIX$^*$' 
and which we refer to as APR, while functional BSk21 was fitted to a still 
stiffer EoS of NeuM, the one labelled `V18' in \citet{ls08}, and which we
refer to as LS2. 

When we came to apply our three EoSs, BSk19-21, to neutron-star structure we
found that the choice of the EoS of NeuM to which we fit our functionals
has a significant impact on the maximum possible mass of neutron stars. In
particular, we found that if we constrained to the EoS of FP then it was
impossible to support the heaviest neutron star that had been 
observed \citep{cfpg11}. On the other hand, there was no such problem with the 
APR and LS2 constraints.

The present paper stems in part from the realization that fitting NeuM and 
nuclear 
masses (along with some other properties of real nuclei such as charge radii) 
does not exhaust completely the degrees of freedom allowed by our generalized 
Skyrme functionals. In particular, it does not tie down the 
symmetry energy of our functionals, defined here as the difference between the 
energy per nucleon in NeuM and the energy per nucleon in SNM,
\beq\label{1.3}
S(n) = e_\mathrm{NeuM}(n) - e_\mathrm{SNM}(n) \quad ,
\eeq
essentially because the fit to nuclear masses does not determine $e_\mathrm{SNM}$
uniquely. We are thus left with a certain flexibility in the choice of the 
symmetry coefficient $J$, defined as follows in an expansion of the energy per 
nucleon of INM of density $n = n_0(1 + \epsilon)$ and charge asymmetry
$\eta = (n_\mathrm{n} - n_\mathrm{p})/n$ (with $n_\mathrm{n}$ and $n_\mathrm{p}$ the neutron and proton number densities respectively) about the equilibrium density $n=n_0$ 
and $\eta = 0$:
\begin{multline}\label{1.4}
e(n, \eta) = a_v + \left(J + \frac{1}{3}L\epsilon\right)\eta^2 +
\frac{1}{18}(K_v + \eta^2K_\mathrm{sym})\epsilon^2\\ + \cdots  \quad .
\end{multline}
(see Eqs.~(5)\,--\,(7) of \citealt*{gcp13}).

Now the functionals BSk19-21 had all been constrained to $J$ = 30 MeV. Thus to 
investigate the degree of freedom remaining on $J$ we constructed a new series 
of functionals constrained to different values of $J$, but fitted
to the same EoS of NeuM, for which we took LS2 \citep{gcp13}. In the meantime 
the 2012 Atomic Mass Evaluation, AME 2012 \citep{ame12}, had become available,
and we fitted the new functionals to the 2353 measured masses of nuclei having 
$N$ and $Z \ge$ 8 given there. In this way we generated four new functionals,
BSk22, BSk23, BSk24 and BSk25 fitted to $J$ = 32, 31, 30 and 29 MeV, 
respectively; we were unable to find acceptable mass fits outside this 
range. 

In addition to the new functionals 
and the new mass data that have become available, another recent 
development has been a significant improvement in our code for computing the 
inner crust, with the inclusion of pairing and the consideration of proton
drip for the first time in any of our work (see Section~\ref{iceos}). This 
makes it worthwhile to revisit 
our earlier work with functionals BSk19-21 relating to the NeuM constraint, 
and so we constructed a fifth new functional, BSk26, constrained like BSk24 to 
$J$ = 30 MeV, but fitted to the softer APR EoS of NeuM. On the other hand,
we do not consider functional BSk23 in the present paper, since it was fitted 
to $J=31$~MeV, and thus is intermediate between BSk22 and BSk24. 

Then to assess the impact of the symmetry energy we compare functionals BSk22,
BSk24 and BSk25, while to assess the impact of the NeuM constraint we compare
BSk24 and BSk26. In Table~\ref{JL} we list for these four functionals the
values of $J$ along with the corresponding values of the symmetry-slope
coefficient $L$, as well as the incompressibility $K_v$ and the
symmetry-incompressibility $K_\mathrm{sym}$. A comparison of BSk22, BSk24 and BSk25
shows the correlation between $J$ and $L$ that has been noted many times
over the last forty years: see, for example, 
\citet*{far78,latt12,ll13,bb16}. This correlation
arises primarily from the fits to nuclear masses, an increase in $J$ being 
offset by the $L$ term in the coefficient of $\eta^2$ in Eq.~(\ref{1.4}) for
subnuclear densities ($\epsilon < 0$), which are dominant for finite nuclei.
However, $J$ cannot determine $L$ uniquely since the EoS must play a role:
while BSk24 and BSk26 have the same $J$, the former has a larger value of $L$, 
as might be expected for a functional fitted to an EoS of NeuM that is stiffer
at higher densities, $\epsilon > 0$. These questions are discussed in greater
detail in Section IIIC1 of \citet{gcp13}.

The quality of the mass fits of the four functionals is shown in 
Table~\ref{tab1.1}. Note particularly that we calculate the deviations with
respect to the data of the 2016 AME \citep{ame16} and not the data to which our
functionals were fitted, those of the 2012 AME~\citep{ame12}. For this reason 
the deviations we show in Table~\ref{tab1.1} are slightly different from those of 
Table IV of \citet{gcp13}. We see that BSk24 gives the best fit
to the masses of the neutron-rich nuclei, which are of greater relevance for 
our purpose. It also is apparent that BSk22 gives a significantly worse fit 
than do the other functionals. 

\begin{table}
\centering
\caption{$J$, $L$, $K_v$ and $K_\mathrm{sym}$ coefficients for the functionals of this
paper (defined in Eq.~(\ref{1.4})).}
\label{JL}
\begin{tabular}{|c|cccc|}\hline
&BSk22&BSk24&BSk25&BSk26\\
\hline
$J$ {[MeV]} &32.0&30.0&29.0&30.0\\
$L$ {[MeV]} &68.5&46.4&36.9&37.5\\
$K_v$ {[MeV]} &245.9&245.5 & 236.0 & 240.8\\
$K_\mathrm{sym}$ {[MeV]} &13.0 & -37.6 & -28.5 & -135.6\\
\hline
\end{tabular}
\end{table}

\begin{table}
\caption{Root-mean-square ($\sigma$) deviations between the 
experimental nuclear mass data (AME 2016, \citealt{ame16}) 
and the predictions for the 
models based on the functionals of this 
paper. The 
first line refers to the 2408 measured masses $M$ with $Z$ and $N \ge$ 8, and 
the second to the masses $M_\mathrm{nr}$ of the subset of 286 neutron-rich nuclei
(neutron separation energy $\le $ 5.0 MeV).}
\label{tab1.1}
\begin{tabular}{|c|cccc|}
\hline\noalign{\smallskip}
&HFB-22&HFB-24&HFB-25&HFB-26\\
\noalign{\smallskip}\hline\noalign{\smallskip}
$\sigma(M)$ {[MeV]}    &0.648 &0.565 &0.556&0.580 \\
$\sigma(M_\mathrm{nr})$ {[MeV]} &0.904 &0.781 &0.829&0.811 \\
\noalign{\smallskip}\hline
\end{tabular}
\end{table}

\begin{figure}
\includegraphics[width=\columnwidth]{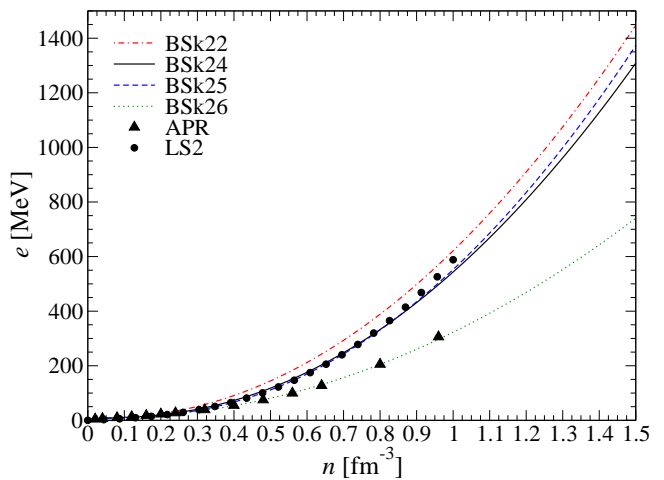}
\caption{(Color online.) 
EoSs for completely degenerate 
neutron matter with functionals of this paper. The points APR refer to the calculations of 
\citet{apr98} labelled as `A18 + $\delta\,v$ + UIX$^*$', while the points LS2 refer to those 
labelled as `V18'  in \citet{ls08}.} 
\label{fig1}
\end{figure}

Fig.~\ref{fig1} shows the EoSs of completely degenerate NeuM for our four 
functionals; it will be seen that while the fits of BSk24 and BSk25 to the LS2 
EoS are excellent, that of BSk22 is less so at high densities, being somewhat
stiffer than LS2. The fit of BSk26 to the APR EoS is very good. 

\begin{figure}
\includegraphics[width=\columnwidth]{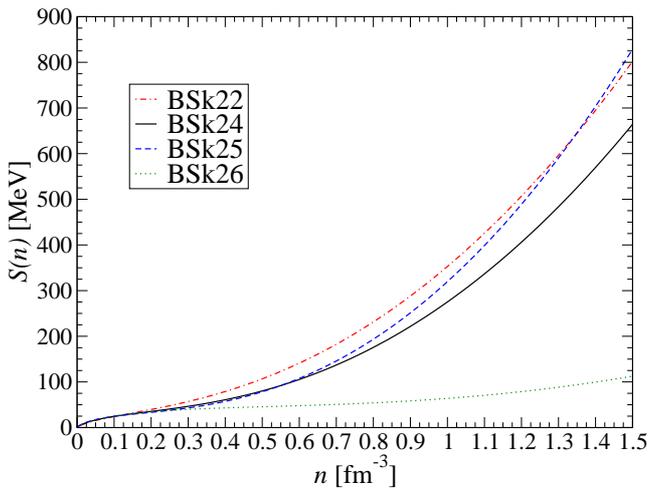}
\caption{(Color online.) Variation of symmetry energy $S(\bar{n})$ (defined in
Eq.~(\ref{1.3})) with density $n$ for functionals of this paper.}
\label{fig2}
\end{figure}

\begin{figure}
\includegraphics[width=\columnwidth]{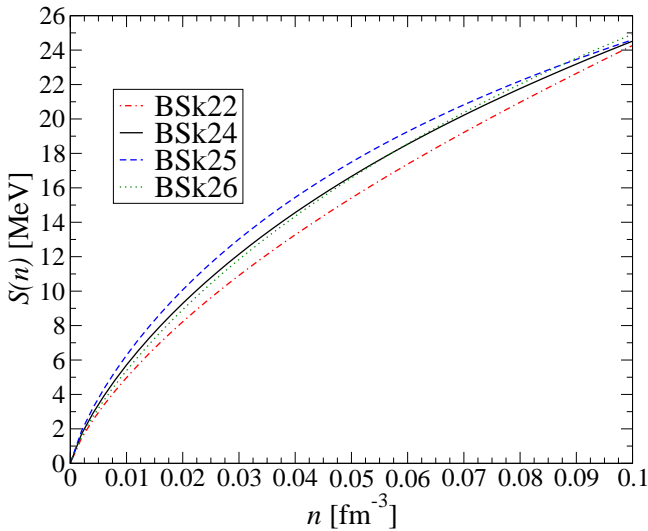}
\caption{(Color online.) Low-density zoom of Fig.~\ref{fig2}.} 
\label{fig3}
\end{figure}

In Figs.~\ref{fig2} and~\ref{fig3} we plot for our four functionals the symmetry
energy defined in Eq.~(\ref{1.3}). Comparing BSk24 and BSk26, we see how at high
densities the symmetry energy of the latter increases much less steeply, as is
appropriate, given the much softer EoS of NeuM to which it has been fitted. On
the other hand, at subnuclear densities the two functionals have virtually 
identical symmetry energy (this is particularly apparent in fig.~2b of 
\citealt{gcp13}), showing that the constraining EoS of NeuM is irrelevant at 
these densities, and that it is the symmetry coefficient $J$ that is the 
determining factor. Comparing now BSk22, BSk24 and BSk25 to investigate the
role of $J$ we see that while there is a correlation between $S(n)$ and $J$ at 
super-nuclear densities (at least up to around 4$n_0$), these quantities are 
anticorrelated at lower densities ($n < n_0$). Moreover, while the functionals
BSk24 and BSk25 behave very similarly in NeuM their symmetry energies start to 
diverge at higher densities. This can only be a result of differing 
$e_\mathrm{SNM}$, a difference
that is easily understood given that the same mass data are being fitted with 
different values of $J$. It will be seen in the course of this paper how far 
these symmetry properties of INM are reflected in the EoSs for the different 
functionals in the various regions of the neutron star.

While the LS2 EoS of NeuM to which we fit functionals BSk22-25 can be 
regarded as typically hard, and the APR EoS to which we fit BSk26 as typically
soft, the question arises as to what extent the `real' EoS of NeuM could lie
outside these limits. Certainly, an EoS much softer than APR, such as FP, 
can be ruled out as long as we neglect the possibility of neutron stars with 
exotic cores~\citep{cfpg13}. Concerning the possibility of an EoS that is
stiffer than LS2 we recall the case of the EoS labelled BOB by \citet{ls08} and
LS3 by \citet{gcp13} and \citet*{gcp16}, but we ruled it out of our 
considerations as probably being too soft at low densities to satisfy the 
constraints imposed by chiral effective field theory \citep{tews13}: see 
the discussion in Section IVA of \citet{gcp16}. At the same time we recall from
the discussion in Section IIIA of \citet{gcp13} that the functionals of this 
paper are consistent with the quantum Monte Carlo calculations 
of~\citet*{gcr12}. In the time since our functionals~\citep{gcp13} were constructed in 2013, a number of new calculations of NeuM using chiral effective field 
theory~\citep*{rogg14,rrap16,dri16,tews16} have appeared. However, 
these calculations were restricted to densities below 0.3 fm$^{-3}$; two 
of them at least~\citep{rogg14,rrap16} appear to favour LS2 
over APR in this low-density regime, where LS2 is \emph{softer} than APR. 
At higher densities we are not aware of any later calculations indicating 
that the real EoS of NeuM is stiffer than LS2.
This is corroborated by recent gravitational-wave observations (see 
Section~\ref{massrad}).

Concerning the value of the symmetry coefficient $J$, we have been unable to 
fit, with the same accuracy, our generalized Skyrme density functionals to the 
mass data with values of $J$ lying outside the range 32 MeV $\ge J \ge $29 MeV;
indeed, $J$ = 32 MeV might be too high. On the other hand, many relativistic 
mean-field (RMF) functionals are characterized with significantly higher values
of $J$. However, the high values quoted do not come from mass fits at all, but 
rather from measurements on giant dipole resonances, pygmy resonances or 
neutron-skin thicknesses. The point is that so far it has not been possible to 
fit masses in the RMF framework with a precision at all comparable to what has 
been achieved with Skyrme functionals, and as a result the quality of the
fits becomes relatively insensitive to $J$. It should nevertheless be noted 
that the best RMF mass fits, those of Ref.~\citet*{pen16}, for which rms 
deviations of about 1.2 MeV were found for 2353 masses, have $J$ = 30 or 31
MeV.

Our methods of calculation for the three different regions of neutron stars are
presented in Section~\ref{methods}, where we summarize much of our work from 
earlier papers and describe our current refinements, particularly with regards
to proton drip. Section~\ref{results} presents our numerical results for the 
three different regions of neutron stars in both graphical and tabular form
(extensive tables relating to the inner crust will be found in the 
supplementary material). 
In Section~\ref{gross} we apply our EoSs to some gross properties of neutron 
stars, namely the mass-radius relation, the maximum mass and the direct Urca 
process. Our conclusions are summarized in Section~\ref{concl}. In     
Appendix~\ref{app:mung} we derive a result used to calculate the chemical 
potentials of neutrons and protons, while our treatment of leptonic gases 
appears in Appendix~\ref{app:lepton}. Finally, Appendix~\ref{analytic} presents
what is in many ways the central feature of this paper: in order to facilitate 
the application of our results to the modelling of neutron-star structure they 
are fitted by analytic parametrizations.

\section{Methods of calculation}
\label{methods}

\subsection{Outer crust}\label{outer}

As discussed in the Introduction, we calculate the outer crust only for
densities $\rho \gtrsim 10^6$~g~cm$^{-3}$ ($\bar{n} \gtrsim  10^{-9}$ 
nucleons~fm$^{-3}$). We also 
suppose that the temperature is not only low enough for complete degeneracy
to hold, but is also lower than the crystallization temperature, so that the
nuclei of the outer crust can be supposed to be arranged in a regular crystal 
lattice.
It is easy to check (see, e.g., Chapter~2 of \citealt{hae07}) that for $^{56}$Fe
this is true at $\rho\gtrsim3\times10^6\,(T/10^8\mbox{~K})^3$
g~cm$^{-3}$.
 For simplicity, we suppose the crystal lattice to be made of only 
one type of ion ($A, Z$) with proton number $Z$ and mass number
$A$, for which the most stable crystal structure is the body-centred cubic
(bcc) lattice (see \citealt{chf2016mix} for a recent discussion on multinary
compounds in the outer crust of neutron stars).

We calculate the EoS and the composition of the outer crust mainly as described
in detail by \citet{pgc11}. To recall the essentials, in a region of mean
density $\bar{n}$ let the equilibrium nucleus have mass number
$A$ and atomic number $Z$. Then the energy per nucleon in that region is
\beq\label{2.1}
e_\mathrm{eq} = M^\prime(A,Z)c^2/A + \mathcal{E}_\mathrm{e}(n_\mathrm{e})/\bar{n} +E_L(A, Z)/A  - 
M_\mathrm{n}c^2   . \!\!\!
\eeq
Here $M^\prime(A,Z)$ is the \emph{atomic} mass for the element ($A, Z$) with 
the binding energy of the atomic electrons subtracted, $\mathcal{E}_\mathrm{e}(n_\mathrm{e})$ is
the energy density of the electrons (without the electron mass), $E_L(A, Z)$ 
is the lattice energy per nucleus, and the constant term in the neutron mass 
$M_\mathrm{n}$ is subtracted out for convenience, as in our treatment of the inner 
crust and the core. Note particularly that this missing neutron mass has to be
restored if we want the corresponding total mass-energy density, thus
\beq\label{2.2}
\rho = \bar{n}\left(e_\mathrm{eq}/c^2 + M_\mathrm{n}\right)  \quad ;
\eeq
and likewise in Eq. (\ref{geq}) below.

For the first of the quantities in Eq.~(\ref{2.1}) we have, in units of MeV,
\begin{multline}
\label{2.3}
M^\prime(A,Z)c^2 = M(A,Z)c^2 + 1.44381\times 10^{-5}\,Z^{2.39}\\ +
1.55468\times 10^{-12}\,Z^{5.35}
\end{multline}
(see Eq.~(A4) of \citealt*{lpt03}), in which $M(A,Z)$ is just the usual atomic 
mass. We use for this latter quantity the HFB-22, HFB-24 HFB-25 or HFB-26 
tables available on the \textsc{bruslib} database~\citep{bruslib}, except when 
experimental values are available, for which we use the 2016 Atomic Mass 
Evaluation \citep{ame16}, supplemented by the very recent measurements of 
copper isotopes \citep{welker2017}. 

For the energy density 
$\mathcal{E}_\mathrm{e}(n_\mathrm{e})$ of the electrons ($n_\mathrm{e} = Z\bar{n}/A$), we now, unlike 
\citet{pgc11}, use complete expressions for the electron exchange and 
screening (also sometimes referred to as 
`polarization') corrections to the electron energy density, as discussed in
Appendix~\ref{app:lepton}. The electron-correlation energy is included as in 
\citet{pgc11}. The lattice energy $E_L(A, Z)$ is calculated as in 
\citet{pgc11}, with quantum zero-point and finite nuclear-size corrections.

The EoS and the composition are now determined by minimizing, at fixed pressure
$P$, the Gibbs free energy per nucleon
\begin{equation}\label{geq}
   g = e_\mathrm{eq} + M_\mathrm{n}c^2 + \frac{P}{\bar{n}} \quad .
\end{equation}
 The 
pressure is determined from the energy $e_\mathrm{eq}$, as described 
by \citet{pgc11} and in Appendix~\ref{app:lepton}.

We calculate the neutron chemical potential using the identity
\beq\label{mung}
\mu_\mathrm{n} = g    \quad ,
\eeq
valid only because of the beta equilibrium that holds throughout the
neutron star (see \citealt*{bps71} and Appendix~\ref{app:mung}). Because of this 
beta equilibrium we can then calculate the proton chemical potential through
\beq\label{mupg}
\mu_\mathrm{p} = \mu_\mathrm{n} - \mu_\mathrm{e}   \quad , 
\eeq
where $\mu_\mathrm{e}$ is the electron chemical potential (see 
Appendix~\ref{app:lepton}). 

The minimization at constant pressure of the Gibbs free energy per nucleon $g$
is repeated with increasing value of the pressure until neutron drip sets in,
the condition for which is 
\beq \label{drip}
\mu_\mathrm{n} = g = M_\mathrm{n} c^2
\eeq
(this is equivalent to Eq.~(2.7.3) of \citealt{st83}). 

\subsection{Inner crust}
\label{iceos}

Since the theoretical nuclear masses used in Section~\ref{outer} for the outer 
crust were derived from our functionals by application of the HFB method, it 
might seem appropriate to use this same method for the inner-crust calculations
also. As in the pioneer HF calculations of \citet{nv73},
such calculations are generally performed within the framework of the 
spherical Wigner-Seitz (WS) approximation \citep*{mmm04,bst07,grill}. But an 
inevitable consequence of the WS approximation is to introduce shell effects 
in the spectrum of unbound neutron states, which dominate the properties of 
the inner crust. Such shell effects are to a large extent spurious, since in 
reality the unbound neutron states form a quasi-continuum, as shown by 
band-structure calculations \citep{ch05,ch06,ch12}. 
This difficulty is analysed in detail by \citet*{cha07,marg07}, the latter 
reference showing that the error thereby introduced in the energy per nucleon 
cannot easily be reduced below 50 keV, which is incompatible with a reliable 
calculation of the composition of the inner crust. 
\citet{pastore2017} have recently proposed a new method to 
reduce the errors to a few keV by considering a supercell with a fixed radius 
of 80 fm. But they also stressed that the WS treatment becomes unreliable at 
densities above 0.02 fm$^{-3}$. 
In the last few years 3D calculations (mostly at finite temperatures and  
fixed proton fractions) have been carried out \citep*{mag02,gog07,new09,
pais12,skimmr13,skimmr14,pais14,skismmr15,sffph16}. 
However, the use of a cubic box with periodic boundary conditions can still 
lead to spurious shell effects (see, for example, Section IIC2 in 
\citealt{new09}) that can only be removed by choosing a large enough 
box \citep*{seb09,seb11} or imposing Bloch boundary conditions \citep{ch12,sn15}; 
moreover such calculations require computer times that are quite impractical 
for the applications that we are undertaking here. 

In view of these problems it is not surprising that a more popular approach to
the calculation of the inner crust has been to use the much simpler
compressible liquid-drop model; a typical such calculation is that of
\citet{dh01}. Within each WS cell this method makes a clear
separation of INM into two distinct homogeneous phases, the
densities of which are free parameters of the model. The bulk properties of
the two phases are calculated microscopically using the adopted functional,
as are the surface properties (preferably including curvature corrections) of
the interface between them. A more accurate treatment of spatial
inhomogeneities is to employ semi-classical methods such as the Thomas-Fermi
(TF) approximation, as for instance in \citet{oya07,oka13,bcpm}, 
or its extension to second order in $\hbar^2$ by~\citet{mu15} (also private
communication from N.~Martin). 
However, none of these calculations makes any shell correction.

The shortcomings of all these different methods are avoided in our application 
of the ETFSI (fourth-order extended Thomas-Fermi plus Strutinsky integral) method to the 
calculation of the EoS of the inner crust \citep*{dut04,ons08,pcgd12}. In these 
papers we extend a method that was originally developed as a fast approximation
to a Skyrme-HF-BCS treatment of finite nuclei \citep{dut86}: it was based on a 
full ETF treatment of the kinetic-energy and spin current densities, with 
neutron and proton shell corrections added perturbatively. When compared to 
exact HF-BCS calculations performed with the same Skyrme and pairing forces it 
was found to overbind by 170 keV/nucleon at most (for $^{16}$O), and only a few 
tens of keV/nucleon for heavier elements. For comparison, zero-order TF 
calculations with parametrized distributions
(whose parameters are fully determined by $Z$, $A$, and nuclear-matter properties)
lead to errors of order 0.5-1 MeV/nucleon \citep{papa13}. The extension to
second order reduces the errors below 300 keV/nucleon \citep*{aymard14}. 

However, for most purposes,
what counts is not the absolute binding energy but the \emph{differences}
in energy between different configurations, e.g., different nuclei, or WS cells
of different composition. Viewed in this light the ETFSI method emerges 
as a much better approximation to the Skyrme-HF-BCS method for finite nuclei.
Referring to the 69 spherical nuclei listed in Table 2 of \citet{dut86}, 
we took the differences of 48 pairs (in most cases giving thereby nucleon 
separation or beta-decay energies), and found that the greatest discrepancy 
between the ETFSI and HF results was 5 keV per nucleon, and generally was much 
less. 

When the ETFSI method is applied, as here, to the EoS problem, it allows, like
the TF method, for a continuous variation of the density of nuclear matter
within each spherical WS cell, without any artificial separation into two 
distinct phases. However it is expected to provide a much better description of
nuclear clusters than does the TF method, since the semi-classical expressions 
for the kinetic-energy and spin current densities include density-gradient 
terms up to the fourth order. Moreover, our application of the ETFSI 
method to the EoS problem \citep{dut04,ons08,pcgd12} includes proton shell 
corrections. On the other hand, we do not calculate the neutron shell corrections  
since they are known to be much smaller than the proton 
shell corrections when there are unbound neutrons \citep{oy94,ch06,cha07}; the 
error thereby introduced is negligible compared to the spuriously large neutron
shell corrections that arise in current implementations of the HF equations, 
as noted above. Also, while our earlier ETFSI calculations of the 
EoS \citep{dut04,ons08,pcgd12} did not include pairing, \citet{pcpg15} 
showed how to include, for zero temperature, a BCS treatment of proton
pairing, and  we follow that procedure here.  
It is to be noted that the errors of a few keV per nucleon incurred by adopting 
the BCS approximation (instead of solving the full HFB equations) lie within 
the errors of the ETFSI approach \citep*{pastore16}. 

To recall the main features of our method, we write the total energy density at 
any point in the inner crust as
\begin{multline}
\label{3.-1}
{\mathcal E}(\bm{r}) = {\mathcal E}_\mathrm{nuc}(\bm{r}) + 
{\mathcal E}_\mathrm{ee,ep}^\mathrm{c}(\bm{r}) + {\mathcal E}_\mathrm{e}(\bm{r})\\
 + 
\Big[\bar{n}_\mathrm{n}M_\mathrm{n} + \bar{n}_\mathrm{p}(M_\mathrm{p} + m_\mathrm{e})\Big]c^2  \quad ,
\end{multline}
where $M_\mathrm{n}, M_\mathrm{p}$ and $m_\mathrm{e}$ denote the masses of the neutron, proton and
electron, respectively. The 
first term here can be broken up into a term corresponding to the 
generalized Skyrme force (in which we include the nucleonic kinetic
energy) and a proton-proton Coulomb term,
\beq\label{3.-2}
{\mathcal E}_\mathrm{nuc} = {\mathcal E}_\mathrm{Sky} + {\mathcal E}_\mathrm{pp}^\mathrm{c} 
\quad ,
\eeq  
while the second term represents the total direct part of the Coulomb 
interactions of the electrons, i.e., the $e - e$ and $e - p$ interactions.
The third term, which is purely electronic, contains both kinetic-energy and 
Coulomb-exchange terms: see Appendix~\ref{app:lepton}. Bracketing together the two 
Coulomb terms,
${\mathcal E}_\mathrm{pp}^\mathrm{c}+{\mathcal E}_\mathrm{ee,ep}^\mathrm{c}={\mathcal E}_\mathrm{c}$,
we can rewrite Eq.~(\ref{3.-1}) as 
\beq\label{3.-3}
{\mathcal E}(\bm{r}) = {\mathcal E}_\mathrm{Sky}(\bm{r}) + 
{\mathcal E}_\mathrm{c}(\bm{r}) + {\mathcal E}_\mathrm{e}(\bm{r})
+ \bar{n}(M_\mathrm{n}c^2 - Y_\mathrm{p}Q_\mathrm{n,\beta}) \quad ,
\eeq
where $Y_\mathrm{p} = Z/A$ with $Z$ and $A$ the total number of protons and nucleons in 
the WS cell respectively, and $Q_\mathrm{n,\beta}$ is the $\beta$-decay energy of the neutron 
(= 0.782 MeV). Then for the total energy per nucleon we have, after integrating
over the WS cell of volume $V_\mathrm{cell}$,
\beq\label{3.-4}
e=\frac{1}{A} \int_{V_\mathrm{cell}}{\mathcal E}(\bm{r})\,\mathrm{d}^3r
 = e_\mathrm{Sky} + e_\mathrm{c} + e_\mathrm{e} - Y_\mathrm{p}\,Q_\mathrm{n,\beta} \quad ,
\eeq
in which we have dropped for convenience the constant term $M_\mathrm{n}c^2$, as we do for 
the outer crust and the core.

The following salient points are to be noted for the calculation of the first 
two terms. With spherical symmetry being assumed the neutron
and proton density distributions within the WS cells are parametrized as the 
sum of a constant `background' term and a `cluster' term according to 
\beq\label{3.1}
n_q(r) = n_{\mathrm{B}q} + n_{\Lambda q}f_q(r)  \quad ,
\eeq
in which, with $q = n$ or $p$, 
\beq\label{3.2}
f_q(r) = \frac{1}{1 + \exp \left[\Big(\frac{C_q - R}
{r - R}\Big)^2 - 1\right] \exp \Big(\frac{r-C_q}{a_q}\Big) }
\eeq
(if $n_{\Lambda q}$ is negative the cluster becomes a bubble).
This `damped' modification of the usual simple Fermi profile is introduced in
order to satisfy the requirement of the usual implementation of the ETF method 
that the first three derivatives of the density vanish at the cell surface
(see section~II of \citealt{pcgd12}, where other relevant details
will be found). At the same time, a smooth matching of the nucleonic 
distributions between adjacent cells is ensured.

With $Z$ protons and $N$ neutrons in the cell, the cell radius $R$ is
determined by the mean density $\bar{n}$ through
\beq\label{3.2A}
A =V_\mathrm{cell}\bar{n}=
 \frac{4\pi}{3}R^3\,\bar{n}      \quad ,
\eeq
where $A = Z + N$. The eight independent cell parameters that are left, four
for each charge type, will then be constrained by
\bmlet
\beq\label{3.2Ba}
Z = 4\pi\int_0^R r^2n_\mathrm{p}(r)\,\mathrm{d}r  
\eeq
and 
\beq\label{3.2Bb}
N = 4\pi\int_0^R r^2n_\mathrm{n}(r)\,\mathrm{d}r  \quad   .
\eeq
\emlet
It is convenient now to define a proton-cluster number and a neutron-cluster
number through
\bmlet
\beq\label{3.2Ca}
Z_\mathrm{cl} = 4\pi\,n_\mathrm{\Lambda p}\int_0^R r^2f_\mathrm{p}(r)\,\mathrm{d}r
\eeq
and
\beq\label{3.2Cb}
N_\mathrm{cl} = 4\pi\,n_\mathrm{\Lambda n}\int_0^R r^2f_\mathrm{n}(r)\,\mathrm{d}r   \quad   .
\eeq
\emlet 

Using the density distribution (\ref{3.1}),  
one can calculate \citep{cgp09} an ETF energy density,
${\mathcal E}_\mathrm{Sky}^\mathrm{ETF}(r)$
for the 
Skyrme force, whence for the ETF energy per nucleon we have
\beq\label{3.3}
e_\mathrm{Sky}^\mathrm{ETF} = \frac{4\pi}{A}\int_0^R r^2
{\mathcal E}_\mathrm{Sky}^\mathrm{ETF}(r)\,\mathrm{d}r  \quad . 
\eeq
Since we now include pairing as well as shell corrections we have 
\beq\label{3.4}
e_\mathrm{Sky} = e_\mathrm{Sky}^\mathrm{ETF} + \frac{1}{A}\left(E^\mathrm{sc, pair}_\mathrm{p}  + 
E_\mathrm{pair}\right) \quad ,
\eeq
in which $E^\mathrm{sc, pair}_\mathrm{p}$ is the Strutinsky-integral shell correction, as 
modified by pairing, and $E_\mathrm{pair}$ is the BCS energy (see Eqs.~(5) and~(6),
respectively, of \citealt{pcpg15}).

For the direct part of the Coulomb term $e_\mathrm{c}$ in Eq.~(\ref{3.-4}) we have
\beq\label{3.5}
{\mathcal E}^\mathrm{c} =  \frac{e^2}{2} (n_\mathrm{p}(\bm{r}) - n_\mathrm{e})
\int\frac{n_\mathrm{p}(\bm{r}^\prime) - n_\mathrm{e}}
{|\bm{r} - \bm{r^{\prime}}|}\,\mathrm{d}^3\bm{r^{\prime}}  \quad   ; 
\eeq
there is no protonic Coulomb exchange term for the functionals of this paper 
(see Sect~II of \citealt{gcp13}). Both the direct and exchange parts of 
the electronic term $e_\mathrm{c}$ are discussed in Appendix~\ref{app:lepton}. 

For a given mean density $\bar{n}$ nucleons per unit volume the total energy 
$e$ per nucleon has to be minimized with respect to eight parameters: $N$, $Z$
and six of the eight geometric parameters defined in Eqs.~(\ref{3.1}) and 
(\ref{3.2}). In making this minimization care has to be taken that in
allowing for bubble configurations we do not encounter negative densities at
any point for either charge type of nucleon: see \citet*{opp97}. 

Insofar as shell corrections are included for protons, $Z$ has to 
be discretized to integral values in the minimization, but $N$ is treated as 
a continuous variable. Even though the total number of neutrons in the crustal
layer is integral, the notion of a fractional number of neutrons per WS cell 
corresponds to the physical reality, since the neutrons are delocalized. Then 
with $Z$ being held fixed at different integral values automatic minimization 
is performed with respect to $N$ and six geometric parameters. 

Since our 2012 paper \citep{pcgd12} we have made significant improvements to the
code that calculates the inner crust. It will be recalled that at higher 
densities close to the point of transition to a homogeneous medium our search 
routine broke down and was unable to find any solution, when minimizing with
respect to all seven parameters. We have now rectified 
this problem with our code, and can in fact find solutions at densities known 
to correspond to homogeneous nuclear matter. To quantify this capability it is
convenient to use as a global measure of the departure from homogeneity
the `inhomogeneity factor'
\beq\label{3.15}
\Lambda = \frac{1}{V_\mathrm{cell}}\int \mathrm{d}^3\bm{r}\left(\frac{n(\bm{r})}{\bar{n}}
-1\right)^2 \quad ,
\eeq
where the integration goes over
the cell. At the interface between the inner and outer crusts we find values of
around 400, but at densities corresponding to homogeneous nuclear matter we find
values of $\Lambda$ of the order of 10$^{-7}$, or even less, values that
are appropriately small, suggesting that our code is working correctly even
at densities higher than for which it was intended.

\emph{Proton drip.} Now that our inner-crust code works properly right into the
homogeneous region we can address the problem of proton drip, which we 
neglected in our earlier work, but which becomes significant at higher 
densities. At all values of $\bar{n}$ in the inner crust,
there is a non-vanishing proton density at the surface of the WS 
cell, but it rises sharply when $\bar{n}$
is sufficiently high for the motion of the highest-energy proton
to become infinite in the classicle single-particle approximation,
that is when
\beq\label{3.6a}
t(r=0) > U_\mathrm{p}(r=R) - U_\mathrm{p}(r=0)   \quad  ,
\eeq
where $t(r=0)$ is the maximum proton kinetic energy at the centre 
of the cell 
and $U_\mathrm{p}(r)$ is the value of the proton
single particle effective potential
  at the point $r$. For the
former quantity we can safely use the TF expression at the centre of the cell,
thus
\beq\label{3.6b}
t(r=0) = \frac{\hbar^2}{2M^*_\mathrm{p}(r=0)}\left[3\pi^2n_\mathrm{p}(r=0)\right]^{2/3} 
\quad ,
\eeq
where $M^*_\mathrm{p}(r=0)$ is the effective proton mass at the centre of the 
cell (see the Appendix of \citealt{cgp09}). 
Once this proton drip has begun the
unbound proton single particle states will form a quasi-continuum, and proton shell
effects will largely vanish, exactly as do neutron shell effects at all
densities in the inner crust, i.e., beyond the neutron-drip point. For values
of $\bar{n}$ below the proton drip point, as defined by the condition 
(\ref{3.6a}), the shell correction $E^\mathrm{sc, pair}_\mathrm{p}$ of Eq.~(\ref{3.4}) is kept
in its entirety, along with the BCS energy $E_\mathrm{pair}$; otherwise \emph{both}
terms are dropped completely. This sharp cutoff of the shell correction
probably exaggerates the actual situation, but it is not devoid of all
physical sense, and we see no simple way of smoothing it that would not 
introduce an arbitrary element. Simultaneously dropping $E_\mathrm{pair}$ is more 
contentious, and is done for simplicity; nevertheless, since pairing 
contributes to the shell correction $E^\mathrm{sc, pair}_\mathrm{p}$ it would not be entirely 
logical to drop the latter and not $E_\mathrm{pair}$. In any case, $E_\mathrm{pair}$ 
will be quite small, since the gas of free protons will be much more dilute 
than the system of bound protons, with the result that the pairing field is 
much weaker.

Once protons begin to unbind and we drop shell corrections, $Z$ is no longer
restricted to integral values and in principle should be treated, like $N$, as a
continuous variable. However, since it is more reliable to minimize with respect
to seven than eight variables, we discretize $Z$ in intervals of 0.1. 

\emph{Pressure.} The pressure in the inner crust at any mean density $\bar{n}$ 
is calculated as described in Appendix B of \citet{pcgd12}. Briefly, we can
write the pressure as a sum of nucleonic and electronic terms, 
\beq\label{B25}
P = P_\mathrm{nuc}+P_\mathrm{e}   \quad ,
\eeq
in which both the direct and exchange contributions to the electronic term are 
found in Appendix B of the present paper, while the nuclear term, which 
contains contributions from both neutrons and protons, is given by  
the simple expression~(B28) of \citet{pcgd12}. We point out here that there 
is a misprint in Eq.~(B30b) of this latter reference, which should read
\beq
C_1^n=-\frac{1}{4}t_0\left(\frac{1}{2}+x_0\right)-
\frac{1}{24}t_3\left(\frac{1}{2}+x_3\right)n_{B0}^\alpha     \quad ;
\eeq
the calculations of \citet{pcgd12} were made with this correct expression.
Note that the second term of Eq.~(B27) of \citet{pcgd12} vanishes here
because we drop the Coulomb-exchange terms for protons.    

This procedure is more reliable and much faster than 
applying numerical differentiation to the identity
\beq\label{3.7}
P = \bar{n}^2\left(\frac{\partial e}{\partial \bar{n}}\right)_{T, Y_\mathrm{p}} \quad ,
\eeq
in which it is understood that electrical neutrality is maintained.

\emph{Chemical potentials.} The nucleonic chemical potentials $\mu_\mathrm{n}$ and 
$\mu_\mathrm{p}$ are calculated as in the outer crust, i.e., by using Eqs.~(\ref{mung}) 
and~(\ref{mupg}). Note that these equations can only be applied when
beta equilibrium holds.

\emph{Non-spherical configurations.} It should be stressed that in both this and
our previous papers we restrict ourselves to WS cells that are spherically 
symmetric. However, many, but not all, calculations, beginning with the work
of \citet*{rpw83} and \citet*{hashi84} (see also Section 3.3 of \citealt{lrr} 
and Section 3.4.2 of \citealt{hae07})
suggest that as the interface with the core is approached 
there might be a slight energetic preference for non-spherical shapes, referred
to collectively as `pasta'. Of especial relevance to our own work is that of 
\citet{mu15}, in which ETF calculations without shell corrections were 
performed over a wide range of densities with a number of Skyrme functionals, 
including BSk22 and BSk24. These authors report a transition to non-spherical shapes at densities slightly below 0.06 fm$^{-3}$ in the inner crust, but their
conclusions are sensitive to the approximations made since the energy 
differences involved are very small. In particular, the parametrization that 
they adopted for the nucleon distributions does not satisfy the necessary 
boundary condition that the first three derivatives vanish at the cell surface 
(see the discussion after Eq.~(\ref{3.2}), and also in Section~\ref{cci}). 
Moreover, their ETF calculations are limited to second 
order \citep{mu15}, and it is known from work on fission barriers that the 
fourth-order terms can make departure from sphericity energetically expensive: 
see section~4.3 of \citet*{bgh85}. We have confirmed that the fourth-order 
terms become increasingly positive as the deformation increases, using a code 
based on \citet{tond87} that aims to extend the present work to deformed WS
cells but which has not yet been fully developed.

Adding further doubt to the existence of pasta,
\citet{vinas17} have recently performed self-consistent TF calculations in 
which the Euler-Lagrange equations were solved without any parametrization of 
the density distributions, and have found no pasta for several Skyrme 
functionals. In particular, their calculations predict that spherical clusters 
remain stable throughout the inner crust for SLy4, in disagreement with the 
results obtained by \citet{mu15}. However, our code for deformed WS cells
shows that the second-order ETF terms, unlike the fourth-order terms, can 
reduce the deformation energy. Thus it is not clear that the conclusion of 
\citet{vinas17} would survive a full ETF calculation. 

\subsection{Core}
\label{corecalc}

Because of translational invariance a considerable simplification occurs in the
N*M that defines the core, even though muons 
may now be present. To take account of the latter we denote by $Y_\mu$ the 
number of muons per nucleon, where, because of overall charge neutrality,
\beq\label{corea.1}
Y_\mathrm{e} + Y_\mu = Y_\mathrm{p} \quad .
\eeq
Then in place of Eq.~(\ref{3.-4}) we now have for the total energy per nucleon
\beq\label{corea.2}
e = \frac{1}{n}\left(\mathcal{E}_\mathrm{Sky} + {\mathcal E}_\mathrm{lept}\right)
- Y_\mathrm{p}Q_\mathrm{n,\beta} + Y_\mu(m_\mu-m_\mathrm{e})c^2   \quad ,
\eeq
where the leptonic term
 ${\mathcal E}_\mathrm{lept}={\mathcal E}_\mathrm{e}+{\mathcal E}_\mu$
now includes the muonic contribution ${\mathcal E}_\mu$,
 and we have again dropped the neutron mass.
Note that because of exact neutrality at all points there is no direct Coulomb 
term ${\mathcal E}_\mathrm{c}$, while the exchange terms are included in 
${\mathcal E}_\mathrm{lept}$ (see Appendix~\ref{app:lepton}), there being no 
protonic Coulomb exchange term for the functionals of this paper.

A great simplification in the $\mathcal{E}_\mathrm{Sky}$ term of 
Eq.~(\ref{corea.2}) arises, since in the zero-temperature approximation 
considered here it is given
by an analytic expression. Assuming the system to be unpolarized, i.e., if 
time-reversal invariance holds, we have from Eq.~(A13) of \citet{cgp09} 
\begin{multline}\label{corea.3}
\mathcal{E}_\mathrm{Sky} = 
\frac{3\hbar^2}{20}\left[\frac{1}{M_\mathrm{n}}(1+\eta)^{5/3} +
\frac{1}{M_\mathrm{p}}(1-\eta)^{5/3}\right]nk_\mathrm{F}^2    \\
 + \frac{1}{8}t_0\Biggl[3 - (1+2x_0)\eta^2\Biggr]n^2  \\
 + \frac{3}{40}t_1\Biggl[(2+x_1)F_{5/3}(\eta) -
\left(\frac{1}{2}+x_1\right)F_{8/3}(\eta) \Biggr]n^2k_\mathrm{F}^2 \\
 + \frac{3}{40}t_2\Biggl[(2+x_2)F_{5/3}(\eta) +
\left(\frac{1}{2}+x_2\right)F_{8/3}(\eta) \Biggr]n^2k_\mathrm{F}^2 \\
 + \frac{1}{48}t_3\Biggl[3-(1+2x_3)\eta^2\Biggr]n^{\alpha+2} \\
 + \frac{3}{40}t_4\Biggl[(2+x_4)F_{5/3}(\eta) - 
\left(\frac{1}{2}+x_4\right)F_{8/3}(\eta) \Biggr]n^{\beta+2}k_\mathrm{F}^2 \\
 + \frac{3}{40}t_5\Biggl[(2+x_5)F_{5/3}(\eta) + 
\left(\frac{1}{2}+x_5\right)F_{8/3}(\eta) \Biggr]n^{\gamma+2}\,k_\mathrm{F}^2 \quad , 
\end{multline}
where
\beq
\label{corea.4}
k_\mathrm{F}= \left(\frac{3\pi^2n}{2}\right)^{1/3} \quad  ,
\eeq
\beq\label{corea.5}
\eta = \frac{n_\mathrm{n} - n_\mathrm{p}}{n} = 1 - 2Y_\mathrm{p}
\eeq
and
\beq
\label{corea.6}
F_x(\eta)= \frac{1}{2}\Biggl[(1+\eta)^x+(1-\eta)^x\Biggr]\quad .
\eeq
The leptonic term in Eq.~(\ref{corea.2}) is
$
{\mathcal E}_\mathrm{lept} = {\mathcal E}^\mathrm{kin}_\mathrm{e} + {\mathcal E}^\mathrm{kin}_\mu +
{\mathcal E}^\mathrm{ex}_\mathrm{e} + {\mathcal E}^\mathrm{ex}_\mu
$,
in which all quantities are calculated as described in 
Appendix~\ref{app:lepton}.

For the total pressure we can now write
\beq\label{corea.8}
P = P_\mathrm{Sky} + P_\mathrm{lept} \quad .
\eeq
For the first term here we have
\begin{multline}\label{corea.9}
P_\mathrm{Sky} \equiv - \mathcal{E}_\mathrm{Sky} + 
n\left(\frac{\partial \mathcal{E}_\mathrm{Sky}}{\partial n}\right)_{\eta}  \\ 
=\frac{\hbar^2}{10}\left[\frac{1}{M_\mathrm{n}}(1+\eta)^{5/3}+
\frac{1}{M_\mathrm{p}}(1-\eta)^{5/3}\right]nk_\mathrm{F}^2    \\
+ \frac{1}{8}t_0\Biggl[3 - (1+2x_0)\eta^2\Biggr]n^2  \\
+ \frac{1}{8}t_1\Biggl[(2+x_1)F_{5/3}(\eta) -
\left(\frac{1}{2}+x_1\right)F_{8/3}(\eta) \Biggr]n^2k_\mathrm{F}^2 \\
 + \frac{1}{8}t_2\Biggl[(2+x_2)F_{5/3}(\eta) +
\left(\frac{1}{2}+x_2\right)F_{8/3}(\eta) \Biggr]n^2k_\mathrm{F}^2 \\
+ \frac{(\alpha + 1)}{48}t_3\Biggl[3-(1+2x_3)\eta^2\Biggr]n^{\alpha+2} \\
+ \frac{3\beta + 5}{40}t_4\Biggl[(2+x_4)F_{5/3}(\eta) -
\left(\frac{1}{2}+x_4\right)F_{8/3}(\eta) \Biggr]n^{\beta+2}k_\mathrm{F}^2 \\
+ \frac{3\gamma + 5}{40}t_5\Biggl[(2+x_5)F_{5/3}(\eta) + 
\left(\frac{1}{2}+x_5\right)F_{8/3}(\eta) \Biggr]n^{\gamma+2}\,k_\mathrm{F}^2 \quad ,
\\
\end{multline}
and for the second
\beq\label{corea.10}
P_\mathrm{lept} = P_\mathrm{e}^\mathrm{kin} + P_{\mu}^\mathrm{kin}
 + P_\mathrm{e}^\mathrm{ex} + P_{\mu}^\mathrm{ex} \quad ,
\eeq
where the first two terms are given by Eq.~(\ref{lep.4}) and the
last two terms by Eq.~(\ref{eq:p-exc}).

Because there is no Coulomb term coupling leptons with protons the
nucleonic chemical potentials in the core will depend only on 
$\mathcal{E}_\mathrm{Sky}$. Given that we are dealing with a homogeneous system, 
the neutron chemical potential becomes
\bmlet
\begin{align}\label{corea.11a}
\mu_\mathrm{n}^\mathrm{Sky} - M_\mathrm{n}c^2 &=
\left(\frac{\partial \mathcal{E}_\mathrm{Sky}}{\partial n_\mathrm{n}}\right)_{n_\mathrm{p}} 
\nonumber\\
&= e_\mathrm{Sky}(n, \eta) + 
n\left(\frac{\partial e_\mathrm{Sky}(n, \eta)}{\partial n_\mathrm{n}}\right)_{n_\mathrm{p}}
\nonumber \\
& = e_\mathrm{Sky} + 
n\left[\left(\frac{\partial e_\mathrm{Sky}}{\partial n}\right)_{\eta} 
+ \left(\frac{\partial e_\mathrm{Sky}}{\partial \eta}\right)_n
\left(\frac{\partial \eta}{\partial n_\mathrm{n}}\right)_{n_\mathrm{p}}\right] \nonumber \\
& = e_\mathrm{Sky} + \frac{1}{n}P_\mathrm{Sky} + 
(1 - \eta)\left(\frac{\partial e_\mathrm{Sky}}{\partial \eta}\right)_n  \quad .
\end{align}
Likewise for protons we have
\beq\label{corea.11b}
\mu_\mathrm{p}^\mathrm{Sky} - M_\mathrm{p}c^2 = e_\mathrm{Sky} + \frac{1}{n}P_\mathrm{Sky} -
(1 + \eta)\left(\frac{\partial e_\mathrm{Sky}}{\partial \eta}\right)_n \quad .
\eeq
\emlet
(These equations can be obtained as the homogeneous limit of Eq.~(B20) of
\citealt{pcgd12}.) In these expressions
 there appears the following partial derivative: 
\begin{multline}\label{corea.12}
\left(\frac{\partial e_\mathrm{Sky}}{\partial \eta}\right)_n =
\frac{\hbar^2}{4}\left[\frac{1}{M_\mathrm{n}}(1+\eta)^{2/3}-
\frac{1}{M_\mathrm{p}}(1-\eta)^{2/3}\right] k_\mathrm{F}^2\\
-\frac{1}{4}t_0\,(1+2x_0)\eta\,n \\
+\frac{1}{40}t_1\,\Big[-(10+5x_1)F_{2/3}(\eta) +
(14+13x_1)F_{5/3}(\eta)\\~~~
- (4+8x_1)F_{8/3}(\eta) \Big]\,nk_\mathrm{F}^2/\eta \\
+\frac{1}{40}t_2\,\Big[-(10+5x_2)F_{2/3}(\eta) +
(6-3x_2)F_{5/3}(\eta) \\~~~
 + (4+8x_2)F_{8/3}(\eta) \Big]\,nk_\mathrm{F}^2/\eta \\
-\frac{1}{24}t_3(1+2x_3)\eta\,n^{\alpha+1} \\
+\frac{1}{40}t_4\,\Big[-(10+5x_4)F_{2/3}(\eta) 
+(14+13x_4)F_{5/3}(\eta) \\~~~
-(4+8x_4)F_{8/3}(\eta) \Big]\,n^{\beta+1}k_\mathrm{F}^2/\eta \\
+\frac{1}{40}t_5\,\Big[-(10+5x_5)F_{2/3}(\eta) +
(6-3x_5)F_{5/3}(\eta)  \\~~~
+(4+8x_5)F_{8/3}(\eta) \Big]\,n^{\gamma+1}k_\mathrm{F}^2/\eta \quad .
\end{multline}
For the leptonic chemical potentials in the core see Appendix~\ref{app:lepton}.

The equilibrium composition at any given nucleonic density $n$ is characterized
by the parameters $Y_\mathrm{e}$ and $Y_\mu$, but rather than minimize numerically the 
total energy per nucleon $e$, as given by Eq.~(\ref{corea.2}), with respect to
these two parameters we solve the equations of beta equilibrium
\bmlet
\beq\label{corea.13a}
\mu_\mathrm{p}(n, Y_\mathrm{p}) + \mu_\mathrm{e}(n,Y_\mathrm{e}) = \mu_\mathrm{n}(n, Y_\mathrm{p})
\eeq
and 
\beq\label{corea.13b}
\mu_\mathrm{p}(n, Y_\mathrm{p}) + \mu_\mu(n,Y_\mu) = \mu_\mathrm{n}(n, Y_\mathrm{p})  \quad ,
\eeq
\emlet
where $Y_\mathrm{p}$ is given by the neutrality condition~(\ref{corea.1}).

\section{Results}
\label{results}

\subsection{Outer crust}
\label{outres}

\begin{figure}
\includegraphics[width=\columnwidth]{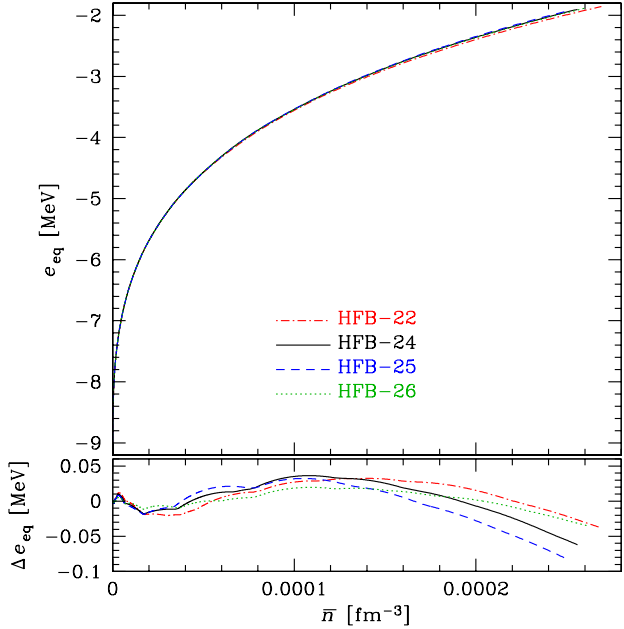}
\caption{(Color online.) Upper panel: Computed energy per 
nucleon $e_\mathrm{eq}$ as a function of mean baryon number density $\bar{n}$
in the outer crust
for the nuclear mass models HFB-22, HFB-24, HFB-25 and HFB-26, corresponding
to functionals BSk22, BSk24, BSk25 and BSk26, respectively. 
Lower panel: Deviations between the computed data and 
the fitted analytic function~(\ref{Efit}) ($\Delta e_\mathrm{eq}$ = fit $-$ data.)
 }
\label{fig:outere}
\end{figure}

\begin{figure}
\includegraphics[width=\columnwidth]{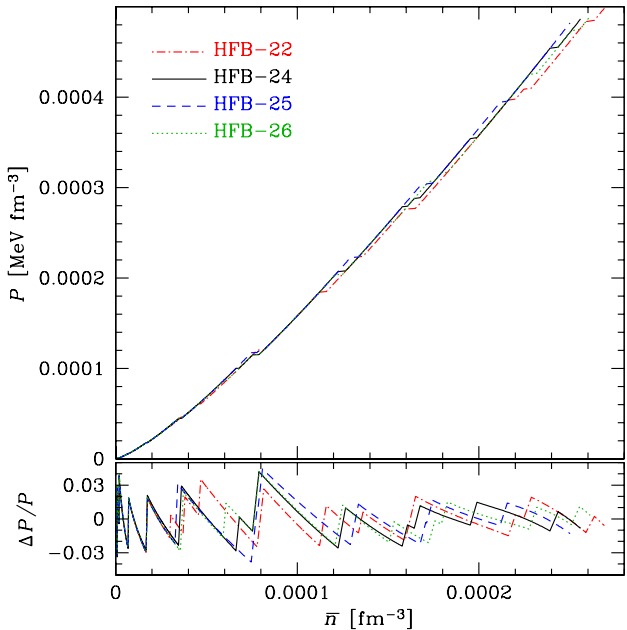}
\caption{(Color online.) Upper panel: Computed pressure $P$ as a function of
$\bar{n}$ in the outer crust for the
 same nuclear mass models as in Fig.~\ref{fig:outere}.
  Lower panel: Fractional deviations between the computed data and the fitted 
analytic function~(\ref{Pfit}) ($\Delta$ P = fit $-$ data).
 }
\label{fig:outer2}
\end{figure}

\begin{figure}
\includegraphics[width=\columnwidth]{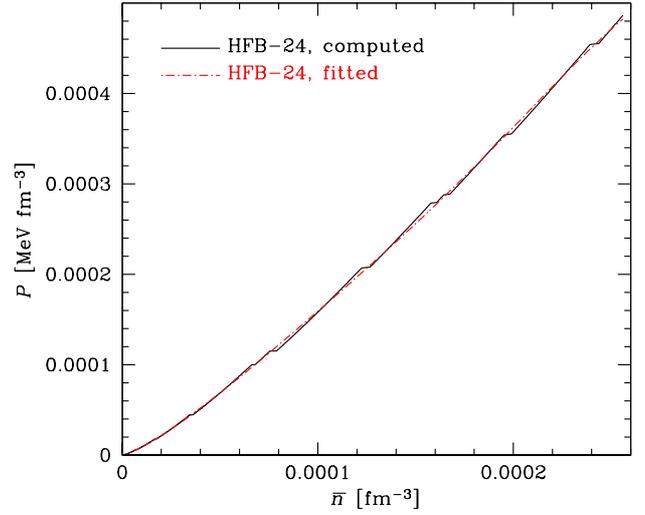}
\caption{(Color online.) Computed pressure in the outer crust for mass model 
HFB-24 (solid line), and the fit to
these data points with the analytic parametrization~(\ref{Pfit})
(dot-dashed line).
} 
\label{fig:outerP24}
\end{figure}

\begin{figure}
\includegraphics[width=\columnwidth]{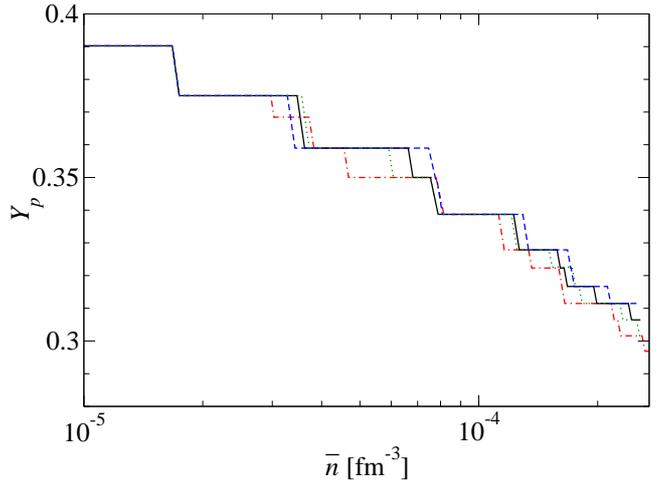}
\caption{(Color online.) Variation of the proton fraction $Y_\mathrm{p}$ as a function 
of the mean baryon number density in the outer crust for the nuclear mass 
models HFB-22, HFB-24, HFB-25 and HFB-26. }
\label{fig:Yp_out}
\end{figure}

\begin{figure}
\includegraphics[width=\columnwidth]{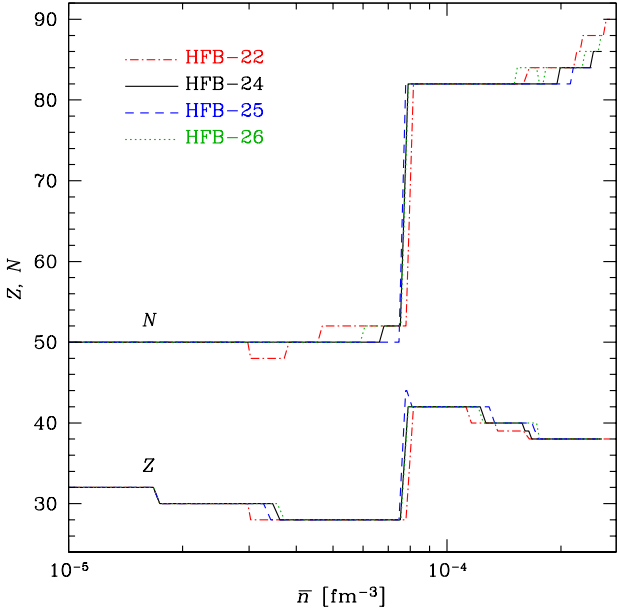}
\caption{(Color online.) Variation of $Z$ and $N$ as a function of mean baryon
number density in the outer crust of the neutron star, for the nuclear mass
models HFB-22, HFB-24, HFB-25, and HFB-26. }
\label{fig:outer1}
\end{figure}

As explained in Section~\ref{outer}, the EoS and the composition of the outer
crust are determined by minimizing the Gibbs free energy per nucleon $g$ at
constant pressure $P$. We begin this process at
$P = 9 \times 10^{-12}$ MeV~fm$^{-3}$, thereby ensuring a density greater than
10$^{6}$ g cm$^{-3}$, a sufficient condition for complete ionization and
degeneracy of the electron gas. The process is then repeated with
$P$ increasing in steps of  $\Delta P = 0.003P$ until neutron drip sets in, as 
given by Eq.~(\ref{drip}). This marks the passage to the inner crust.

The numerical results are summarized in 
Tables~\ref{tab:hfb22-outer}-\ref{tab:hfb26-outer}, where we list the nuclei 
that minimize $g$, the minimum and maximum densities $\bar{n}_\mathrm{min}$ and 
$\bar{n}_\mathrm{max}$ at which that nucleus is present, the pressure $P_\mathrm{max}$ 
and the chemical potentials corresponding to $\bar{n}_\mathrm{max}$.

The variation of the energy per nucleon $e_\mathrm{eq}$, as given by
Eq.~(\ref{2.1}), with the mean number density $\bar{n}$ is shown in
Fig.~\ref{fig:outere}. The four models coincide up to
$\bar{n} \approx 3 \times 10^{-5}$~fm$^{-3}$, since in this low-density
regime we use the masses of experimentally known nuclei, while for higher
densities the differences are only just discernible in the figure. 
The lower panel shows the deviations
of the calculated energies $e_\mathrm{eq}$ from the analytic parametrization
(\ref{Efit}).
Given that this continuous analytic function was fitted to the
calculated energies over the entire density range encountered in neutron stars,
the deviations are remarkably small.

The variation of the pressure $P$ with the mean number density $\bar{n}$ is 
shown in Fig.~\ref{fig:outer2}. Again, the four models coincide up to
$\bar{n} \approx 3 \times 10^{-5}$~fm$^{-3}$, while for higher densities 
the differences are barely discernible in the figure. The steps at the 
transition from one nucleus to another are, however, clearly visible, the pressure 
remaining constant across the transition. Similar steps exist for the chemical potential, because at thermodynamic equilibrium the chemical potential, like
the pressure, should be the same on both sides of the interface between two 
phases in contact, where the density has a discontinuity. (No such steps are 
visible in Fig.~\ref{fig:outere} since the
energy per nucleon continues to rise across the transition.) The lower panel 
of Fig.~\ref{fig:outer2} shows the deviations of the calculated pressures from
the analytic parametrization (\ref{Pfit}). The smallness of these deviations 
is made explicit in Fig.~\ref{fig:outerP24} for the case of the functional 
BSk24. The continuous parametrizing function clearly fails to fit the steps, 
but otherwise the agreement is excellent, especially in view of the fact that 
the function was fitted to the calculated pressures over the entire density 
range encountered in neutron stars.

A more accurate (but discontinuous) parametrization that fits even the steps in the $P - \bar{n}$ curves is given by the analytical thermodynamic model of a fully ionized Coulomb plasma 
(e.g., \citealt{pc10}) with the composition determined by
Tables~\ref{tab:hfb22-outer}\,--\,\ref{tab:hfb26-outer}.

Turning now to the composition, we plot in Fig.~\ref{fig:Yp_out} the variation
of the proton fraction $Y_\mathrm{p} = Z/A$ as a function of the mean density 
$\bar{n}$, and see again that the differences between the four functionals are 
very small. The dominant role of nuclear-structure effects is evident in this 
figure, but certain weak trends can be discerned in the differences between the
four functionals. Comparing BSk22, BSk24 and
BSk25 shows generally, but not everywhere, a tendency for higher $J$ to be
associated with nuclei that are slightly more neutron-rich. This could be a
consequence of the anticorrelation between symmetry energy and $J$ at
subnuclear densities, apparent in Fig.~\ref{fig3}. On the other hand, we see
that BSk24 and BSk26, which have the same value of
$J$, follow each other very closely, suggesting that the NeuM 
constraint is of even less significance than the constraint to $J$.  

Fig.~\ref{fig:Yp_out} also makes apparent a general feature found in all 
calculations on the outer crust, regardless of the nuclear mass model that is 
adopted: as we pass from one layer to another with $\bar{n}$ increasing,
$Y_\mathrm{p}$ always decreases. This is a consequence of the fact that in the outer 
crust the pressure is determined mainly by the electrons, there being no
free neutrons, and the lattice and ion-electron contributions being small.
Then since mechanical equilibrium requires that the pressure
be continuous across the interface between two layers it follows that the 
electron density must be likewise almost continuous, and with it the
mean proton density $\bar{n}_\mathrm{p}$. 
Then, in an obvious notation
\beq\label{nico1}
\bar{n}_\mathrm{p} = \bar{n}_1\frac{Z_1}{A_1} \approx \bar{n}_2\frac{Z_2}{A_2} \quad .
\eeq
Thus if $\bar{n}_2 > \bar{n}_1$ it follows that $Z_2/A_2 < Z_1/A_1$.

The variation of $Z$ and $N$ is plotted as a function of $\bar{n}$ in 
Fig.~\ref{fig:outer1}. The composition is identical for the four models up to 
$\bar{n} \approx 3 \times 10^{-5}$~fm$^{-3}$, since in the low-density 
regime we use the masses of experimentally known nuclei. At higher densities, 
the sequence of nuclei in Tables~\ref{tab:hfb22-outer}--\ref{tab:hfb26-outer} 
becomes model dependent, but it is similar for the different models, apart from
some missing nuclides. 
For example, HFB-22 (HFB-25) predicts the presence of the nucleus $^{76}$Ni 
($^{126}$Ru), unlike the other models. Without the inclusion of the mass 
measurements of \citet{welker2017}, $^{76}$Ni is replaced by the odd-$A$ nucleus 
$^{79}$Cu from densities  $\bar{n}_\mathrm{min}=2.78 \times 10^{-5}$~fm$^{-3}$ to 
$\bar{n}_\mathrm{max}=4.21 \times 10^{-5}$~fm$^{-3}$ ($P_\mathrm{max}=5.63 \times 
10^{-5}$~MeV~fm$^{-3}$). 
On the other hand, HFB-25 (HFB-22) does not support the presence of $^{80}$Ni 
($^{120}$Sr), unlike the other models. Comparing HFB-24 and HFB-26, both having  the same $J$ but different $L$, HFB-26 predicts the presence of $^{124}$Zr, unlike 
HFB-24, and a different Sr isotope at the neutron drip. The discrepancies in the predicted 
nuclei arise from the uncertainties in the masses of neutron-rich nuclei. 
Indeed, the masses of $^{76}$Ni and $^{120}$Sr calculated with 
HFB-22 and HFB-24 differ by 650 keV and 3.58 MeV respectively, while
for $^{80}$Ni ($^{126}$Ru) the difference in the theoretical mass calculated 
with HFB-25 and HFB-24 amounts to  
750 keV (260 keV).
For $^{124}$Zr, the masses calculated with HFB-26 and HFB-24 differ by 
810 keV.

However, as can be seen from the last lines of Tables~\ref{tab:hfb22-outer}-\ref{tab:hfb26-outer}, 
all four functionals predict the same element (Sr) at their respective drip points,
although not the same isotopes. For convenience, we summarize the relevant 
parameters of the neutron-drip points in Table~\ref{ndrip}. We see that both the 
drip density $\bar{n}_\mathrm{nd}$ and the neutron number $N$ increase as $J$ 
increases. The values for $\bar{n}_\mathrm{nd}$ are slightly different from those given by \citet{fcm16} because of the present inclusion of various corrections, as discussed in Section~\ref{outer}.
\citet{gcp13}
have discussed how the masses of drip-line nuclei are correlated with 
$J$. Since mass is the only nuclear quantity on which the composition of the 
crust depends, one can expect some correlations between $J$ and the 
composition. This correlation may be masked by the noise arising from the 
different errors with which the different functionals fit the data and the 
numerical errors with which masses are calculated with a given functional. 
Nevertheless, it is possible to establish some correlations between the 
properties of the crust at the neutron-drip transition and $J$,
 (or, equivalently, $L$)
as discussed by \citet{fcm16}.

Concerning the corrections, the screening correction to the electron energy 
density may, or may not, change the composition in the outer crust
(see \citealt{chf16-prd,fcpg-press} for a discussion). Indeed, as shown 
by \citet{chf16-prd}, for some models, e.g., HFB-22, the 
differences in the Gibbs free energies for some nuclei may be so small that 
including or not even small corrections leads to different equilibrium nuclei.
Nevertheless, the current uncertainties in nuclear masses are generally more 
important than the effects induced by the different corrections.

Indeed, within the CCM hypothesis, a  
small mass difference can lead to a significant difference in the
equilibrium nuclide at any depth, as exemplified by the measurement of
the mass of $^{82}$Zn a few years ago \citep{wolf13}. It is thus
perhaps remarkable that there is so much agreement between the different mass models. For example, all our HFB models predict the presence of $^{78}$Ni, 
$^{124}$Mo, $^{122}$Zr, and $^{120}$Sr in the deepest layers of the outer crust. These nuclides were also found with our earlier mass models HFB-19, HFB-20, and  HFB-21 \citep{wolf13}  as well  as with the HFB model using the  Gogny  D1M  effective interaction \citep[see][]{pgc11}. The existence of these nuclei in the crust is supported by other models. These include the RMF model using the TMA parametrization \citep[see][]{ruester06}, the model of \citet{DZ} \citep[see][]{roca2008}, the finite-range droplet model of \citet{frdm} \citep[see][]{hemp2013}, and the recent BNN model of \citet{utama2016}.  Descending through the outer crust to the neutron-drip point,
nuclei become more and more exotic (see Fig. \ref{fig:Yp_out}). For this reason, the
composition of the deepest layers remains uncertain. The neutron
enrichment may lead to a profound change in the nuclear structure that
can hardly be anticipated by phenomenological approaches or models
that were adjusted to stable nuclei only. It is therefore not surprising that such models generally agree in predicting the occurrence at the bottom of the outer crust of nuclei with the standard neutron magic number 82, but with a wide spread in the proton number ranging from 32 to 40 \citep[see][]{ruester06}. On the other hand, it is noteworthy that all the aforementioned HFB models, which were 
constrained to reproduce realistic EoSs of NeuM, consistently predict the existence of strontium isotopes at the bottom 
of the outer crust. More importantly, although some mass models such as that of  \citet{DZ}, the finite-range droplet and BNN yield a very good fit to known masses, they are not applicable beyond the neutron-drip point and therefore do not permit a unified treatment of neutron-star interiors.

\begin{table*}
\caption{Composition and EoS of the outer crust of a cold non-accreting neutron  star as obtained using experimental data from the 2016 Atomic Mass Evaluation \citep{ame16} (above the dotted line) supplemented with the nuclear mass model HFB-22 \citep{gcp13}. We have also made use of the very recent measurements of copper isotopes \citep{welker2017}. In the table are listed: the atomic number $Z$, the neutron number $N$, and the mass 
 number $A$ of each nuclide, the minimum and maximum baryon number densities $\bar{n}_\mathrm{min}$ 
 and $\bar{n}_\mathrm{max}$ at which the nuclide is present, the corresponding maximum pressure 
 $P_\mathrm{max}$, and the neutron, proton, and electron chemical potentials 
 at density $\bar{n}_\mathrm{max}$.
 Note that the neutron and proton chemical potentials are given with the rest mass subtracted out. 
 The baryon number densities are measured in units of fm$^{-3}$, pressures are given in units of 
 MeV~fm$^{-3}$ and chemical potentials are given in units of MeV. See text for details.} 
\label{tab:hfb22-outer}
\begin{tabular}{ccccccccc}
\hline 
 $Z$ & $N$ & $A$ & $\bar{n}_\mathrm{min}$   & $\bar{n}_\mathrm{max}$ & $P_\mathrm{max}$ & $\mu_\mathrm{n}-M_\mathrm{n} c^2$ & $\mu_\mathrm{p}-M_\mathrm{p} c^2$ &$ \mu_\mathrm{e}$ \\
\hline 
 26 & 30 & 56 & $-$                 & $4.93 \times 10^{-9}$ & $3.36 \times 10^{-10}$ & $-8.96$ & $-8.62$ & 0.95 \\
 28 & 34 & 62 & $5.08 \times 10^{-9}$  & $1.63 \times 10^{-7}$ & $4.34 \times 10^{-8}$ & $-8.25$ & $-9.56$ & 2.61 \\
 28 & 36 & 64 & $1.68 \times 10^{-7}$  & $8.01 \times 10^{-7}$ & $3.56 \times 10^{-7}$ & $-7.53$ & $-10.57$  & 4.33 \\
 28 & 38 & 66 & $8.28 \times 10^{-7}$  & $8.79 \times 10^{-7}$ & $3.87 \times 10^{-7}$ & $-7.49$ & $-10.62$ & 4.42 \\
 36 & 50 & 86 & $8.98 \times 10^{-7}$  & $1.87 \times 10^{-6}$ & $1.04 \times 10^{-6}$ & $-7.00$ & $-11.36$ & 5.65 \\
 34 & 50 & 84 & $1.94 \times 10^{-6}$  & $6.83 \times 10^{-6}$ & $5.62 \times 10^{-6}$ & $-5.87$ & $-13.16$ & 8.58 \\
 32 & 50 & 82 & $7.09 \times 10^{-6}$  & $1.67 \times 10^{-5}$ & $1.78 \times 10^{-5}$ & $-4.81$ & $-14.94$ & 11.43 \\
 30 & 50 & 80 & $1.74 \times 10^{-5}$  & $2.97 \times 10^{-5}$ & $3.63 \times 10^{-5}$ & $-4.01$ & $-16.37$ & 13.65 
\\[-1ex]\multicolumn{9}{c}{\dotfill}\\
 28 & 48 & 76 & $3.03 \times 10^{-5}$  & $3.72 \times 10^{-5}$ & $4.79 \times 10^{-5}$ & $-3.67$ & $-16.99$ & 14.62 \\
 28 & 50 & 78 & $3.82 \times 10^{-5}$  & $4.55 \times 10^{-5}$ & $6.07 \times 10^{-5}$ & $-3.36$ & $-17.58$ & 15.51 \\
 28 & 52 & 80 & $4.68 \times 10^{-5}$  & $7.79 \times 10^{-5}$ & $1.20 \times 10^{-4}$ & $-2.40$ & $-19.49$ & 18.39 \\
 42 & 82 & 124 & $8.16 \times 10^{-5}$ & $1.12 \times 10^{-4}$ & $1.84 \times 10^{-4}$ & $-1.73$ & $-20.96$ & 20.52 \\
 40 & 82 & 122 & $1.16 \times 10^{-4}$ & $1.34 \times 10^{-4}$ & $2.23 \times 10^{-4}$ & $-1.42$ & $-21.64$ & 21.52 \\
 39 & 82 & 121 & $1.36 \times 10^{-4}$ & $1.59 \times 10^{-4}$ & $2.76 \times 10^{-4}$ & $-1.06$ & $-22.46$ & 22.69 \\
 38 & 84 & 122 & $1.65 \times 10^{-4}$ & $2.16 \times 10^{-4}$ & $3.97 \times 10^{-4}$ & $-0.42$ & $-23.96$ & 24.84 \\
 38 & 86 & 124 & $2.20 \times 10^{-4}$ & $2.25 \times 10^{-4}$ & $4.09 \times 10^{-4}$ & $-0.37$ & $-24.10$ & 25.02 \\
 38 & 88 & 126 & $2.29 \times 10^{-4}$ & $2.59 \times 10^{-4}$ & $4.83 \times 10^{-4}$ & $-0.06$ & $-24.86$ & 26.09 \\
 38 & 90 & 128 & $2.64 \times 10^{-4}$ & $2.69 \times 10^{-4}$ & $4.99 \times 10^{-4}$ & 0.00 & $-25.01$ & 26.30 \\
\hline
\end{tabular}
\end{table*}

\begin{table*}
\caption{Same as in Table~\ref{tab:hfb22-outer}, for the nuclear mass model HFB-24.} 
\label{tab:hfb24-outer}
\begin{tabular}{ccccccccc}
\hline 
 $Z$ & $N$ & $A$ & $\bar{n}_\mathrm{min}$   & $\bar{n}_\mathrm{max}$ & $P_\mathrm{max}$ &$ \mu_\mathrm{n}-M_\mathrm{n} c^2$ &$ \mu_\mathrm{p}-M_\mathrm{p} c^2$ &$ \mu_\mathrm{e}$ \\
\hline 
 26 & 30 & 56 & $-$                   & $4.93 \times 10^{-9}$ & $3.36 \times 10^{-10}$ & $-8.96$ & $-8.62$ & 0.95 \\
 28 & 34 & 62 & $5.08 \times 10^{-9}$  & $1.63 \times 10^{-7}$ & $4.34 \times 10^{-8}$ & $-8.25$ & $-9.56$ & 2.61 \\
 28 & 36 & 64 & $1.68 \times 10^{-7}$  & $8.01 \times 10^{-7}$ & $3.56 \times 10^{-7}$ & $-7.53$ & $-10.57$ & 4.33 \\
 28 & 38 & 66 & $8.28 \times 10^{-7}$  & $8.79 \times 10^{-7}$ & $3.87 \times 10^{-7}$ & $-7.49$ & $-10.62$ & 4.42 \\
 36 & 50 & 86 & $8.98 \times 10^{-7}$  & $1.87 \times 10^{-6}$ & $1.04 \times 10^{-6}$ & $-7.00$ & $-11.36$ & 5.65 \\
 34 & 50 & 84 & $1.94 \times 10^{-6}$  & $6.83 \times 10^{-6}$ & $5.62 \times 10^{-6}$ & $-5.87$ & $-13.16$ & 8.58 \\
 32 & 50 & 82 & $7.09 \times 10^{-6}$  & $1.67 \times 10^{-5}$ & $1.78 \times 10^{-5}$ & $-4.81$ & $-14.94$ & 11.43 \\
 30 & 50 & 80 & $1.74 \times 10^{-5}$  & $3.47 \times 10^{-5}$ & $4.45 \times 10^{-5}$ & $-3.76$ & $-16.83$ & 14.36
\\[-1ex]\multicolumn{9}{c}{\dotfill}\\
 28 & 50 & 78 & $3.62 \times 10^{-5}$  & $6.63 \times 10^{-5}$ & $1.00 \times 10^{-4}$ & $-2.65$ & $-18.93$ & 17.57 \\
 28 & 52 & 80 & $6.81 \times 10^{-5}$  & $7.54 \times 10^{-5}$ & $1.15 \times 10^{-4}$ & $-2.44$ & $-19.34$ & 18.18 \\
 42 & 82 & 124 & $7.89 \times 10^{-5}$ & $1.22 \times 10^{-4}$ & $2.07 \times 10^{-4}$ & $-1.52$ & $-21.36$ & 21.13 \\
 40 & 82 & 122 & $1.27 \times 10^{-4}$ & $1.58 \times 10^{-4}$ & $2.79 \times 10^{-4}$ & $-1.01$ & $-22.47$ & 22.75 \\
 39 & 82 & 121 & $1.61 \times 10^{-4}$ & $1.64 \times 10^{-4}$ & $2.88 \times 10^{-4}$ & $-0.95$ & $-22.59$ & 22.93 \\
 38 & 82 & 120 & $1.68 \times 10^{-4}$ & $1.95 \times 10^{-4}$ & $3.54 \times 10^{-4}$ & $-0.59$ & $-23.44$ & 24.14 \\
 38 & 84 & 122 & $1.99 \times 10^{-4}$ & $2.39 \times 10^{-4}$ & $4.54 \times 10^{-4}$ & $-0.13$ & $-24.53$ & 25.69 \\
 38 & 86 & 124 & $2.44 \times 10^{-4}$ & $2.56 \times 10^{-4}$ & $4.87 \times 10^{-4}$ & 0.00 & $-24.85$ & 26.14 \\

\hline
\end{tabular}
\end{table*}

\begin{table*}
\caption{Same as in Table~\ref{tab:hfb22-outer}, for the nuclear mass model HFB-25.} 
\label{tab:hfb25-outer}
\begin{tabular}{ccccccccc}
\hline 
 $Z$ & $N$ & $A$ & $\bar{n}_\mathrm{min}$   & $\bar{n}_\mathrm{max}$ & $P_\mathrm{max}$ & $\mu_\mathrm{n}-M_\mathrm{n} c^2$ & $\mu_\mathrm{p}-M_\mathrm{p} c^2$ & $\mu_\mathrm{e}$ \\
\hline 
 26 & 30 & 56 & $-$                   & $4.93 \times 10^{-9}$ & $3.36 \times 10^{-10}$ & $-8.96$ & $-8.62$ & 0.95 \\
 28 & 34 & 62 & $5.08 \times 10^{-9}$  & $1.63 \times 10^{-7}$ & $4.34 \times 10^{-8}$ & $-8.25$ & $-9.56$ & 2.61 \\
 28 & 36 & 64 & $1.68 \times 10^{-7}$  & $8.01 \times 10^{-7}$ & $3.56 \times 10^{-7}$ & $-7.53$ & $-10.57$ & 4.33 \\
 28 & 38 & 66 & $8.28 \times 10^{-7}$  & $8.79 \times 10^{-7}$ & $3.87 \times 10^{-7}$ & $-7.49$ & $-10.62$ & 4.42 \\
 36 & 50 & 86 & $8.98 \times 10^{-7}$  & $1.87 \times 10^{-6}$ & $1.04 \times 10^{-6}$ & $-7.00$ & $-11.36$ & 5.65 \\
 34 & 50 & 84 & $1.94 \times 10^{-6}$  & $6.83 \times 10^{-6}$ & $5.62 \times 10^{-6}$ & $-5.87$ & $-13.16$ & 8.58 \\
 32 & 50 & 82 & $7.09 \times 10^{-6}$  & $1.67 \times 10^{-5}$ & $1.78 \times 10^{-5}$ & $-4.81$ & $-14.94$ & 11.43 \\
 30 & 50 & 80 & $1.74 \times 10^{-5}$  & $3.28 \times 10^{-5}$ & $4.13 \times 10^{-5}$ & $-3.85$ & $-16.66$ & 14.10
\\[-1ex]\multicolumn{9}{c}{\dotfill}\\
 28 & 50 & 78 & $3.42 \times 10^{-5}$  & $7.46 \times 10^{-5}$ & $1.17 \times 10^{-4}$ & $-2.41$ & $-19.40$ & 18.28 \\
 44 & 82 & 126 & $7.78 \times 10^{-5}$ & $7.84 \times 10^{-5}$ & $1.19 \times 10^{-4}$ & $-2.39$ & $-19.49$ & 18.39 \\
 42 & 82 & 124 & $8.09 \times 10^{-5}$ & $1.29 \times 10^{-4}$ & $2.23 \times 10^{-4}$ & $-1.39$ & $-21.61$ & 21.51 \\
 40 & 82 & 122 & $1.34 \times 10^{-4}$ & $1.68 \times 10^{-4}$ & $3.02 \times 10^{-4}$ & $-0.86$ & $-22.78$ & 23.21 \\
 39 & 82 & 121 & $1.71 \times 10^{-4}$ & $1.71 \times 10^{-4}$ & $3.04 \times 10^{-4}$ & $-0.85$ & $-22.80$ & 23.24 \\
 38 & 82 & 120 & $1.75 \times 10^{-4}$ & $2.12 \times 10^{-4}$ & $3.94 \times 10^{-4}$ & $-0.38$ & $-23.89$ & 24.80 \\
 38 & 84 & 122 & $2.16 \times 10^{-4}$ & $2.50 \times 10^{-4}$ & $4.83 \times 10^{-4}$ & 0.00 & $-24.79$ & 26.08 \\
\hline
\end{tabular}
\end{table*}

\begin{table*}
\caption{Same as in Table~\ref{tab:hfb22-outer}, for the nuclear mass model HFB-26.} 
\label{tab:hfb26-outer}
\begin{tabular}{ccccccccc}
\hline 
 $Z$ & $N$ & $A$ & $\bar{n}_\mathrm{min}$   & $\bar{n}_\mathrm{max}$ & $P_\mathrm{max}$ & $\mu_\mathrm{n}-M_\mathrm{n} c^2$ & $\mu_\mathrm{p}-M_\mathrm{p} c^2$ &$ \mu_\mathrm{e}$ \\
\hline 
 26 & 30 & 56 & $-$                   & $4.93 \times 10^{-9}$ & $3.36 \times 10^{-10}$ & $-8.96$ & $-8.62$ & 0.95 \\
 28 & 34 & 62 & $5.08 \times 10^{-9}$  & $1.63 \times 10^{-7}$ & $4.34 \times 10^{-8}$ & $-8.25$ & $-9.56$ & 2.61 \\
 28 & 36 & 64 & $1.68 \times 10^{-7}$  & $8.01 \times 10^{-7}$ & $3.56 \times 10^{-7}$ & $-7.53$ & $-10.57$ & 4.33 \\
 28 & 38 & 66 & $8.28 \times 10^{-7}$  & $8.79 \times 10^{-7}$ & $3.87 \times 10^{-7}$ & $-7.49$ & $-10.62$ & 4.42 \\
 36 & 50 & 86 & $8.98 \times 10^{-7}$  & $1.87 \times 10^{-6}$ & $1.04 \times 10^{-6}$ & $-7.00$ & $-11.36$ & 5.65 \\
 34 & 50 & 84 & $1.94 \times 10^{-6}$  & $6.83 \times 10^{-6}$ & $5.62 \times 10^{-6}$ & $-5.87$ & $-13.16$ & 8.58 \\
 32 & 50 & 82 & $7.09 \times 10^{-6}$  & $1.67 \times 10^{-5}$ & $1.78 \times 10^{-5}$ & $-4.81$ & $-14.94$ & 11.43 \\
 30 & 50 & 80 & $1.74 \times 10^{-5}$  & $3.56 \times 10^{-5}$ & $4.62 \times 10^{-5}$ & $-3.71$ & $-16.91$ & 14.49
\\[-1ex]\multicolumn{9}{c}{\dotfill}\\
 28 & 50 & 78 & $3.72 \times 10^{-5}$  & $5.91 \times 10^{-5}$ & $8.59 \times 10^{-5}$ & $-2.88$ & $-18.49$ & 16.91 \\
 28 & 52 & 80 & $6.07 \times 10^{-5}$ & $7.55 \times 10^{-5}$ & $1.15 \times 10^{-4}$ & $-2.44$ & $-19.35$ & 18.20 \\
 42 & 82 & 124 & $7.91 \times 10^{-5}$ & $1.21 \times 10^{-4}$ & $2.03 \times 10^{-4}$ & $-1.55$ & $-21.29$ & 21.04 \\
 40 & 82 & 122 & $1.25 \times 10^{-4}$ & $1.51 \times 10^{-4}$ & $2.62 \times 10^{-4}$ & $-1.13$ & $-22.23$ & 22.39 \\
 40 & 84 & 124 & $1.53 \times 10^{-4}$ & $1.72 \times 10^{-4}$ & $3.07 \times 10^{-4}$ & $-0.85$ & $-22.86$ & 23.30 \\
 38 & 82 & 120 & $1.76 \times 10^{-4}$ & $1.79 \times 10^{-4}$ & $3.16 \times 10^{-4}$ & $-0.80$ & $-22.97$ & 23.46 \\
 38 & 84 & 122 & $1.83 \times 10^{-4}$ & $2.28 \times 10^{-4}$ & $4.25 \times 10^{-4}$ & $-0.26$ & $-24.24$ & 25.27 \\
 38 & 86 & 124 & $2.32 \times 10^{-4}$ & $2.51 \times 10^{-4}$ & $4.73 \times 10^{-4}$ & $-0.06$ & $-24.73$ & 25.96 \\
 38 & 88 & 126 & $2.55 \times 10^{-4}$ & $2.61 \times 10^{-4}$ & $4.90 \times 10^{-4}$ & 0.00 & $-24.88$ & 26.18 \\

\hline
\end{tabular}
\end{table*}

\begin{table}
\centering
\caption{Parameters of the neutron drip point for the functionals of this 
paper.}
\label{ndrip}
\begin{tabular}{|c|cccc|}\hline
&BSk22&BSk24&BSk25&BSk26\\
\hline
$J$ {[MeV]} & 32 & 30 & 29 & 30\\
$\bar{n}_\mathrm{nd}$ {[$\times10^{-4}$ fm$^{-3}$]} & $2.69$ & $2.56$ & $2.50$ & $2.61$ \\
$Z$ & 38 & 38 & 38 & 38 \\
$N$ & 90 & 86 & 84 & 88 \\
\hline
\end{tabular}
\end{table}

\subsection{Inner crust}
\label{icres}

The complete inner-crust results for each of our four functionals are
presented in the supplementary 
material. 

\begin{figure}
\includegraphics[width=\columnwidth]{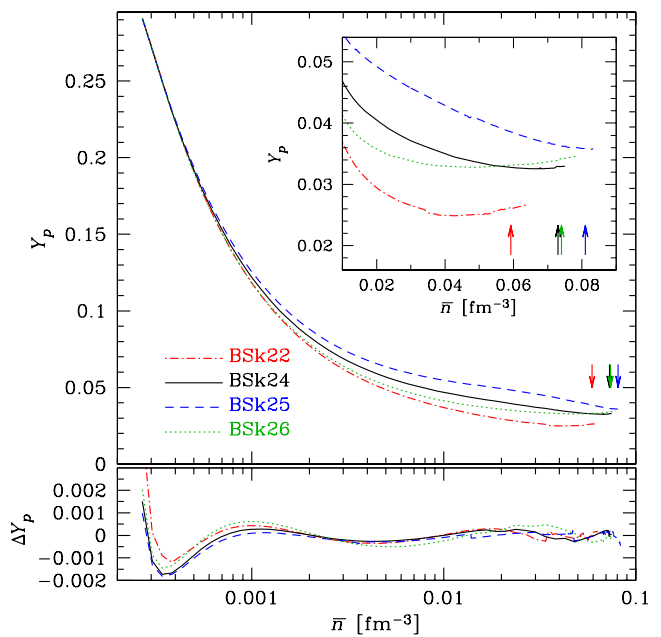}
\caption{(Color online.) Upper panel: Curves show the computed equilibrium 
proton fraction $Y_\mathrm{p} = Z_\mathrm{eq}/A$ in the inner crust as a function of mean 
baryon density $\bar{n}$ for our four functionals; arrows indicate onset of
proton drip. The inset shows the cross-over between BSk24 and BSk26 just
before proton drip. 
Lower panel: Deviations between the computed data and the fitted analytic
function~(\ref{Y_e_crust}) ($\Delta Y_\mathrm{p}$ = fit $-$ data).
(Corrected version.)}
\label{Yic}
\end{figure}

\emph{Proton fraction.} Fig.~\ref{Yic} shows the variation of the equilibrium
value of the proton fraction $Y_\mathrm{p} = Z_\mathrm{eq}/A$ as a function of $\bar{n}$ for
our four functionals. Comparing BSk22, BSk24 and BSk25, we see that $Y_\mathrm{p}$ has a 
modest dependence on the symmetry coefficient $J$, increasing $J$
being associated with a decreasing proton fraction, i.e., increasing asymmetry.
This is in accordance with the anti-correlation shown in Fig.~\ref{fig3}
between $J$ and the symmetry energy $S(\bar{n})$ of INM
at subnuclear densities $\bar{n} < n_0$; it reflects the fact 
that higher symmetry energies imply lower asymmetries at beta equilibrium, 
i.e., higher values of $Y_\mathrm{p}$. On the other hand, BSk26 and BSk24, which
have the same value of $J$, show similar values of $Y_\mathrm{p}$ over the
entire inner crust. Thus the choice of the constraining EoS of NeuM has no
significant impact on the proton fraction $Y_\mathrm{p}$ in the inner crust: if masses
are fitted $Y_\mathrm{p}$ is determined entirely by the symmetry coefficient $J$.

The analytic function~(\ref{Y_e_crust}) has been fitted to our computed values
of $Y_\mathrm{p}$ for all four functionals; the lower panel of Fig.~\ref{Yic} shows the
deviations between this fit and the computed data.

\begin{figure}
\includegraphics[width=\columnwidth]{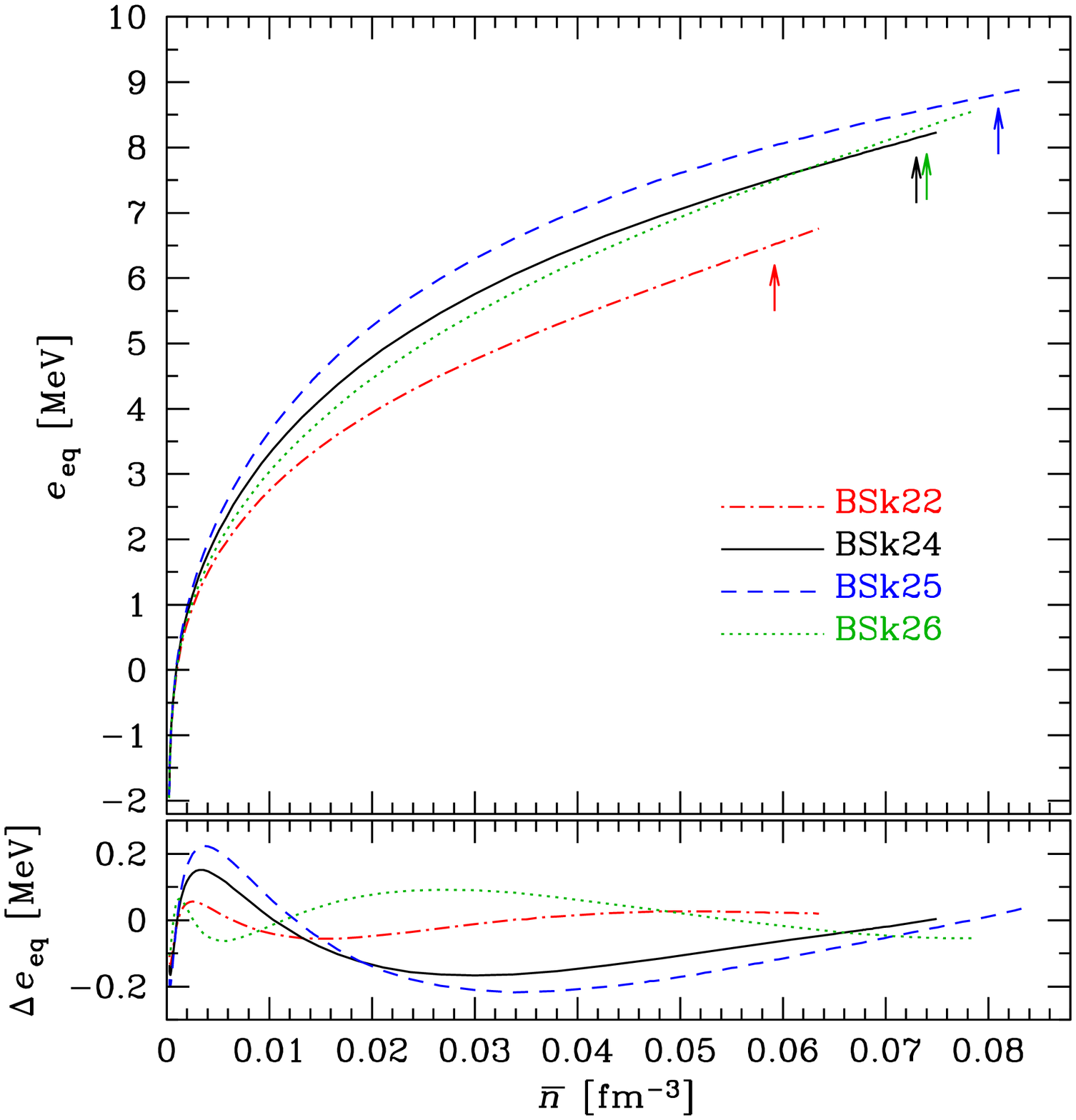}
\caption{(Color online.) Upper panel: Curves show the computed equilibrium 
energy per nucleon $e_\mathrm{eq}$ in the inner crust as a function of mean baryon 
density $\bar{n}$ for our four functionals; arrows indicate onset of
proton drip. 
Lower panel: Deviations between the computed data and the fitted analytic
function~(\ref{Efit}) ($\Delta e_\mathrm{eq}$ = fit $-$ data).
(Corrected version.)}
\label{eic}
\end{figure}

\emph{Energy per nucleon.} In Fig.~\ref{eic} we show the variation of the
equilibrium energy per nucleon $e_\mathrm{eq}$ as a function of $\bar{n}$ for our
four functionals. Comparing functionals BSk22, BSk24 and BSk25, we see from
Fig.~\ref{fig3} that $e_\mathrm{eq}$ is correlated with $J$ in the same way as is the
symmetry energy $S(n)$ at subnuclear densities. That is, the higher the
symmetry energy (or equivalently, the lower the asymmetry) the higher $e_\mathrm{eq}$.
In other words, an increase in the symmetry energy is not entirely offset by
the reduction of the asymmetry, i.e., by the increase in $Y_\mathrm{p}$, at beta
equilibrium (this can be shown analytically). Nevertheless, as core densities 
are approached in the inner crust the gradient of $e_\mathrm{eq}$ is greater the lower
the value of $e_\mathrm{eq}$ ; this may be reflecting the fact that all three 
functionals tend to the same EoS of NeuM.  

On the other hand, the values for $e_\mathrm{eq}$ in the inner crust show very little 
difference between functionals BSk24 and BSk26: once again, it is $J$ rather 
than the constraining EoS of NeuM that is the relevant factor. Nevertheless, it
will be seen that the rate of variation of $e_\mathrm{eq}$ with $\bar{n}$ is somewhat 
steeper for BSk26 than for BSk24. This will have consequences for the pressure,
as we will discuss below.

Our computed values of $e_\mathrm{eq}$ in the inner crust have been fitted by the
same analytic function~(\ref{Efit}) with which we fit the two other regions
of the neutron star; the lower panel of Fig.~\ref{eic} shows the deviation 
between this fit and the computed data.

\begin{figure}
\includegraphics[width=\columnwidth]{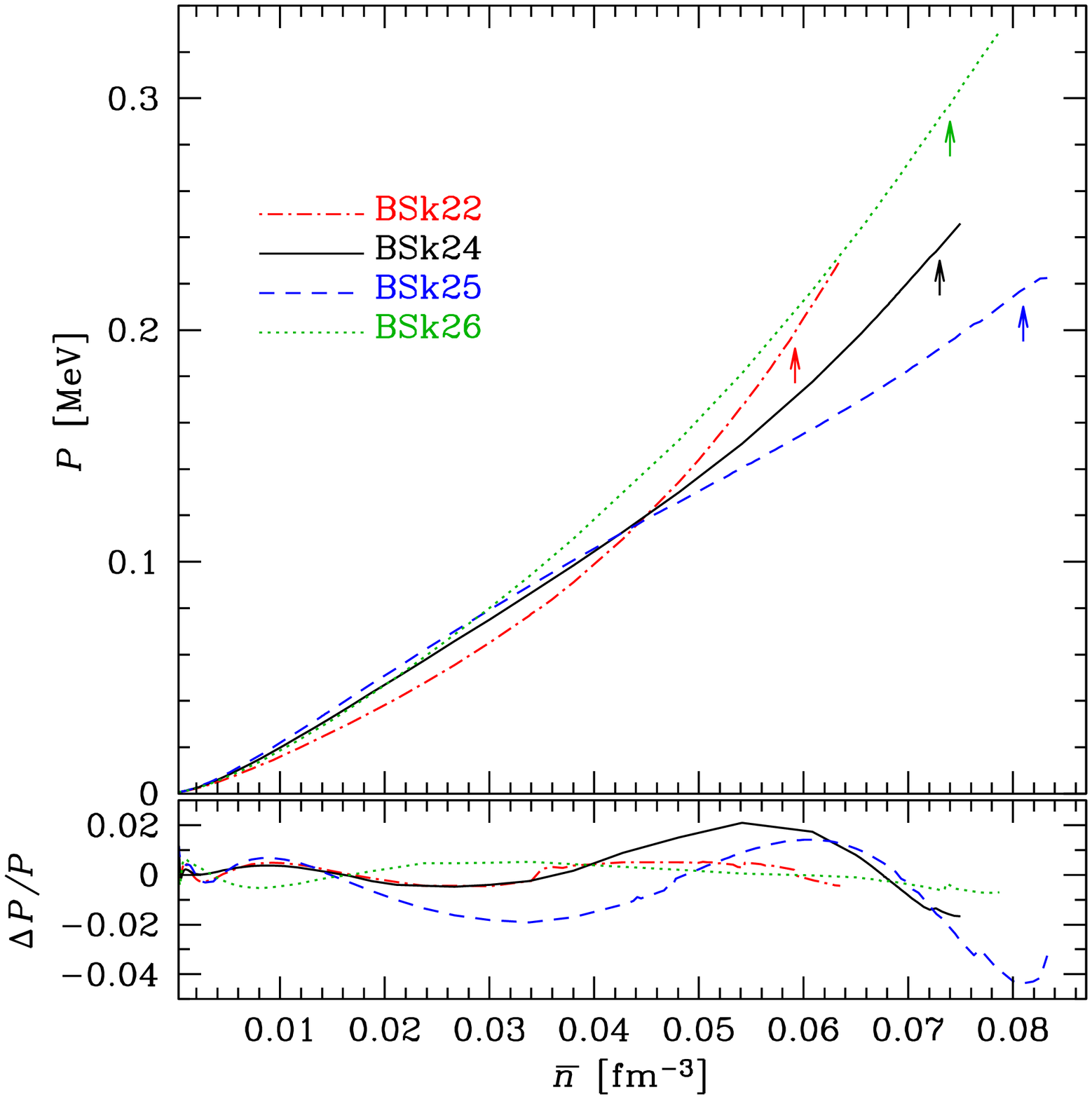}
\caption{(Color online.) Upper panel: Curves show the computed pressure $P$ in 
the inner crust as a function of mean baryon density $\bar{n}$ for our four 
functionals; arrows indicate onset of proton drip.
Lower panel: Fractional deviations between the computed data and the
fitted analytic function~(\ref{Pfit}) ($\Delta P$ = fit $-$ data).
(Corrected version.)}
\label{Pic}
\end{figure}

\emph{Pressure.} The variation of the equilibrium pressure $P$ with $\bar{n}$ is
shown in Fig.~\ref{Pic} for our four functionals. Comparing the first three 
functionals, we see that the higher $J$ the stiffer is the EoS, in the sense that 
the pressure rises more rapidly with $\bar{n}$; the differences in the pressure at any
given density are simply reflecting the differences in the gradients of the
$e_\mathrm{eq}$ curves of Fig.~\ref{eic}.

On the other hand, although $J$ has the same value for BSk24 and BSk26, the 
latter shows throughout most of the inner crust a consistently higher pressure 
than the former; it is only in the homogeneous core that BSk26 begins to exert 
a lower pressure than BSk24. This may cause some surprise, given that BSk26 has
been fitted to a softer EoS for NeuM than has BSk24, but it can be traced to
the purely nuclear properties of INM, as follows. From Eq.~(\ref{1.4}) we have
\beq\label{Peta} 
P_\mathrm{INM} = \frac{n^2}{n_0}\left[\frac{1}{3}L\eta^2 + 
\frac{1}{9}(K_v + \eta^2K_\mathrm{sym})\frac{n-n_0}{n_0} + \cdots \right] \quad .
\eeq
This shows that constraining BSk26 at higher densities, $n > n_0$, to a 
softer EoS of NeuM than BSk24 can only be achieved by having a lower value of 
the factor $K_v + \eta^2K_\mathrm{sym}$, given that $L$ has roughly the same 
value for the two functionals. Since our functionals are all fitted to the
experimental value of the incompressibility $K_v$, the coefficient 
$K_\mathrm{sym}$ will have to be much lower for BSk26 than for BSk24. 
Table~\ref{JL} shows that this is indeed the case. It follows that at 
sufficiently low subnuclear densities, $n < n_0$, $P_\mathrm{INM}$ will be 
higher for BSk26 than for BSk24 (we find that this happens for $n/n_0 < 0.72)$. 

Our computed values of $P$ in the inner crust have been fitted by the
same analytic function~(\ref{Pfit}) with which we fit the two other regions
of the neutron star; the lower panel of Fig.~\ref{Pic} shows the deviations
between this fit and the computed data.

\begin{figure}
\includegraphics[width=\columnwidth]{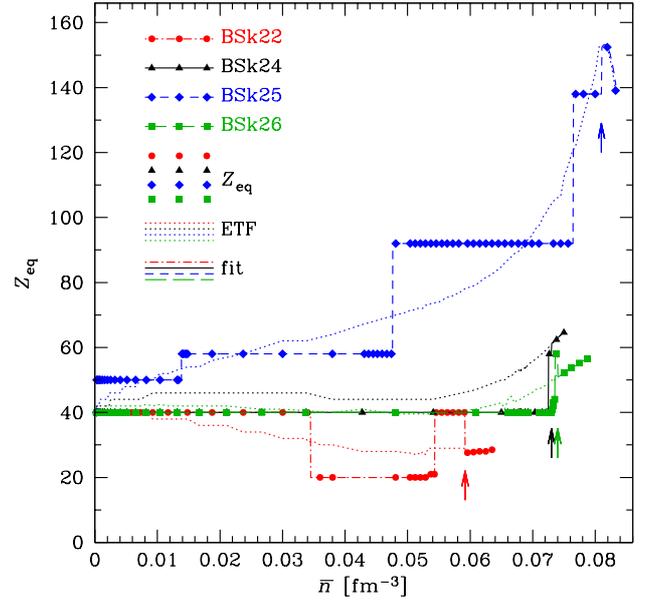}
\caption{(Color online.) Equilibrium value $Z_\mathrm{eq}$ of number of protons in
inner crust as a function of mean baryon density $\bar{n}$ for our four 
functionals. For clarity only every second point is shown. Arrows indicate 
onset of proton drip. The dotted 
curves relate to the values of $Z_\mathrm{eq}$ calculated in the ETF approximation. (A
zoom of the free-proton part of this figure is shown in Figs.~\ref{Zic_zooma}
for functionals BSk22, BSk24 and BSk25 and in Fig.~\ref{Zic_zoomb} for
functionals BSk24 and BSk26.)
(Corrected version.)}
\label{Zic}
\end{figure}

\begin{figure}
\includegraphics[width=\columnwidth]{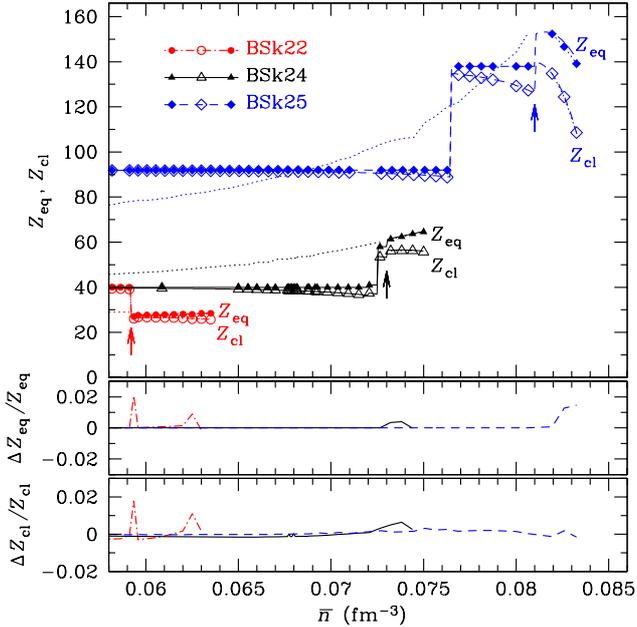}
\caption{(Color online.) Zoom of free-proton region of Fig.~\ref{Zic} for
functionals BSk22, BSk24 and BSk25. Upper panel: Equilibrium values $Z_\mathrm{eq}$ of
number of protons in inner crust (solid symbols) and $Z_\mathrm{cl}$, the cluster 
component of $Z_\mathrm{eq}$, given by Eq.~(\ref{3.2Ca}), (open symbols), as functions
of mean baryon density $\bar{n}$. All data points are shown. The curves in the 
free-proton region represent the fit of the computed values of $Z_\mathrm{eq}$ by 
Eq.~(\ref{Zcell_fit}) and of $Z_\mathrm{cl}$ by Eq.~(\ref{Zfree2}). 
Lower panel: Fractional deviations of these fits from the computed data points
($\Delta \{Z_\mathrm{eq}, Z_\mathrm{cl}\}$ = fit $-$ data).
(Corrected version.)} 
\label{Zic_zooma}
\end{figure}

\begin{figure}
\includegraphics[width=\columnwidth]{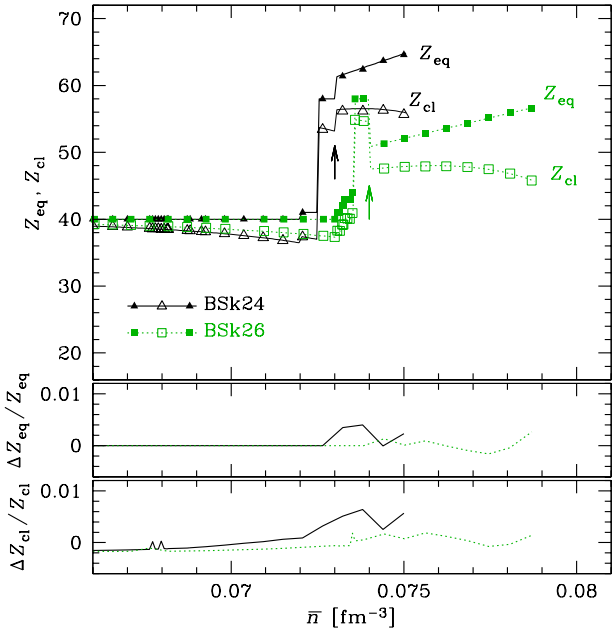}
\caption{(Color online.) As in Fig.~\ref{Zic_zooma}, for  
functionals BSk24 and BSk26.
(Corrected version.)} 
\label{Zic_zoomb}
\end{figure}

\emph{Equilibrium value of $Z$.} Fig.~\ref{Zic} shows how the value of $Z_\mathrm{eq}$ 
varies throughout the inner crust for functionals BSk22, BSk24, BSk25 and 
BSk26; a zoom of this figure in the free-proton region is shown in 
Fig.~\ref{Zic_zooma} for functionals BSk22, BSk24 and BSk25, and in 
Fig.~\ref{Zic_zoomb} for functionals BSk24 and BSk26. 
While there is no evidence of any shell
structure in Figs.~\ref{Yic} -- \ref{Pic} for $Y_\mathrm{p}, e_\mathrm{eq}$ and $P$, 
respectively, we see from Fig.~\ref{Zic} that before proton drip sets in the 
value of $Z_\mathrm{eq}$ expressed as a function of $\bar{n}$ has a strong shell 
structure, which differs significantly according to the value of $J$. 
Furthermore, this $J$-dependent shell structure is superimposed on smoothly 
varying ETF estimates of the optimal values of $Z$ (dotted lines
in Fig.~\ref{Zic}) that themselves are strongly $J$-dependent.

\begin{figure*}
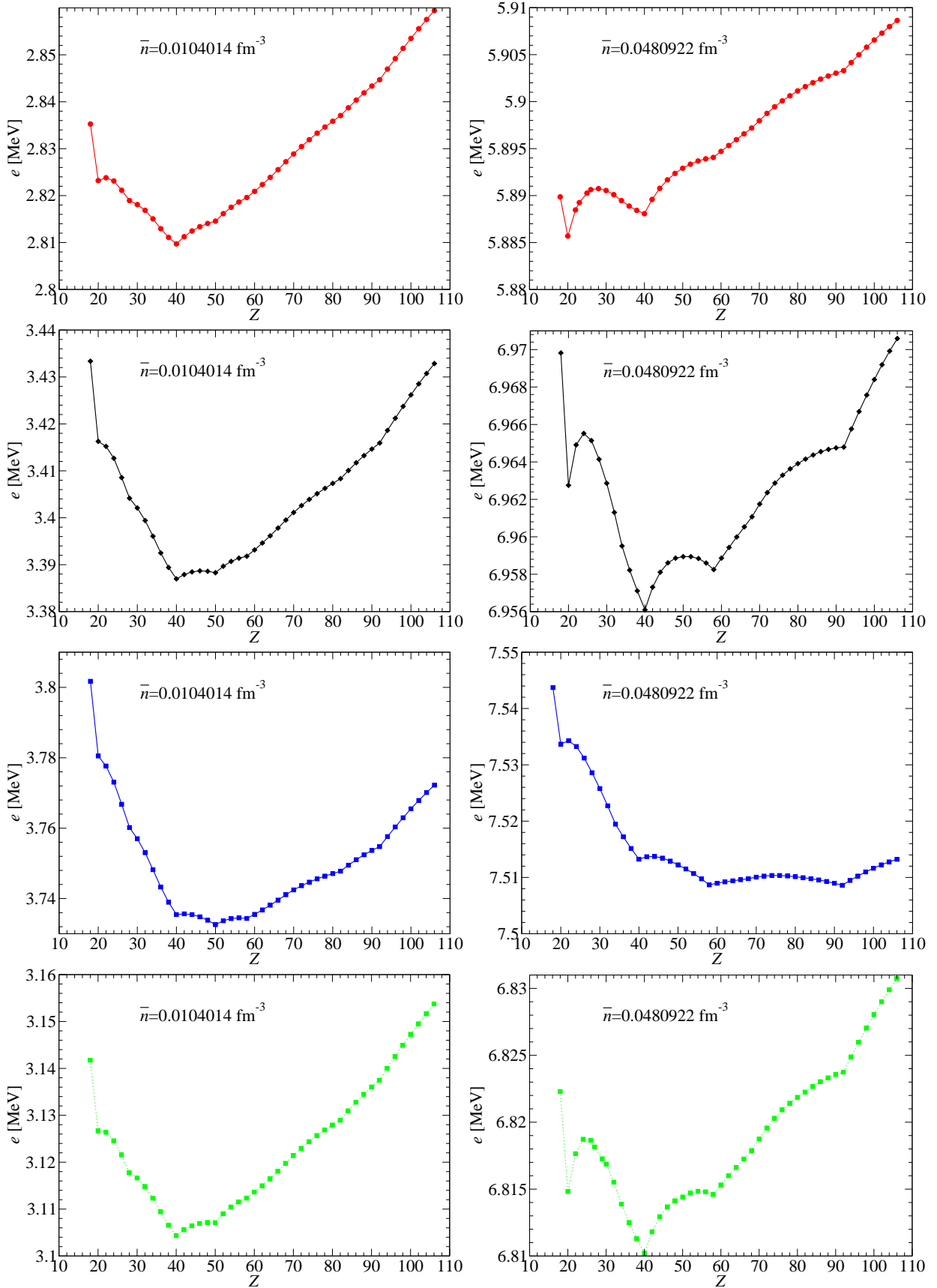

\includegraphics[width=.95\columnwidth]{e_vs_Z_bsk22a.eps}
\includegraphics[width=.95\columnwidth]{e_vs_Z_bsk22b.eps}
\includegraphics[width=.95\columnwidth]{e_vs_Z_bsk24a.eps}
\includegraphics[width=.95\columnwidth]{e_vs_Z_bsk24b.eps}
\includegraphics[width=.95\columnwidth]{e_vs_Z_bsk25a.eps}
\includegraphics[width=.95\columnwidth]{e_vs_Z_bsk25b.eps}
\includegraphics[width=.95\columnwidth]{e_vs_Z_bsk26a.eps}
\includegraphics[width=.95\columnwidth]{e_vs_Z_bsk26b.eps}
\caption{(Color online.) Energy per nucleon as a function of $Z$ at two 
different densities in the bound-proton region. The top two figures refer to
functional BSk22, the next two to BSk24, etc.}
\label{eZic}
\end{figure*}

Exceptional stability is seen in Fig.~\ref{Zic} for $Z$ = 20 and 40 in the case
of functional BSk22, 40 for BSk24 and BSk26, and 40 (at the neutron drip line),
50, 58 and 92 for BSk25 (see Tables~\ref{tabZ22}-\ref{tabZ26}). These 
`magic numbers' are reflected as local minima 
or at least as kinks in the plots of the energy per nucleon $e$ as a function 
of $Z$ that we show in Fig.~\ref{eZic} for two different mean densities
$\bar{n}$; note that at each $Z$ the energy is minimized with respect to the
neutron number $N$. The disappearance of the familiar magic numbers 28 and 82
associated with bound finite nuclei and the appearance of the unfamiliar        
numbers 58 and 92 is related to the very strong quenching of the proton         
spin-orbit splitting that we find in the presence of a large neutron excess.    
Experimental evidence for such a quenching has been found in neutron-rich       
bound finite nuclei \citep{sch04}.

While there are sharp variations in the equilibrium value of $Z$ as a function 
of density, the fact that the proton fraction $Y_\mathrm{p}$ varies smoothly without
any apparent shell effects (see Fig.~\ref{Yic}) means that the shell effects 
seen in $Z_\mathrm{eq}$ are manifesting themselves primarily as sharp variations in 
the \emph{size} of the WS cells. 

We stress that all these calculations assumed complete degeneracy, i.e., 
effectively  zero temperature. However, the calculations of 
\citet{ons08} suggest that the shell effects observed here 
should survive at realistic crust temperatures of 0.01 MeV.

\begin{table}
\centering
\caption{Principal values of $Z_\mathrm{eq}$ in the inner crust below proton drip
for BSk22.}
\label{tabZ22}
\begin{tabular}{|ccc|}\hline
$Z_\mathrm{eq}$&$\bar{n}_\mathrm{min}$ {[fm$^{-3}$]}&
$\bar{n}_\mathrm{max}$ {[fm$^{-3}$]}\\
\hline
40&2.69$\times10^{-4}$&0.0340\\
20&0.0350&0.0533\\
40&0.0544&0.591\\
\hline
\end{tabular}
\end{table}

\begin{table}
\centering
\caption{Principal value of $Z_\mathrm{eq}$ in the inner crust below proton drip
for BSk24 
(corrected as per the Erratum).}
\label{tabZ24}
\begin{tabular}{|ccc|}\hline
$Z_\mathrm{eq}$&$\bar{n}_\mathrm{min}$ {[fm$^{-3}$]}&
$\bar{n}_\mathrm{max}$ {[fm$^{-3}$]}\\
\hline
40&2.56$\times10^{-4}$&0.0715\\
\hline
\end{tabular}
\end{table}

\begin{table}
\centering
\caption{Principal values of $Z_\mathrm{eq}$ in the inner crust below proton drip
for BSk25 
(corrected as per the Erratum).}
\label{tabZ25}
\begin{tabular}{|ccc|}\hline
$Z_\mathrm{eq}$&$\bar{n}_\mathrm{min}$ {[fm$^{-3}$]}&
$\bar{n}_\mathrm{max}$ {[fm$^{-3}$]}\\
\hline
40&2.50$\times10^{-4}$&2.70$\times10^{-4}$\\
50&2.70$\times10^{-4}$&0.0138\\
58&0.0139&0.0474\\
92&0.0478&0.07663\\
138 & 0.0768 & 0.0806 \\
\hline
\end{tabular}
\end{table}

\begin{table}
\centering
\caption{Principal value of $Z_\mathrm{eq}$ in the inner crust below proton drip
for BSk26.}
\label{tabZ26}
\begin{tabular}{|ccc|}\hline
$Z_\mathrm{eq}$&$\bar{n}_\mathrm{min}$ {[fm$^{-3}$]}&
$\bar{n}_\mathrm{max}$ {[fm$^{-3}$]}\\
\hline
40&2.61$\times10^{-4}$&0.0730\\
\hline
\end{tabular}
\end{table}

\begin{table}
\centering
\caption{Onset of proton drip in the inner crust 
(corrected as per the Erratum).}
\label{tabdrip}
\begin{tabular}{|c|cccc|}\hline
&BSk22&BSk24&BSk25&BSk26\\
\hline
$\bar{n}$ {[fm$^{-3}$]}&0.059&0.073&0.081&0.074\\
\hline
\end{tabular}
\end{table}

\emph{Proton drip.} Unlike the calculations of \citet*{bbp}, based on
the  compressible liquid-drop model, we find that proton drip can occur
in the inner crust for all four of our functionals, Eq.~(\ref{3.6a})
becoming satisfied at  the densities indicated in Table~\ref{tabdrip}.
It will be seen from  Table~\ref{JL} that these drip densities are
tightly correlated with $L$, more so than with $J$, despite the loose
correlation that exists between these coefficients (see
\citealt{rogg18}).

To understand why we should 
find proton drip while \citet{bbp} do not, it should be noted that rather
than have a continuously varying density within the WS cell, as we do, the 
model of \citet{bbp} adopts a simpler, but less realistic, picture of just 
two distinct homogeneous phases. The tendency for `protons to uncluster' was 
observed by \citet*{buc71,buc72} from TF calculations. More recently, the 
appearance of free protons was also found in the ETF calculations of 
\citet{mu15}. 

\begin{figure}
\includegraphics[width=\columnwidth]{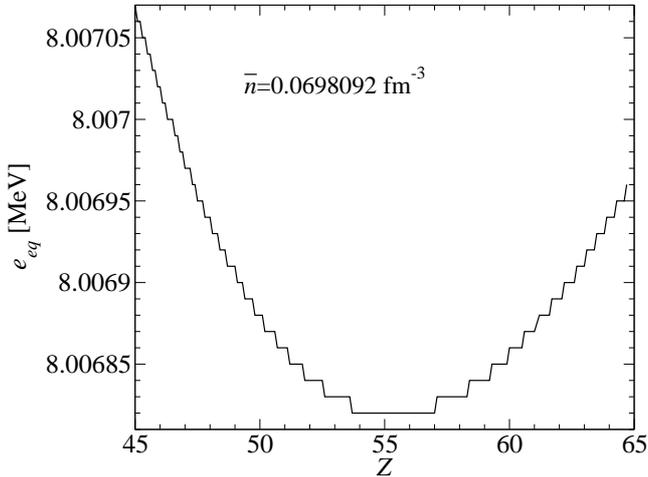}
\caption{(Color online.) Energy per nucleon as a function of $Z$ 
in the free-proton region for functional BSk24 at a mean density of $\bar{n}$ =
0.0698092 fm$^{-3}$.}
\label{eZic2}
\end{figure}

In the region of free protons the optimal value $Z_\mathrm{eq}$ of $Z$ is less well 
determined than in the bound-proton region, as a comparison of Fig.~\ref{eZic2}
with Fig.~\ref{eZic} shows. At any given density we take for our final value of
$Z_\mathrm{eq}$, as displayed in Figs.~\ref{Zic_zooma} and \ref{Zic_zoomb}, the 
arithmetic mean of the lowest and highest values of $Z$ for which the minimum 
value (to six significant figures) of the energy per nucleon $e$ is found. 

The curves in the free-proton region of the upper panels of  
Figs.~\ref{Zic_zooma} and \ref{Zic_zoomb} represent the fit of the computed 
values of $Z_\mathrm{eq}$ to Eq.~(\ref{Zcell_fit}) and of $Z_\mathrm{cl}$ to 
Eq.~(\ref{Zfree2}). The lower panels of these figures
show the deviations between these fits and the computed data.

Consulting the supplementary material 
shows that over the 
range of uncertainty in the value of $Z_\mathrm{eq}$, the proton fraction $Y_\mathrm{p} \equiv Z/A$ 
remains sensibly constant. This translates the uncertainty in $Z_\mathrm{eq}$ into an 
uncertainty in the cell size, a circumstance propitious to the development of 
non-spherical cell shapes, such as pasta. We return to this question in Section~\ref{cci}. 

The absence of shell effects above the proton-drip density is apparent in
Figs.~\ref{Zic_zooma} and~\ref{Zic_zoomb}, and reflects our decision to drop 
the proton shell corrections,
along with the pairing term, in this density domain. The rationale for this
was discussed in Section~\ref{iceos}, and is altogether consistent with our
treatment of the neutronic shell and pairing corrections. In any case, 
referring to the supplementary material 
shows that there is a general tendency for the Strutinsky-integral shell correction 
$E^\mathrm{sc, pair}_\mathrm{p}$ to decrease as the density increases to the point where 
proton drip sets in.

\begin{figure}
\includegraphics[width=\columnwidth]{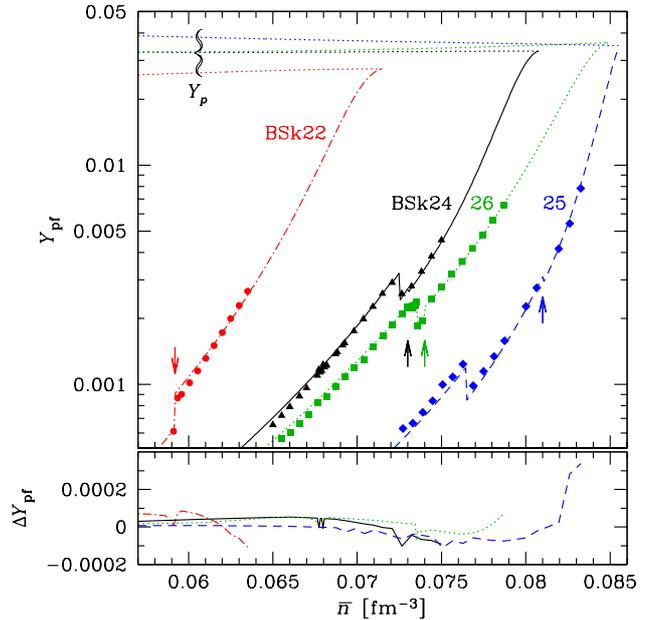}
\caption{(Color online.) Equilibrium fractions of nucleons that are free 
protons $Y_\mathrm{pf} =  Z_f/A$ in the inner crust as a function of mean baryon 
density $\bar{n}$ for our four functionals. 
Upper panel: Curves represent the fit of the computed values of $Y_\mathrm{pf}$ by 
analytic expressions, as described in the text. All data points are shown. 
The dotted lines represent the values of $Y_\mathrm{p}$ (ETF values below the proton
drip point). Lower panel: Deviations of curves from the
computed data points ($\Delta Y_\mathrm{pf}$  = fit $-$ data).
(Corrected version.)}
\label{pfic}
\end{figure}

Consulting now Table~\ref{tabdrip} for the specific  cases of functionals
BSk22, BSk24 and BSk25, we see that the proton drip density increases as the
symmetry energy $J$ decreases. Fig.~\ref{Zic_zooma} likewise shows that $Z$ is
consistently higher the lower $J$. From Fig.~\ref{Yic} it is seen that the 
anti-correlation found between $Y_\mathrm{p}$ and $J$ at lower densities is maintained 
in the free-proton region.  

Open symbols in Fig.~\ref{Zic} and its zooms (Figs.~\ref{Zic_zooma}
and ~\ref{Zic_zoomb}) denote the cluster component $Z_\mathrm{cl}$ of $Z_\mathrm{eq}$, given 
by Eq.~(\ref{3.2Ca}). These data points have been fitted by the analytic 
function (\ref{Zfree2}). The difference
\beq\label{eqpf}
Z_f = Z_\mathrm{eq} - Z_\mathrm{cl}
\eeq
represents the number of what can be described as free protons, i.e., those 
associated with the background term in Eq. (\ref{3.1}). It should be
realized that this number is non-vanishing, although small, even at densities
 below proton drip. 
In any case, the free proton fraction $Y_\mathrm{pf} = Z_f/A$  rises very rapidly with 
$\bar{n}$ above proton drip. The curves in Fig.~\ref{pfic} represent analytic 
fits to the computed $Y_\mathrm{pf}$, in which we combine Eq.~(\ref{eqpf}) with 
Eqs.~(\ref{Zcell_fit}) and (\ref{Zfree2}), and use Eqs.~(\ref{Zcell_fit}) and 
(\ref{Y_e_crust}) for $A = Z_\mathrm{eq}/Y_\mathrm{p}$. 

\begin{figure}
\includegraphics[width=\columnwidth]{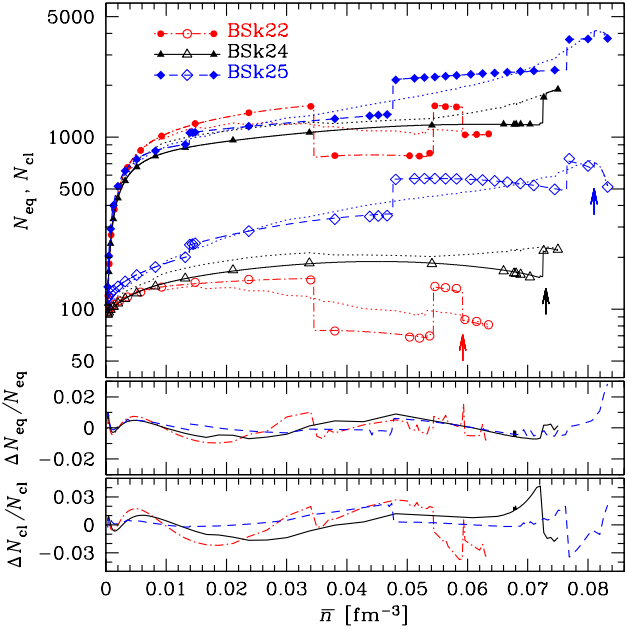}
\caption{(Color online.) Equilibrium values $N_\mathrm{eq}$ of number of neutrons and 
of cluster component $N_\mathrm{cl}$ of $N_\mathrm{eq}$ (defined in Eq.~(\ref{3.2Cb})) in 
inner crust as a function of mean 
baryon density $\bar{n}$ for functionals BSk22, BSk24 and BSk25. 
Upper panel: Solid symbols represent $N_\mathrm{eq}$, open symbols $N_\mathrm{cl}$. For 
clarity only every fourth data point is shown. Arrows indicate onset of proton 
drip. Upper set of solid curves represent analytic fit to computed values of
$N_\mathrm{eq}$; lower set to analytic fit to computed values of $N_\mathrm{cl}$ 
(see text for details). Dotted curves refer to ETF values.
Lower panels: Deviations between analytically fitted curves and computed data
points ($\Delta N_\mathrm{eq}$ and $\Delta N_\mathrm{cl}$ = fit $-$ data).
(Corrected version.)}
\label{Nica}
\end{figure}

\begin{figure}
\includegraphics[width=\columnwidth]{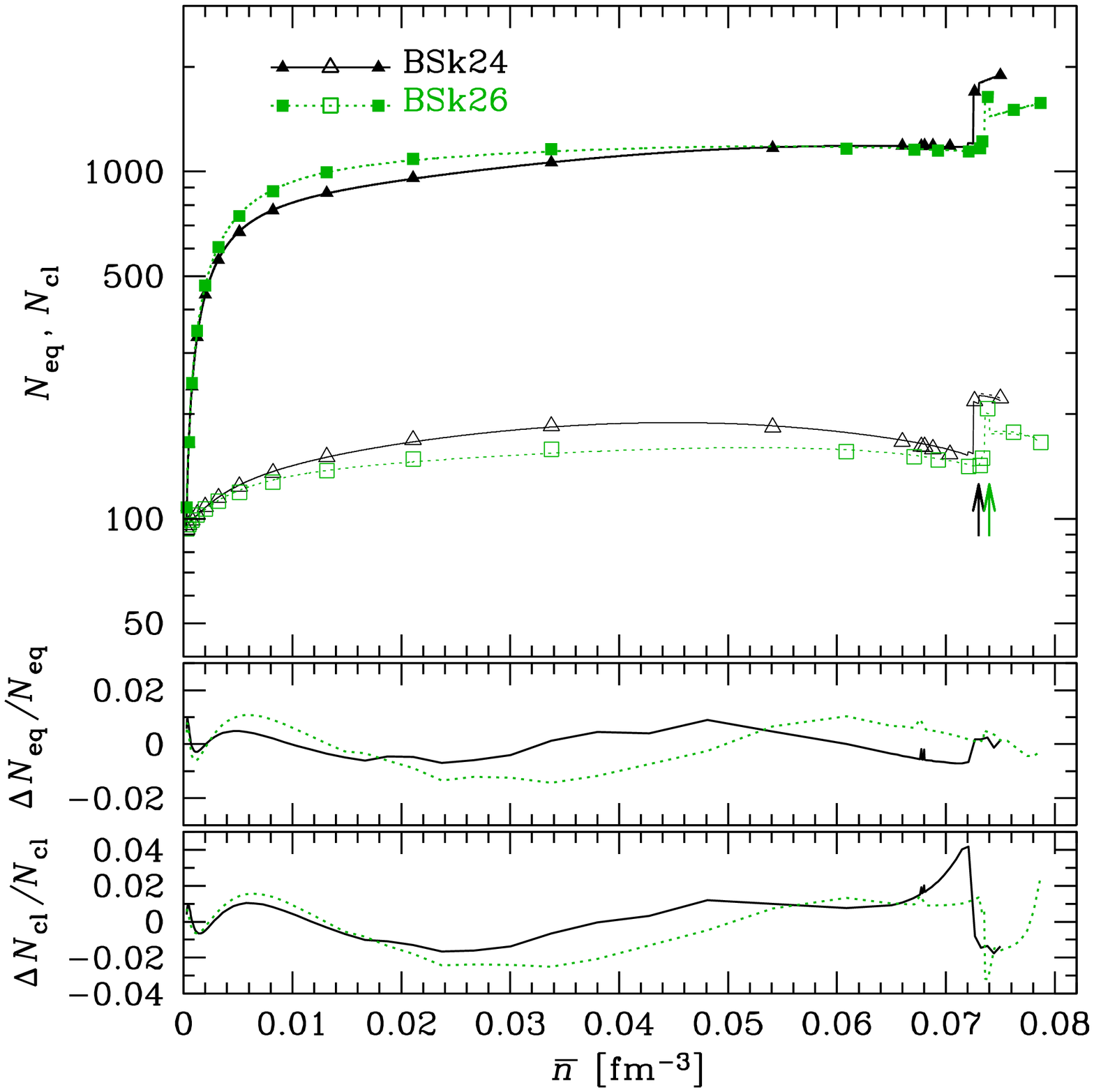}
\caption{(Color online.) Equilibrium values $N_\mathrm{eq}$ of number of neutrons and
of cluster component $N_\mathrm{cl}$ of $N_\mathrm{eq}$ (defined in Eq.~(\ref{3.2Cb})) in
inner crust as a function of mean baryon density $\bar{n}$ for functionals BSk24 
and BSk26. Upper panel: Solid symbols represent $N_\mathrm{eq}$, open symbols $N_\mathrm{cl}$. For
clarity only every fourth data point is shown. Arrows indicate onset of proton
drip. Upper set of solid curves represent analytic fit to computed values of
$N_\mathrm{eq}$; lower set to computed values of $N_\mathrm{cl}$ (see text for details).
Dotted curves refer to ETF values. Lower panels: Deviations between analytically 
fitted curves and computed data points ($\Delta N_\mathrm{eq}$ and 
$\Delta N_\mathrm{cl}$ = fit $-$ data).
(Corrected version.)}  
\label{Nicb}
\end{figure}

\begin{figure}
\includegraphics[width=\columnwidth]{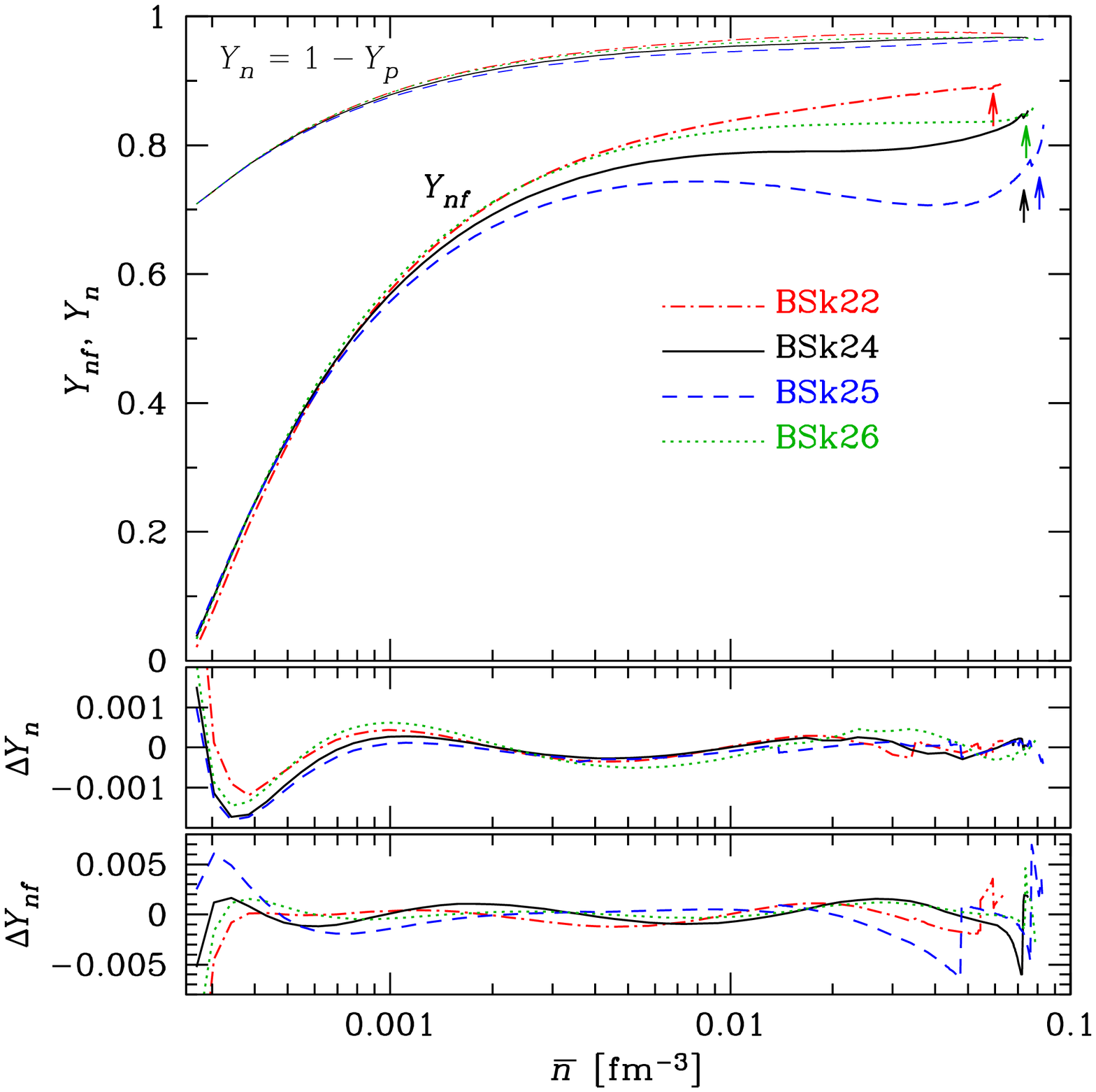}
\caption{(Color online.) Variation of equilibrium fraction of nucleons that
are free neutrons $Y_\mathrm{nf} = N_f/A$ in the inner crust as a function of mean 
baryon density $\bar{n}$ for our four functionals. Upper panel: Curves labelled
$Y_\mathrm{nf}$ represent the computed data for $Y_\mathrm{nf}$; arrows indicate onset of 
proton drip. This panel also shows the neutron fraction $Y_\mathrm{n} = N_\mathrm{eq}/A =
1 - Y_\mathrm{p}$. Lower panels: deviations of fitted curves, Eq.~(\ref{Ynf}),
from the computed data points ($\Delta Y_\mathrm{n}$ and $\Delta Y_\mathrm{nf}$ = fit $-$ data).
(Corrected version.)}
\label{nfic}
\end{figure}

\emph{Equilibrium value of $N$.} Given that the proton fraction $Y_\mathrm{p}$ varies
smoothly it follows that the number of neutrons $N_\mathrm{eq}$ per WS cell in the
inner crust at equilibrium must display the same shell effects as does the
proton number $Z_\mathrm{eq}$. This is confirmed in Figs.~\ref{Nica} and~\ref{Nicb}. 
The discontinuities seen in $N_\mathrm{eq}$ are a direct consequence of the proton
shell effects, given that $Y_\mathrm{p}$ varies continuously.
The analytic expressions for $N_\mathrm{eq}$ represented by the
upper set of curves in Figs.~\ref{Nica} and~\ref{Nicb} correspond to the
quantities appearing in Eq.~(\ref{jmp1}).   

We can also define $N_\mathrm{cl}$, the cluster component of $N_\mathrm{eq}$, given by 
Eq.~(\ref{3.2Cb}), and then 
\beq\label{eqnf}
N_f = N_\mathrm{eq} - N_\mathrm{cl}  \quad  ,
\eeq
the number of free neutrons in the WS cell. With the ratio $Y_\mathrm{nf} = N_f/A$
parametrized according to Eq.~(\ref{Ynf}), $N_\mathrm{cl}$ becomes parametrized
through Eq.~(\ref{jmp3}), which defines the lower set of curves in 
Figs.~\ref{Nica} and~\ref{Nicb}. 

The computed ratio $Y_\mathrm{nf} = N_f/A$ is plotted separately in Fig.~\ref{nfic}; 
this figure also shows the ratio $Y_\mathrm{n} =  N_\mathrm{eq}/A =1 - Y_\mathrm{p}$.  

\begin{figure}
\includegraphics[width=\columnwidth]{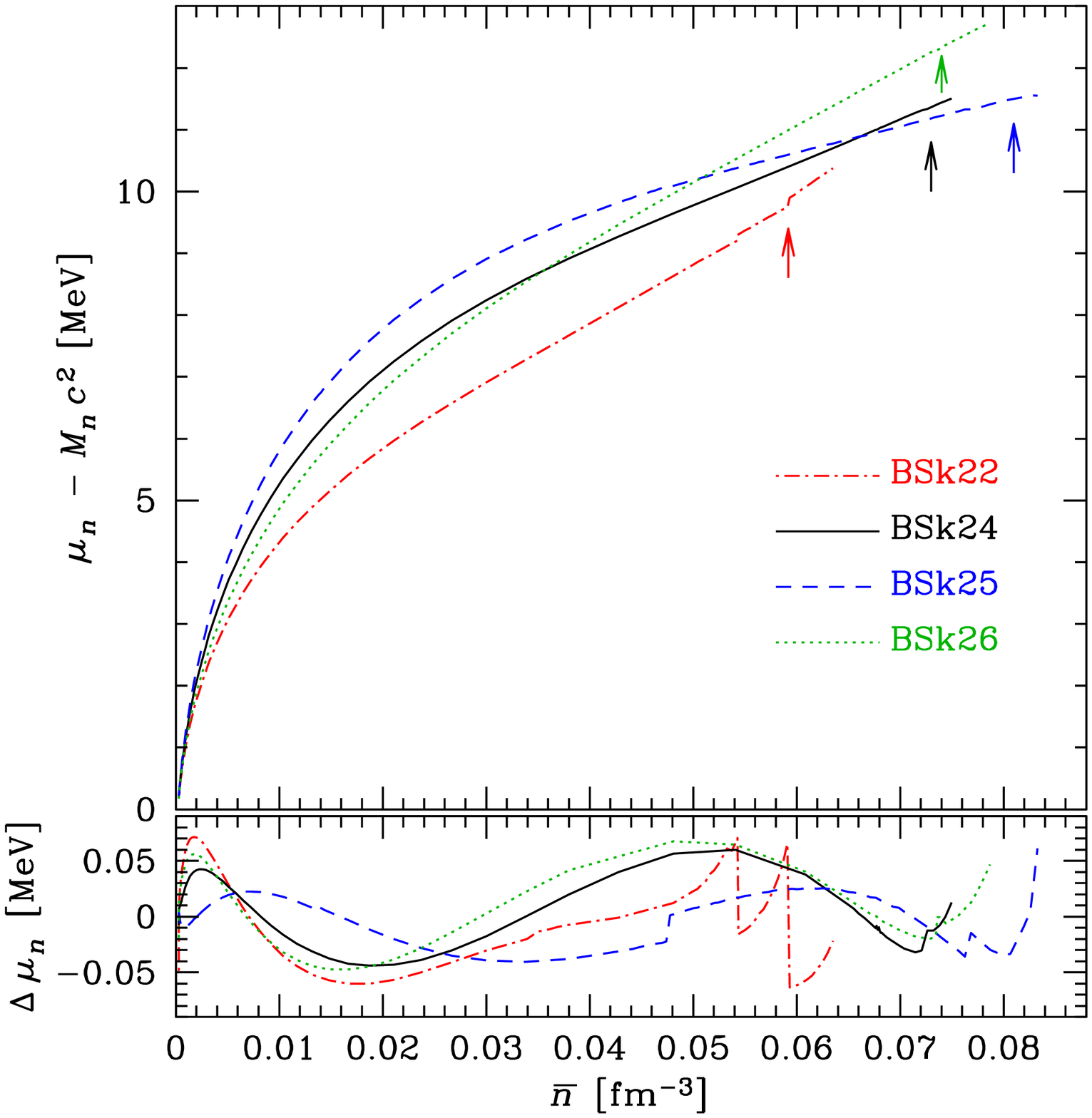}
\caption{(Color online.) Upper panel: Curves show the computed values of the
neutron chemical potential $\mu_\mathrm{n}$ in the inner crust as a function of mean 
baryon density $\bar{n}$ for our four functionals; arrows indicate onset of 
proton drip. Lower panel: Deviations between the computed data and the 
analytic function (\ref{munsfit1}) ($\Delta \mu_\mathrm{n}$ = fit $-$ data).
(Corrected version.)}
\label{munic}
\end{figure}

\begin{figure}
\includegraphics[width=\columnwidth]{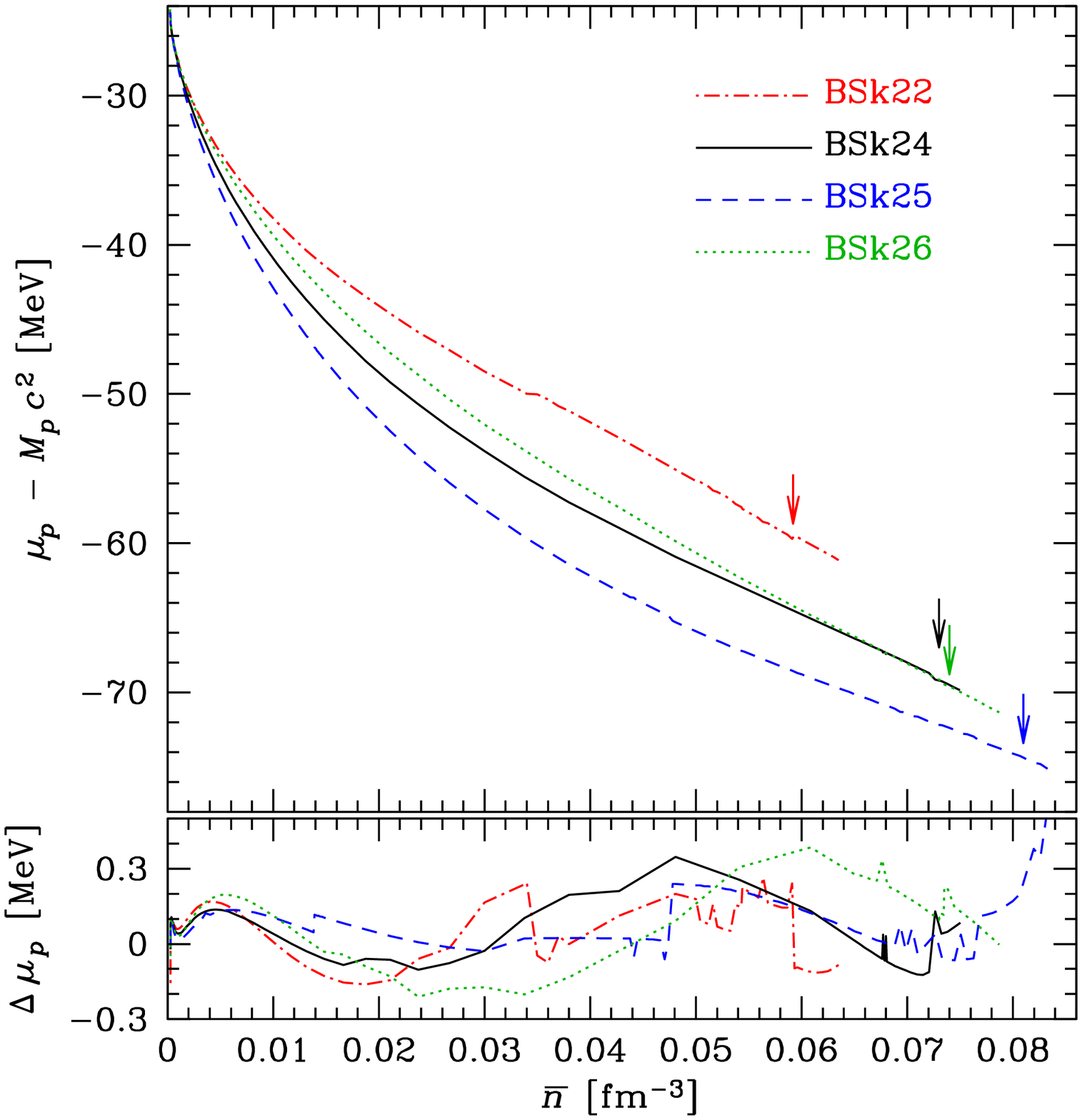}
\caption{(Color online.) Upper panel: Curves show the computed values of the
proton chemical potential $\mu_\mathrm{p}$ in the inner crust as a function of mean
baryon density $\bar{n}$ for our four functionals; arrows indicate onset of
proton drip. Lower panel: Deviations between the computed data and the
analytic fit described in the text ($\Delta \mu_\mathrm{p}$ = fit $-$ data).
(Corrected version.)}
\label{mupic}
\end{figure}

\emph{Neutron chemical potential.} The neutron chemical potential $\mu_\mathrm{n}$ is 
calculated (for configurations in beta equilibrium) using Eq.~(\ref{mung}), and
its variation over the inner crust is shown in Fig.~\ref{munic} for our four 
functionals (with the rest-mass energy of the neutron subtracted). For all four
functionals the trend is for $\mu_\mathrm{n}$ to increase monotonically with $\bar{n}$, 
following the growing neutron excess. 

The analytic function (\ref{munsfit1}) has been fitted to our computed values 
of $\mu_\mathrm{n}$; the lower panel of Fig.~\ref{munic} shows the deviations between
this fit and the computed data. 

\emph{Proton chemical potential.} This is calculated (for configurations in beta
equilibrium) from the calculated values of $\mu_\mathrm{n}$ and $\mu_\mathrm{e}$, using 
Eq.~(\ref{mupg}), and its variation over the inner crust is shown in 
Fig.~\ref{mupic} for all four of our functionals (with the rest-mass energy of 
the proton subtracted).  
The general trend for $\mu_\mathrm{p}$ is the opposite to that for $\mu_\mathrm{n}$: it becomes 
more and more negative with increasing $\bar{n}$, following the growing neutron
excess. The $J$-dependence also is reversed, with higher $J$ being associated 
with higher $\mu_\mathrm{p}$, simply because $Y_\mathrm{p}$ and hence $\mu_\mathrm{e}$ are lower for
higher values of $J$.  

\begin{table*}
\centering
\caption{Comparison of inner-crust and outer-crust codes at the neutron
drip-point density $\bar{n}_\mathrm{nd}$ of the functional in question; results 
for latter code in
parentheses (except for the neutron-drip density). $e_\mathrm{eq}$ is the internal 
energy per nucleon (with the neutron rest-mass
energy $M_\mathrm{n}c^2$ subtracted), and $P$ the pressure.}
\label{tabinout}
\begin{tabular}{|c|c|c|c|c|c|}
\hline
Force &$\bar{n}_\mathrm{nd}$ (fm$^{-3}$) &$Z$ & $N$ & $e_\mathrm{eq}$ (MeV)& $P$ (MeV fm$^{-3}$)\\
\hline
BSk22 & $2.69\times10^{-4}$& 40 (38)& 98 (90)& $-1.78$ ($-1.85$) & $5.10\times10^{-4}$ ($4.99\times10^{-4}$)\\
BSk24 & $2.56\times10^{-4}$& 40 (38)& 94 (86)&$-1.83$ ($-1.90$)& $4.95\times10^{-4}$ ($4.87\times10^{-4}$)\\
BSk25 & $2.50\times10^{-4}$& 40 (38)& 93 (84)&$-1.85$ ($-1.93$)& $4.85\times10^{-4}$ ($4.83\times10^{-4}$)\\
BSk26 & $2.61\times10^{-4}$& 40 (38)& 95 (88)& $-1.80$ ($-1.87$) & $5.04\times10^{-4}$ ($4.90\times10^{-4}$)\\
\hline
\end{tabular}
\end{table*}

\subsection{Matching between inner and outer crust.}

Our inner-crust code, as used here, is in principle applicable to the outer
crust, and it is thus meaningful to compare this code with the HFB code that we
used for the outer-crust calculation of Section~\ref{outres}. In 
Table~\ref{tabinout} we make this comparison at the drip-point density for the 
functional in question; the results for the outer-crust code are shown in 
parentheses. The values for $\bar{n}_\mathrm{nd}$ and $P$ from our outer-crust code are slightly different from those previously obtained by \citet{fcm16} without all the corrections considered here. 

It will be seen that there is a slight disagreement between our inner- and 
outer-crust codes 
concerning the values of $Z$ and $N$ at the drip point: for all four 
functionals we find $Z$ = 40 rather than 38, and somewhat bigger discrepancies
for $N$. This can be attributed to the several approximations made in our 
ETFSI method for calculating the internal energy per nucleon $e$ in the inner
crust; we see from  Table~\ref{tabinout} that the inner-crust code (ETFSI) 
underbinds with respect to the outer-crust code (HFB) by around 4 \%; this
rises to about 5\% if we take the same values of $Z$ and $N$ in the two codes.
(The corresponding figures for the pressure are 2\% and 5\%.)

We recall that the approximations made in the ETFSI method, but not in the HFB 
method adopted in our outer-crust calculations, are as follows. i) The
kinetic energy and spin currents are calculated with the semiclassical ETF 
method. ii) Proton shell corrections are put in perturbatively, and neutron 
shell corrections (shown to be much smaller than proton shell corrections as 
soon as neutron drip sets in \citep{ch06,cha07}, but obviously not zero, in the 
outer crust) are neglected completely. This source of discrepancy between the
two codes will be maximal at the interface between the inner and outer crust,
since the neglected neutron shell effects will decrease as the density 
increases. iii) Rather than allowing arbitrary density variations when 
minimizing the total energy, the density is parametrized according to 
Eqs.~(\ref{3.1}) and (\ref{3.2}). iv) Proton pairing is treated approximately,
while neutron pairing is neglected completely. v) The collective and Wigner 
correction terms that are added to the HFB energy in the outer-crust calculations 
are neglected in the inner crust. 

On the other hand, we have checked that errors arising
from the assumption of sphericity in the inner-crust code are negligible 
in this region of the nuclear chart.

We must stress that even if there is a disagreement of around 4\% between the
two codes the energy \emph{differences} between adjacent values of $Z$ and $N$
calculated by the inner-crust code are much more precise: the perceived
regularities suggest at least 5-figure, and possibly 6-figure accuracy: see,
for example, Fig.~\ref{eZic}.

\subsection{Core}
\label{coreres}

\begin{figure}
\includegraphics[width=\columnwidth]{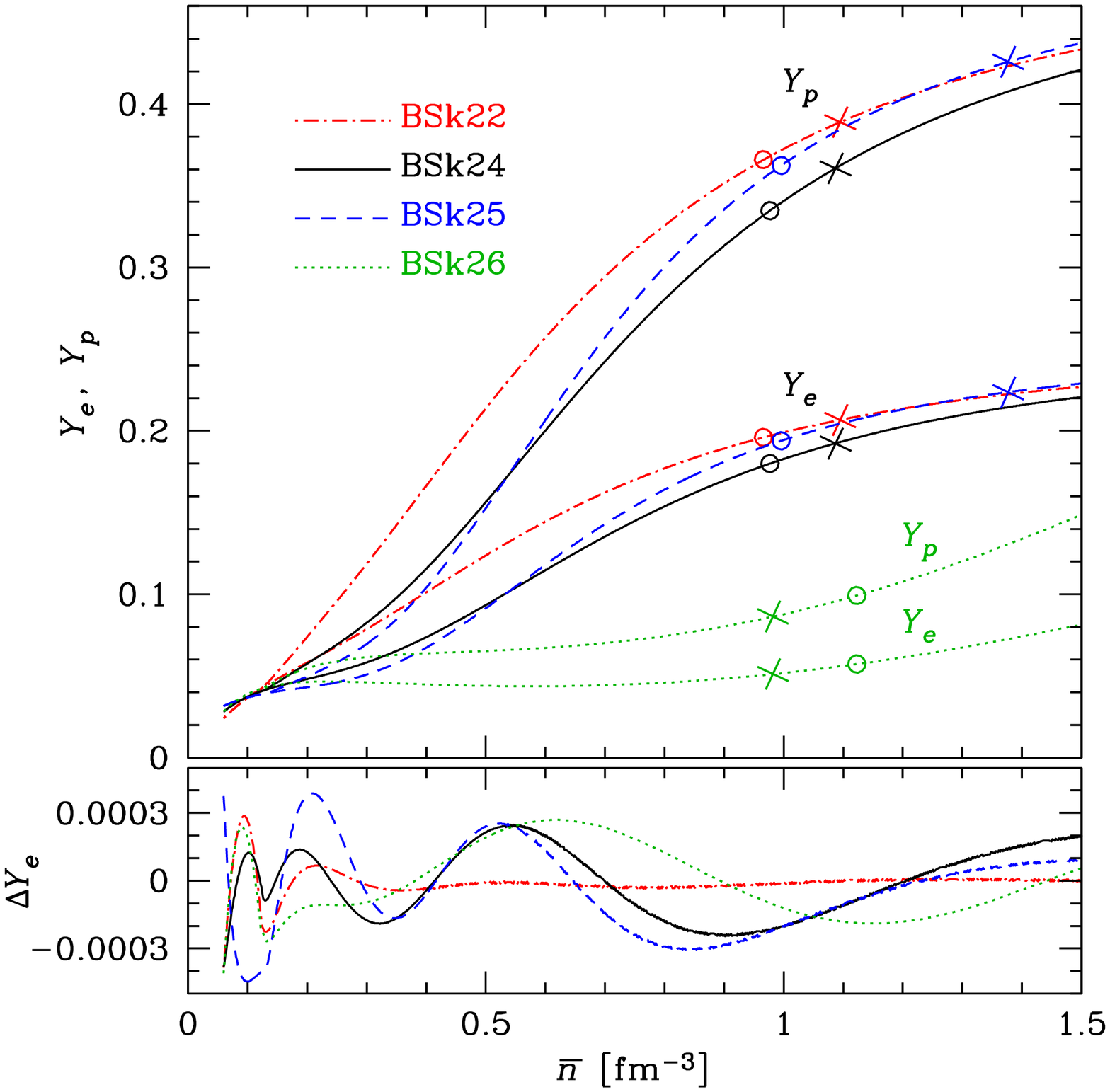}
\caption{(Color online.) Upper panel: Curves show the computed values of $Y_\mathrm{p}$ 
and $Y_\mathrm{e}$ in the core as a function of mean baryon density $\bar{n}$ for our 
four functionals. Circles represent the central density for the neutron star 
with the maximum possible mass; crosses represent the causality limit. Lower 
panel: Deviations between computed data points and the analytic fits of 
Eq.~(\ref{Y_e_core}) ($\Delta Y$  = fit $-$ data).}
\label{ycore}
\end{figure}

Our results for the equilibrating values in the core of the number of protons 
$Y_\mathrm{p}$ and of the number of electrons $Y_\mathrm{e}$ per nucleon, calculated as described
in Section~\ref{corecalc}, are shown for our four functionals in 
Fig.~\ref{ycore}; the difference 
between the two sets of curves represents the number of muons 
$Y_{\mu} = Y_\mathrm{p} - Y_\mathrm{e}$ per nucleon. This figure also indicates the densities 
corresponding to the breakdown of causality and the maximum neutron-star mass
for the different functionals (see Section~\ref{gross}). The lower panel 
indicates the deviations
between our calculated results and the analytic fits to these results given
by Eq.~(\ref{Y_e_core}). The most striking feature of this figure is the fact 
that in the core both 
$Y_\mathrm{p}$ and $Y_\mathrm{e}$ increase with increasing density, in contrast to what happens 
in both the outer and inner crusts (see Figs.~\ref{fig:Yp_out} and~\ref{Yic}).

Comparing functionals BSk22, BSk24 and BSk25 to look for a possible 
$J$-dependence, we see from Fig.~\ref{fig2} that at all core densities there is
a close correlation of $Y_\mathrm{p}$ (and $Y_\mathrm{e}$) with the symmetry energy $S(n)$, which
is only partially correlated with $J$. There is, however, a tendency for the 
three functionals (especially BSk22 and BSk25) to converge at high densities, 
reflecting the fact that all three have been constrained to the same EoS of 
NeuM. In fact, the BSk26 curve in Fig.~\ref{ycore} shows that the NeuM 
constraint has much more influence than the choice of $J$: the softer EoS of 
NeuM favours higher neutron excesses, and thus lower values of $Y_\mathrm{p}$ (and $Y_\mathrm{e}$).  

\begin{figure}
\includegraphics[width=\columnwidth]{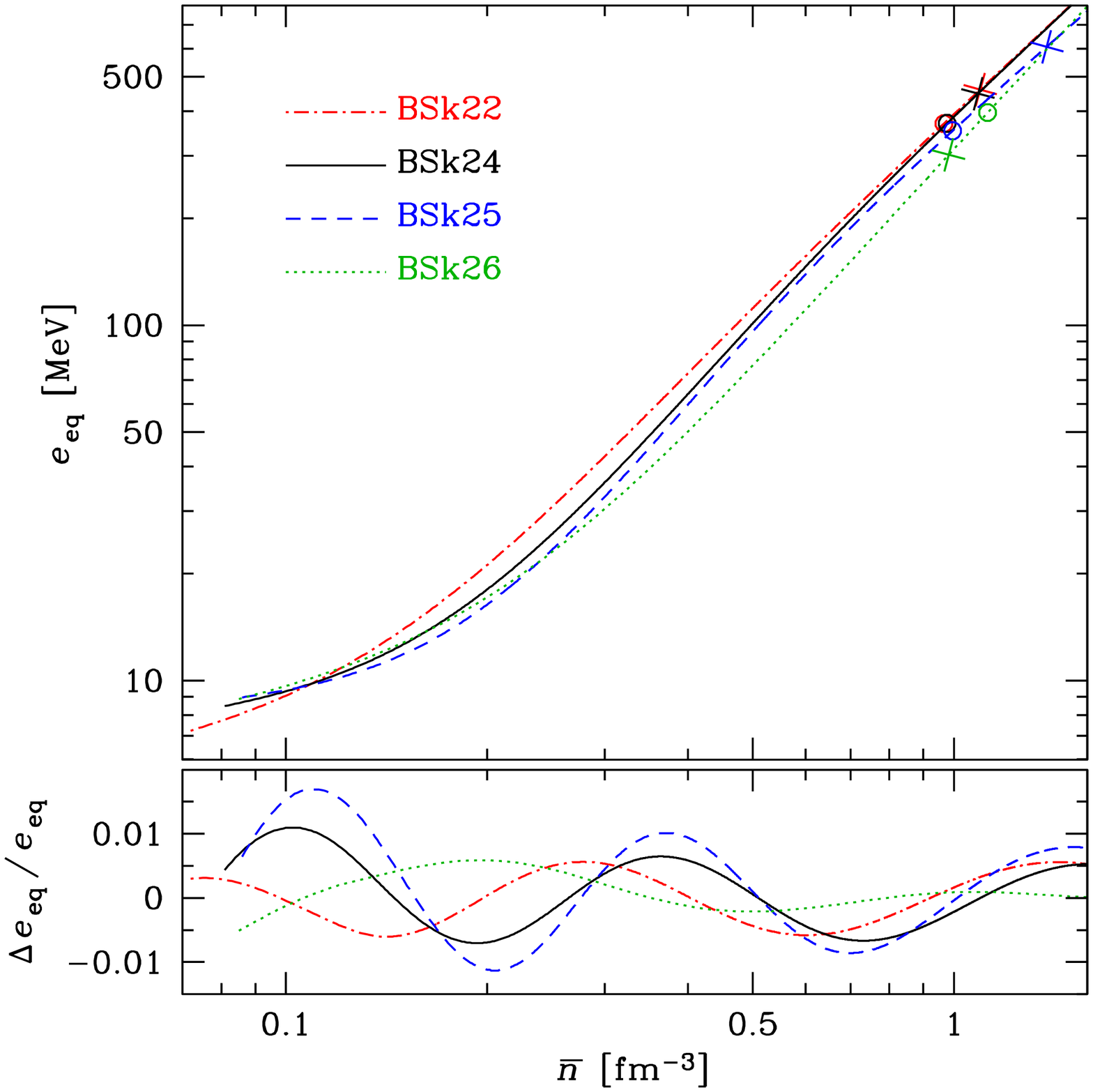}
\caption{(Color online.) Upper panel: Curves show the calculated values of 
$e_\mathrm{eq}$ in the core as a function of mean baryon density $\bar{n}$ for our 
four functionals. Circles represent the central density for the neutron star 
with the maximum possible mass; crosses represent the causality limit. Lower 
panel: Deviations between computed data points and the 
analytic fit of Eq.~(\ref{Efit}) ($\Delta e_\mathrm{eq}$  = fit $-$ data).}
\label{ecore}
\end{figure}

\begin{figure}
\includegraphics[width=\columnwidth]{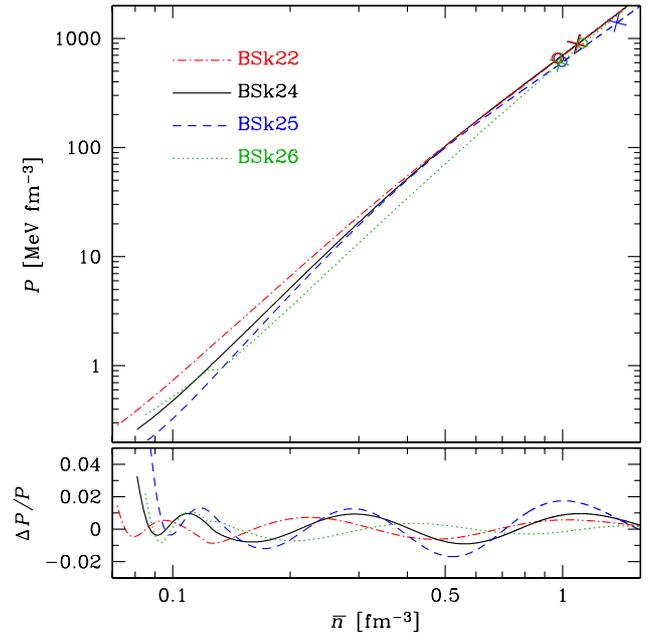}
\caption{(Color online.) Upper panel: Curves show calculated values of $P$ in 
the core as a function of mean baryon density $\bar{n}$ for our four 
functionals. Circles represent the central density for the neutron star 
with the maximum possible mass; crosses represent the causality limit.
Lower panel: Fractional deviations between computed data points and the 
analytic fit of Eq.~(\ref{Pfit}) ($\Delta P$  = fit $-$ data).}
\label{Pcore}
\end{figure}

The energy per nucleon $e_\mathrm{eq}$ and the pressure $P$ at equilibrium in the core
are shown for our four functionals 
in Figs.~\ref{ecore} and~\ref{Pcore}, respectively. In the lower panels of 
these two figures we see the deviations between the calculated values and the 
analytic fits of Eqs.~(\ref{Efit}) and ~(\ref{Pfit}), respectively. We stress 
once again that these analytic fits to $e_\mathrm{eq}$ and  $P$ are valid over the 
entire star. Comparing these two figures with Figs.~\ref{eic} and~\ref{Pic}, 
respectively, we see that BSk22, BSk24 and BSk25 behave much more similarly in 
the core than in the inner crust; this simply reflects the fact that all three 
have been constrained to the same EoS of NeuM. Not surprisingly, in both 
Figs.~\ref{ecore} and~\ref{Pcore}, BSk26 is seen to be somewhat softer
than the other three functionals, but one might have expected a stronger 
dependence on the NeuM constraint. The point is that the softer EoS of NeuM is
partially offset by the higher asymmetries.   

\begin{figure}
\includegraphics[width=\columnwidth]{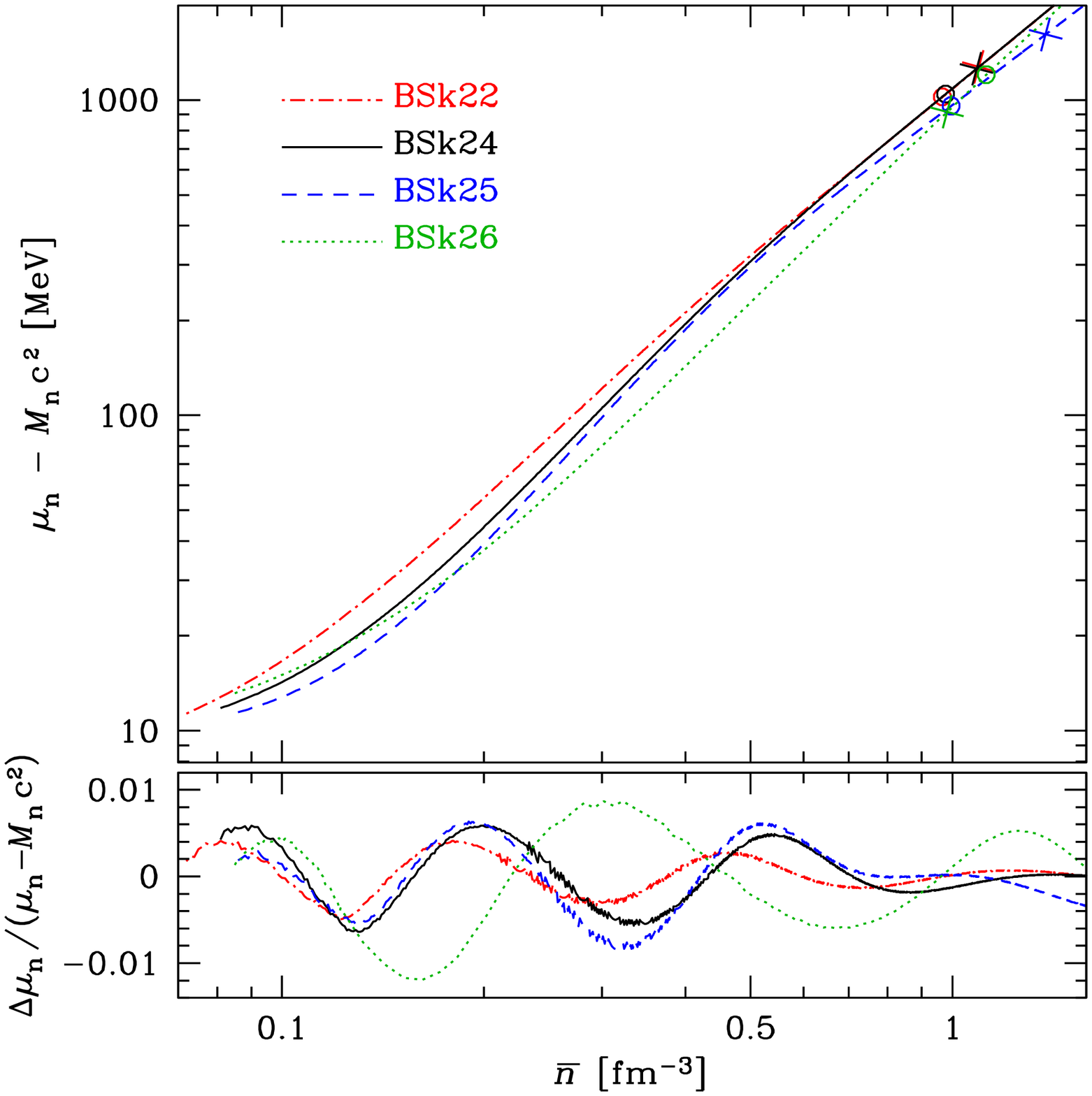}
\caption{(Color online.) Upper panel: Curves show calculated values of 
$\mu_\mathrm{n}$ in the core as a function of mean baryon density $\bar{n}$ for our four 
functionals. Circles represent the central density for 
the neutron star with the maximum possible mass; crosses represent the 
causality limit. Lower panel: Deviations between the computed data and the 
analytic fit of Eq.~(\ref{mu_core_fit}) ($\Delta \mu_\mathrm{n}$  = fit $-$ data).}
\label{muncore}
\end{figure}

\begin{figure}
\includegraphics[width=\columnwidth]{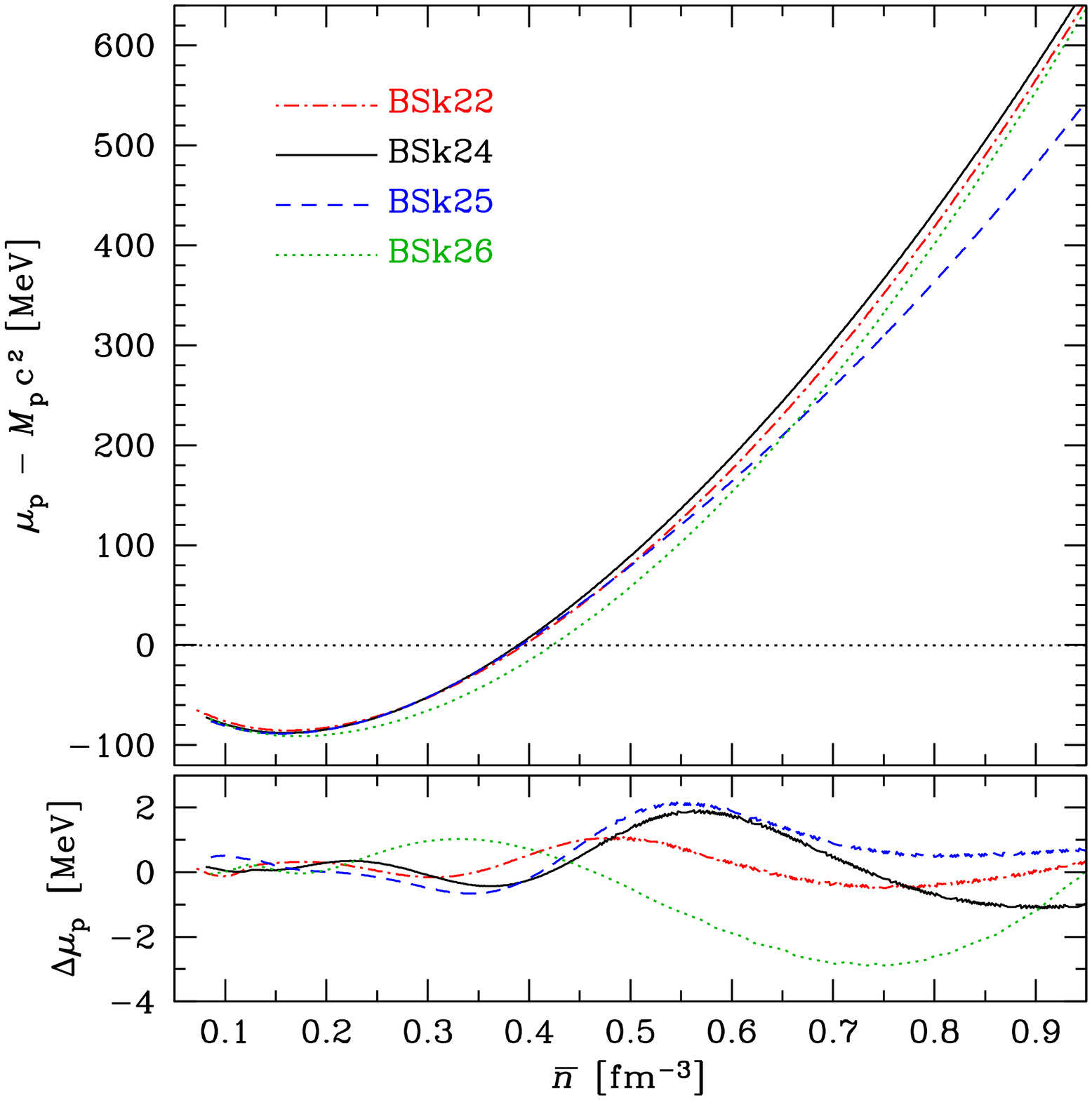}
\caption{(Color online.) Upper panel: Curves show calculated values of
$\mu_\mathrm{p}$ in the core as a function of mean baryon density $\bar{n}$ for our four
functionals. Circles represent the central density for
the neutron star with the maximum possible mass; crosses represent the
causality limit. Lower panel: Deviations between computed data and the
analytic fit given by Eqs.~(\ref{mu_core_fit}) and (\ref{mupg}) 
($\Delta \mu_\mathrm{p}$  = fit $-$ data).}
\label{mupcore}
\end{figure}

Figs.~\ref{muncore} and~\ref{mupcore} show for our four functionals the 
variation over the core of the neutron chemical potential $\mu_\mathrm{n}$ and the 
proton chemical potential $\mu_\mathrm{p}$, respectively.
It will be seen that the slope of $\mu_\mathrm{p}$ turns positive in this region of the star. 
A comparison of
BSk22, BSk24 and BSk25 shows that the choice of symmetry coefficient $J$ has 
little systematic impact on the values of the chemical potentials in the core.
On the other hand, both $\mu_\mathrm{n}$ and $\mu_\mathrm{p}$ tend to be lower for BSk26, the
functional fitted to the softer EoS of NeuM. In the lower panel of 
Fig.~\ref{muncore} we see the deviations between the calculated values of
$\mu_\mathrm{n}$ and the analytic fit of Eq.~(\ref{mu_core_fit}). The deviation shown
in the lower panel of Fig.~\ref{mupcore} for $\mu_\mathrm{p}$ is calculated from the
analytic expression (\ref{mu_core_fit}) for neutrons using the beta-equilibrium
condition (\ref{mupg}), as explained in Appendix~\ref{analytic}.   

\subsection{Crust-core interface}
\label{cci}

\begin{table*}
\centering
\caption{Parameters relating to the crust-core transition.}
\label{tabcd}
\begin{tabular}{|c|cccc|}
\hline
Functional &$n_\mathrm{cc}$ (fm$^{-3}$)&$Y_\mathrm{e, cc}$ &$P_{cc}$ (MeV fm$^{-3}$)
&$n_\mathrm{cc}^\prime$ (fm$^{-3}$)\\
\hline
BSk22 &0.0716068&0.028294&0.290934&0.0705000\\
BSk24 &0.0807555&0.033671&0.267902&0.0790740\\
BSk25 &0.0855534&0.035829&0.210878&0.0852758\\
BSk26 &0.0849477&0.035721&0.363049&0.0835000\\
\hline
\end{tabular}
\end{table*}

In Table~\ref{tabcd} we tabulate for each of our four functionals the values
for the density of the crust-core transition $n_\mathrm{cc}$, along with the
corresponding values of $Y_\mathrm{e, cc}$ and the pressure $P_{cc}$, as 
calculated by the method of \citet*{duc07}. This is the same method that was 
used by \citet{pcgd12}, (where $n_\mathrm{cc}$ was denoted by
 $n_\mathrm{trans}^{N*M}$),  and it consists of a determination of the 
conditions for homogeneous N*M to be unstable against breakup into finite-size clusters (rather than the conditions for equilibrium between infinite homogeneous phases). The last column of Table~\ref{tabcd} shows 
$n_\mathrm{cc}^\prime$, the lowest density for which we have found no 
inhomogeneous solutions with our inner-crust code, by which
we mean solutions with values of the inhomogeneity parameter $\Lambda$, defined
in Eq.~(\ref{3.15}), greater than 10$^{-7}$ (typically, this value of $\Lambda$
is found by our code for densities known to correspond to N*M). For all four
functionals we find $n_\mathrm{cc}^\prime$ to be slightly smaller than $n_\mathrm{cc}$, which
means that N*M remains stable against breakup down to slightly lower densities
than predicted by the method of \citet{duc07}. In this respect it should be
pointed out that this method is perturbative, assuming small-amplitude
fluctuations of the density. On the other hand it is possible that we are
missing some inhomogeneous solutions because of the discrete grid that we take
for the proton number $Z$ (see Section~\ref{iceos}), although it is unlikely
that this would happen for all four functionals.

For all four functionals we begin to find, as the density approaches 
$n_\mathrm{cc}^\prime$, equilibrium solutions that are still sensibly inhomogeneous, 
with values of $\Lambda$ of the order of 10$^{-2}$), but that are mechanically 
unstable, in the sense that the calculated pressure decreases with an increase 
in $\bar{n}$. The situation is resumed in Table~\ref{tabanom}, where we show 
for each functional the highest density beyond which instability sets in.
Instability of this sort is typical of what happens at a
first-order phase change when minimizing the Helmholtz free energy, rather than 
the Gibbs energy.  The actual physical situation could be restored by a Maxwell 
construction, but this is beyond the scope of the present paper. Thus all our
inner-crust results for densities higher than those indicated in 
Table~\ref{tabanom} are excluded from the supplementary tables 
and from the results shown in Section~\ref{icres}.

\begin{table*}
\centering
\caption{Highest densities in inner crust beyond which instabilities set    
in.}
\label{tabanom}
\begin{tabular}{|c|ccccc|}
\hline
Functional &$\bar{n}$ (fm$^{-3}$)&$Z_\mathrm{eq}$ & $e_\mathrm{eq}$ (MeV)& $\Lambda$ &$P$
(MeV fm$^{-3}$) \\
\hline
BSk22 &0.0635018 & 28.5 & 6.75925 & 0.02798 & 0.23010\\
BSk24 &0.0749994 & 64.6 & 8.23139 & 0.02504 &  0.24594\\
BSk25 &0.0832615 & 139.1 &8.88852 & 0.01830  & 0.22250\\
BSk26 &0.0786984 & 56.5  &8.56597 & 0.02355& 0.32862\\
\hline
\end{tabular}
\end{table*}

It is nevertheless of interest to speculate on the nature of these 
instabilities in our inner-crust results. Phase changes occur throughout the 
inner crust, at every point where the number of protons $Z$ changes, but only 
at higher densities near the transition to homogeneous matter do we see any 
instabilities. Thus the phase changes associated with changes in $Z$ must 
either be of higher order, or else, if of first order, too weak to be visible 
in our calculations. This suggests that the instabilities that we do see in our
calculations involve something more significant than a change in $Z$, and an 
obvious possibility is that changes to non-spherical pasta phases are being 
signalled, in the sense that this is what we would see if our code allowed it. 
This interpretation is strengthened by the fact that some of our solutions in 
the unstable region correspond to spherical bubbles, which means that 
non-spherical configurations such as lasagna or spaghetti might be favoured at 
slightly lower densities, since spherical bubbles are generally the densest 
types of pastas to appear \citep{rpw83,hashi84}. However, with the WS cells of 
our code being limited to spherical shapes we are unable to investigate this 
question any further here.

It is, of course, possible that the apparent bubbles we find may simply be a
result of numerical error, since it will be seen from the supplementary
material 
that in this region the pressure can vary significantly as $Z$ varies while
the energy per nucleon $e$ remains more or less constant. Thus a slight error
in the calculation of the latter could lead to a much larger error in the
pressure. Another possibility is that the apparent instabilities and bubbles
result from our restricted parametrization of the density profile.
Nevertheless, the fact that we find bubble-like solutions emphasizes the need
for a generalization of the present work to include the possibility of
non-spherical pasta phases within the full fourth-order ETFSI framework.

\section{Gross properties of neutron stars}
\label{gross}

\subsection{Mass-radius relation}
\label{massrad}
\begin{figure*}
\centering
\includegraphics[width=\textwidth]{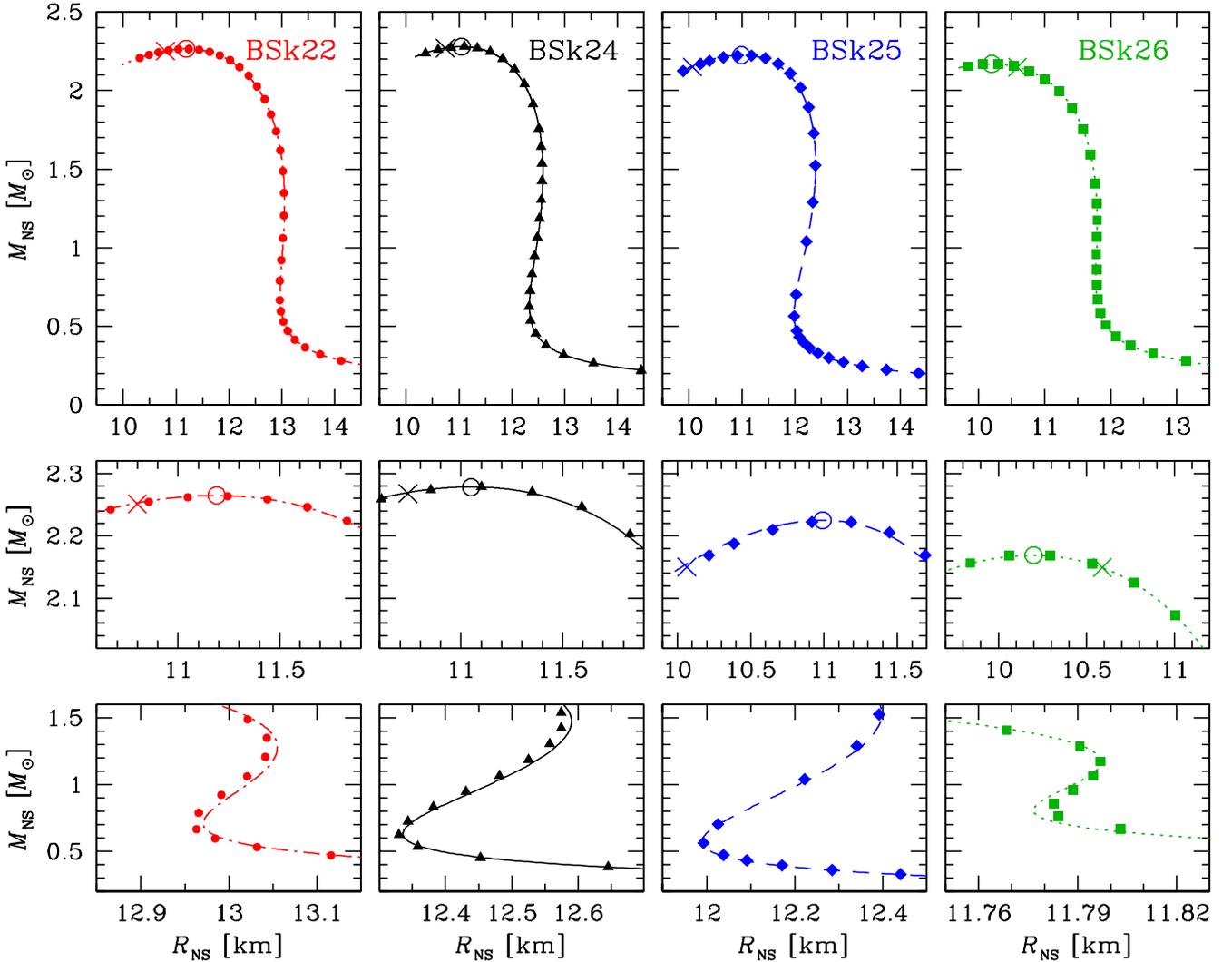}
\caption{Top panels: gravitational mass
$M_\mathrm{NS}$ (in solar masses)
versus circumferential radius $R_\mathrm{NS}$
 of nonrotating
neutron stars, calculated using the tabular data
for the EoSs (symbols) and using the
analytical EoS representations from Eq.~(\ref{Pfit}) (lines). 
The crosses (circles) mark the threshold beyond which the EoS becomes
superluminal (hydrostatically unstable). Middle panels: zoom around the maximum neutron-star mass. 
Bottom panels: zoom around the low mass region.}
\label{fig:MRBSk}
\end{figure*}

We have compared the mass $M_\mathrm{NS}$ and radius $R_\mathrm{NS}$ of a neutron star, calculated using the
tabulated EoSs and their  analytical representations (\ref{Pfit}).  We
integrated the Tolman-Oppenheimer-Volkoff (TOV) equation  from the centre, with
the central mass-energy density $\rho_\mathrm{c}$ as a free parameter, outward to
$\rho=10^6$ g~cm$^{-3}$, using the fourth-order Runge-Kutta method with an
adaptive step and controlled accuracy. 
Since the adaptive step does not generally coincide with the interval
between the density points that are given in the 
EoS tables, we used points obtained by interpolation in the tables. Two
different methods of interpolation were used, linear and cubic, the two
methods agreeing to 5 significant figures in $M_\mathrm{NS}$.
The star was assumed spherically symmetric, nonrotating and nonmagnetized.
The outermost layer of $\rho<10^6$ g cm$^{-3}$, which is excluded from
consideration, is unimportant for the gross properties of a neutron star, because
its thickness does not exceed a few meters, and its mass is only
$\sim10^{-12}\,M_\odot$.

\begin{table}
\caption{Hydrostatic stability limits}
\label{table:Mmax}
\centering
\begin{tabular}{c c c c c }
\hline
EoS  & $M_\mathrm{NS}^\mathrm{stab}$ & $ R_\mathrm{NS}^\mathrm{stab}$ &
 $ n_\mathrm{c}^\mathrm{stab}$ & $\rho_\mathrm{c}^\mathrm{stab}$
\\
 & $(M_\odot)$ &  (km) &
 (fm$^{-3}$) & (g cm$^{-3}$)
 \\
\hline
BSk22 & $2.264$ & 11.20 & 0.967 & $2.26\times10^{15}$ \\
BSk24 & $2.279$ & 11.08 & 0.973 & $2.26\times10^{15}$ \\
BSk25 & $2.224$ & 11.05 & 0.987 & $2.26\times10^{15}$  \\
BSk26 & $2.170$ & 10.20 & 1.123 & $2.67\times10^{15}$ \\
\hline
\end{tabular}
\end{table}

\begin{table}
\caption{Radius, central density of baryons and central mass-energy density
of a neutron star with $M_\mathrm{NS}=1.4\,M_\odot$.}
\label{table:M1.4}
\centering
\begin{tabular}{c | c c c }
\hline
EoS  & $R_\mathrm{NS}^\mathrm{1.4}$~(km) &
 ~$n_\mathrm{c}^\mathrm{1.4}$ (fm$^{-3}$)~ & $\rho_\mathrm{c}^\mathrm{1.4}$ (g cm$^{-3}$)
\\
\hline
BSk22 & 13.04 & 0.385 & $6.92\times10^{14}$ \\
BSk24 & 12.57 & 0.408 & $7.31\times10^{14}$ \\
BSk25 & 12.37 & 0.416 & $7.46\times10^{14}$ \\
BSk26 & 11.77 & 0.506 & $9.19\times10^{14}$ \\
\hline
\end{tabular}
\end{table}

Figure~\ref{fig:MRBSk} shows the mass-radius relation for the EoSs BSk22, BSk24,
BSk25, and BSk26. The neutron-star configurations obtained  with the original
EoSs and with their analytical representations are drawn as symbols and lines,
respectively. 
The maximum neutron-star masses $M_\mathrm{NS}^\mathrm{stab}$, the 
corresponding radii $R_\mathrm{NS}^\mathrm{stab}$, central
number densities of baryons $n_\mathrm{c}^\mathrm{stab}$, and central mass 
densities $\rho_\mathrm{c}^\mathrm{stab}$  are
listed in Table~\ref{table:Mmax}. The values of $M_\mathrm{NS}^\mathrm{stab}$, 
calculated using the tables and the fits, differ by less than 0.06\% for 
each of the four EoSs. For configurations with $n_\mathrm{c}>n_\mathrm{c}^\mathrm{stab}$ 
($\rho_\mathrm{c}>\rho_\mathrm{c}^\mathrm{stab}$), the condition of hydrostatic stability 
$\mathrm{d} M_\mathrm{NS} / \mathrm{d} \rho_\mathrm{c}>0$ is violated;
the unstable configurations are shown by the parts of the curves to the left of
the maxima in Fig.~\ref{fig:MRBSk}, marked by the empty circles.

Comparing the data listed in Table~\ref{table:Mmax} for BSk22, BSk24 and
BSk25, we see that a decrease of $J$ is accompanied by an increase of 
the number density at the centre of the most massive stable configuration,
$n_\mathrm{c}^\mathrm{stab}$, although the corresponding mass-energy density 
$\rho_\mathrm{c}^\mathrm{stab}$ is almost unaffected. 
This can be traced back to the correlation between $J$ and the behavior of the 
symmetry energy at suprasaturation densities, as shown in Fig.~\ref{fig2}.  
Now looking at the last line of the table, we see that the NeuM constraint has a
bigger influence on the most massive stable configuration, as expected: 
the softer EoS BSk26 corresponds to the smaller $M_\mathrm{NS}^\mathrm{stab}$ 
and larger number density and mass-energy density. Very recently, 
gravitational-wave observations have been interpreted as indicating that the 
maximum mass of non-rotating neutron stars cannot exceed $2.16^{+0.17}_{-0.15}$ 
solar masses \citep*{rez18}, supporting our assumption that the real EoS cannot
be much stiffer than LS2.

In Table~\ref{table:M1.4} we list the radii $R_\mathrm{NS}^\mathrm{1.4}$, 
central number densities $n_\mathrm{c}^\mathrm{1.4}$ and central mass 
densities $\rho_\mathrm{c}^\mathrm{1.4}$ of configurations with 
$M_\mathrm{NS}=1.4\,M_\odot$. All the radii except that of functional BSk26
agree very well with the recent estimate by \citet*{mcg18} of 12.7$\pm 0.4$ km.
The first three lines of the table show that for a given       
constraining EoS of NeuM decreasing $J$ (and thus decreasing $L$) is        
accompanied with decreasing $R_\mathrm{NS}$ and increasing central density. 
This reflects the correlation between $L$ and $R_\mathrm{NS}$ known from       
previous studies \citep[e.g.,][]{fortin16,mcg18}.                             
Comparing BSk24 and BSk26 in order to assess the role of the EoS of NeuM    
shows a bigger difference in the radius, a difference large enough to be    
observable \citep*{psaltis2014}. As expected, this pattern is found not only 
for the radii at a fixed mass, but also for the radii of the most massive 
stable configurations, $R_\mathrm{NS}^\mathrm{stab}$ (Table~\ref{table:Mmax}).
All models are consistent with the radius constraint inferred from the recent detection of gravitational waves from the binary neutron star merger GW170817~\citep{annala18,de18,fatt18}.

\subsection{Causality limits}

Table~\ref{table:caus} lists the largest values of baryon
number density ($n^\mathrm{caus}$) and mass-energy density 
($\rho^\mathrm{caus}$), for which the condition 
$\mathrm{d} P/\mathrm{d}\rho < c^2$ is satisfied, i.e., for which the speed 
of sound is smaller than the speed of light. At higher densities the EoS 
becomes superluminal and causality breaks down; for this reason 
$n^\mathrm{caus}$ and $\rho^\mathrm{caus}$ are 
often named causality limits (see, e.g., Section~5.15 of \citealt{hae07}).

\begin{table}
\caption{Causality limits}
\label{table:caus}
\centering
\begin{tabular}{c  c c }
\hline
EoS   & $n^\mathrm{caus}$ (fm$^{-3}$) &
                        $\rho^\mathrm{caus}$ (g cm$^{-3}$) \\
\hline
BSk22  &  1.095 & $2.74\times10^{15}$ \\
BSk24  &  1.088 & $2.69\times10^{15}$ \\
BSk25  &  1.378 & $3.81\times10^{15}$  \\
BSk26  &  0.982 & $2.17\times10^{15}$ \\
\hline
\end{tabular}
\end{table}

The crosses in Fig.~\ref{fig:MRBSk} correspond to
$\rho_\mathrm{c}=\rho^\mathrm{caus}$. The line segments to the left 
of the crosses correspond to configurations where the central part of the star 
has $\rho > \rho^\mathrm{caus}$. We see that for the functionals 
BSk22, BSk24, and BSk25 these segments correspond to configurations that are 
unstable anyway, since $n_\mathrm{c}^\mathrm{stab} < 
n^\mathrm{caus}$, so there is no breakdown of causality in any
situation that would otherwise be physically meaningful. On the other hand, for
functional BSk26 we have $n_\mathrm{c}^\mathrm{stab} > 
n_\mathrm{c}^\mathrm{caus}$, which means that over the range of
central densities $n_\mathrm{c}$ satisfying $ n_\mathrm{c}^\mathrm{stab} > n_\mathrm{c}
> n^\mathrm{caus}$ there will be stable solutions for which causality 
breaks down in the central region of the star. The minimum mass for which this 
can happen, corresponding to $n_\mathrm{c} = n^\mathrm{caus}$, is 
$M_\mathrm{NS} = 2.15\,M_\odot$ (see the BSk26 panel in Fig.~\ref{fig:MRBSk}). The most 
massive possible star (for BSk26) has $M_\mathrm{NS}= 2.17\,M_\odot$, and the central 
region over which causality fails has a maximum radius of 3.6 km and a maximum mass of 
$0.23\,M_\odot$.

Such a breakdown of causality for model BSk26 is, of course, quite unphysical,
and it is a consequence of its failure in the realistic APR EoS. The problem is
discussed in detail by \citet{apr98}, where it is shown that the
reduction in the total stellar mass resulting from a restoration of causality 
is quite small. It should be safe to suppose that the correct limiting mass for
functional BSk26 lies somewhere between 2.15 and 2.17 $M_\odot$. 

\subsection{Direct Urca process}

\begin{table}
\caption{Direct Urca thresholds}
\label{table:DU}
\centering
\begin{tabular}{c c c c}
\hline
EoS  & $n_\mathrm{DU}$ (fm$^{-3}$) & $\rho_\mathrm{DU}$ (g cm$^{-3}$) & $M_\mathrm{DU}/M_\odot$ \\
\hline
BSk22 & 0.333 & $5.88\times10^{14}$ & $1.151$ \\
BSk24 & 0.453 & $8.25\times10^{14}$ & $1.595$ \\
BSk25 & 0.469 & $8.56\times10^{14}$ & $1.612$ \\
BSk26 & 1.458 & $4.19\times10^{15}$ & $(2.115)$ \\
\hline
\end{tabular}
\end{table}

Number fractions of the electrons and muons in the core of a neutron star are
important in the neutron-star cooling theory, because they determine whether the
extremely powerful direct Urca processes of neutrino emission operate or not
(e.g., \citealt{Haensel95} and references therein). For strongly degenerate
particles, the energy-momentum conservation law requires the condition
$n_\mathrm{e}^{1/3} + n_\mathrm{p}^{1/3} > n_\mathrm{n}^{1/3}$ to be fulfilled, in order for these 
processes to work. For the theoretical models where $Y_\mathrm{e}$ monotonically
increases with the increase of density, the direct Urca processes can work in 
the central regions of sufficiently massive neutron stars. Using
Eqs.~(\ref{mu_e_equil})\,--\,(\ref{Y_e_core}), we obtain the corresponding
threshold values of central number densities $n_\mathrm{DU}$. Using the
results of Sec.~\ref{sect:rho_vs_n}, we find the respective mass-energy density values
$\rho_\mathrm{DU}$. Using Eq.~(\ref{Pfit}) and solving the
TOV equation with these central mass densities, we
find the minimum mass $M_\mathrm{DU}$ for a neutron star to cool rapidly via the
direct Urca processes. The values of $n_\mathrm{DU}$,
$\rho_\mathrm{DU}$, and $M_\mathrm{DU}$ for the considered EoS
models are listed in Table~\ref{table:DU}. 
In the case of BSk26, the
$M_\mathrm{DU}$ value is enclosed in brackets, which indicates that 
in this case  $n_\mathrm{DU} > n_\mathrm{c}^\mathrm{stab}$, so that
the corresponding spherically symmetric configuration belongs to the unstable 
branch in Fig.~\ref{fig:MRBSk}. Thus the direct Urca processes cannot work in stable neutron stars described by the BSk26 functional.

Observations indicate that the direct Urca processes operate in a           
relatively small number of neutron stars. None of the around forty          
isolated neutron stars with measured ages and thermal luminosities are as   
cool as required by the ``rapid cooling'' models (see, for example,         
\citealt{pc18} and references therein). Most of them can be explained       
within the so called ``minimal cooling paradigm'', not involving the direct 
Urca process \citep{page04,gusakov04}. However, a few objects show thermal  
luminosities between the minimal and rapid cooling predictions, while most  
of the magnetars are considerably hotter than predicted by both models.     
These exceptions may be explained by internal heating, which can be         
provided by a number of physical mechanisms (see, e.g., \citealt*{ppp15},   
for review and references). Thermal luminosities of accreting neutron       
stars in soft X-ray transients in quiescence can be explained only if the   
direct Urca processes are forbidden in the hottest of them (Aql X-1,        
4U~1608$-$52) but allowed in the coldest one (SAX J1808.4$-$3658)           
\citep{yakovlev04}. Recently, observations of thermal relaxation of the     
neutron star in the transient system MXB 1659$-$29 have delivered an        
evidence of the direct Urca processes operating in its core                 
\citep{brown18}.                                                         

The low value of $M_\mathrm{DU}$ given by the BSk22 functional implies that the
direct Urca processes would operate in most neutron stars, which      
is hardly compatible with the fact that at least a large fraction of them   
are well described by the minimal cooling model. Thus the BSk22 model,      
which in any case gives the worst atomic mass fit of our four functionals, is 
disfavored by neutron-star observations. On the other hand, the presence    
of the direct Urca processes in some neutron stars rules out the BSk26      
functional. Thus, among our four functionals, only BSk24 and BSk25 provide  
$n_\mathrm{DU}$ values compatible with observations. 

\section{Concluding remarks}
\label{concl}

We have presented a microscopic treatment of the nuclear physics of the 
outer crust, the inner crust and the core of neutron stars. Our treatment is
unified in the sense that the EoS (i.e., the pressure as a function of the 
density), the composition and the chemical potentials of all three regions 
are calculated with the same energy-density functional. We have performed 
calculations of the entire neutron star (assumed to be completely degenerate 
and non-accreting) with four different functionals, BSk22, BSk24, BSk25 and 
BSk26 \citep{gcp13}, which are based on generalized Skyrme-type forces and 
contact pairing forces. These functionals were precision fitted to essentially 
all the available atomic mass data (with $Z,N\geq 8$) by using the HFB method. 
In addition, these functionals were constrained to fit, up to the densities 
prevailing in neutron-star cores, the EoS of homogeneous pure NeuM. Since this 
latter EoS is by no means uniquely determined by our present knowledge of 
nuclear physics at high densities we considered two quite different EoSs, the 
hard EoS `V18'  of \citet{ls08} that we label LS2 here, and the soft EoS 
`A18 + $\delta\,v$ + UIX$^*$' of \citet{apr98} that we label APR: functionals 
BSk22, BSk24 and BSk25 were all fitted to LS2, while BSk26 was fitted to APR. 
We have argued that the real EoS of NeuM can probably not be much stiffer than
LS2, and certainly not much softer than APR.

The NeuM constraint, even combined with the atomic mass fit, does not completely
determine the symmetry energy, allowing us thereby some flexibility on the
symmetry coefficient $J$. Accordingly, BSk22, BSk24 and BSk25 were fitted to
values of $J = $ 32, 30 and 29 MeV, respectively. The quality of the mass fits 
deteriorates rapidly outside this range and indeed the quality of the fits 
indicates a 
\emph{slight} preference for $J$ = 30 MeV, i.e., BSk24. We did not exploit this
element of flexibility in $J$ under the constraint of the APR EoS of NeuM, but
instead imposed the unique value of 30 MeV, constructing thereby functional 
BSk26. Thus to study the role of the NeuM constraint one has to compare BSk24
with BSk26, while for the role of the symmetry coefficient it is  BSk22, BSk24 
and BSk25 that have to be compared. 

To calculate the properties of neutron stars from the given functionals we use 
the HFB method in the outer crust (except when the appropriate atomic mass data 
are available), while for the locally homogeneous core the calculation is 
essentially exact. For the inner crust we use the ETFSI approximation to the HF 
method, with pairing handled in the BCS approximation. The first, 
semi-classical, stage of our implementation of the ETFSI method is based on the 
picture of spherically symmetric WS cells, and adopts a parametrization of the 
nucleonic density distributions that respects the boundary conditions necessary 
both for the validity of the fourth-order ETF formalism with which we calculate 
the energy and for continuity between adjacent cells. The second stage 
calculates proton shell effects perturbatively, but neglects the much smaller 
neutron shell effects, and thereby avoids the inevitable problems associated 
with the continuum in the WS approach. 

\begin{figure}
\includegraphics[width=\columnwidth]{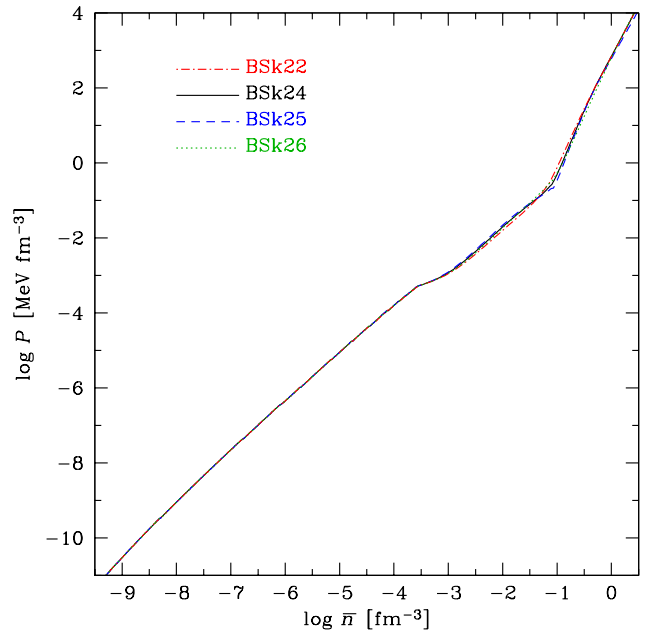}
\caption{(Color online.) Variation over the entire neutron star of pressure $P$
for our four functionals.}
\label{globP}
\end{figure}

\begin{figure}
\includegraphics[width=\columnwidth]{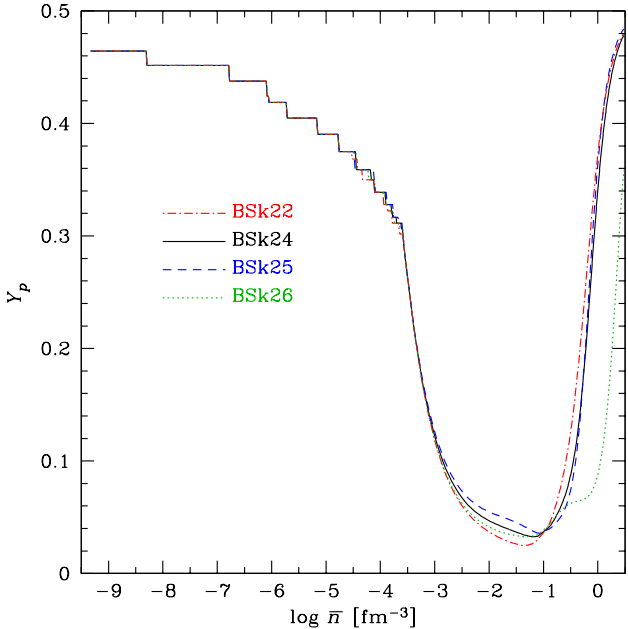}
\caption{(Color online.) Variation over the entire neutron star of proton 
fraction $Y_\mathrm{p}$ for our four functionals.}
\label{globY}
\end{figure}

Concerning our results for the composition, the proton fraction $Y_\mathrm{p}$ 
is nowhere in the star very sensitive to $J$ (see Fig.~\ref{globY}), but in the 
core it depends strongly on the NeuM constraint. While the value of $J$ 
influences only weakly $Y_\mathrm{p}$, the actual values of $N$ and $Z$ in the 
inner crust depend strongly on it, doing so both through the semi-classical ETF
part of the energy and also through the strong proton shell effects: the magic 
numbers that emerge depend strongly on $J$. We stress here that the range of 
values of $J$ that we consider is the widest possible consistent with a good fit
to the atomic mass data: this latter condition constitutes a very stringent 
constraint.   

Turning to our results for the EoS, it might appear from Fig.~\ref{globP} that 
the dependence of the EoS on both $J$ and the NeuM constraint is rather modest.
However, we have shown in Section~\ref{gross} that these apparently small 
differences have a significant and potentially observable impact on the global 
structure of neutron stars.  All our results for the EoS, the composition and 
the chemical potentials have, for the convenience of modellers, been fitted by 
analytic expressions for all four functionals.

A limitation of our calculations is the restriction to spherical cells, which
makes it impossible to study the question of nuclear pasta. The very existence
of these non-spherical cell shapes, which at the most can be favoured only in a 
narrow shell close to the crust-core interface, is controversial because of the
tiny energy differences between spherical and non-spherical cell shapes, but
for precisely this reason the impact of pasta on the EoS is very weak.
Nevertheless, pasta strongly influences other properties of neutron stars, and
it is important to know whether or not it exists. A more detailed study, in 
which the limitation of spherical cell shapes is removed, is desirable.

After the construction of the functionals of this paper \citep{gcp13} a new set
was published, BSk30, BSk31 and BSk32 \citep{gcp16}. These functionals were
characterized by an improved treatment of the pairing, with self-energy effects 
being included. In this way more realistic gaps for INM were obtained, an 
essential condition for the study of superfluidity in neutron stars, but to 
maintain the high quality of the mass fits of our older functionals it was 
necessary to add a phenomenological pairing term dependent on the density 
gradient. The three new functionals were fitted to $J$ = 30, 31 and 32 MeV, 
respectively, all being fitted to the LS2 EoS of NeuM. 
So far, no functional with the new pairing has been mass fitted under the 
softer APR NeuM constraint, which means that the new family is not 
suitable for the sort of study that we have undertaken in the present paper.

However, we can still assess the impact that the improved treatment of the
pairing functional 
would have on the conclusions that we have drawn from the older family of 
functionals considered here. The optimal mass fit with the new pairing is found
with BSk31, for which $J$ = 31 MeV. We therefore compare it with functional 
BSk23 \citep{gcp13}, which also has $J$ = 31 MeV and is constrained to the LS2 
EoS of NeuM, but belongs to the older family of functionals in that it has the 
same form of pairing. Preliminary calculations show that the effect of changing
the pairing functional on the EoS and on the composition is much smaller than 
the effect of changing $J$ from 31 to 29 MeV, or of changing the constraining 
EoS of NeuM from LS2 to APR. Nevertheless, given the importance of the 
improved treatment of pairing for reliable calculations of neutron-superfluid 
properties, we intend to publish a more complete study of the new functionals 
in a later paper.  

In the meantime, we believe that the results we have presented here for
functionals BSk22, BSk24 and BSk25 span the current range of uncertainty
associated with the gaps in our knowledge of the symmetry coefficient $J$ (and
of the symmetry-slope coefficient $L$, which is roughly correlated to $J$ by
the fits to nuclear masses). Comparing, on the other hand, functionals BSk24
and BSk26 gives some insight into the impact of the uncertainty in the EoS of
NeuM. Not surprisingly, the influence of the uncertainty in $J$ is greatest
in the crust and diminishes as we move to the core of the star, while the
contrary is the case for the uncertainty in the EoS. Thus we can conclude with 
some confidence that we have tied down the nuclear uncertainties (aside from 
those related to pairing) in the crust of the star. This confidence decreases
somewhat as we approach the centre of the core, but it is unlikely that the 
actual properties lie far outside the limits that we have considered here.

\section*{Acknowledgements}
We wish to thank M.~Onsi for her invaluable contributions to earlier stages of 
this project, and the referee for valuable remarks. J. M. P.{} acknowledges the
partial support of the NSERC (Canada). The work of A.Y.P.{} was partially 
supported by the RFBR grant 16-29-13009-ofi-m. The work of N. C. was partially 
supported by the Fonds de la Recherche Scientifique - FNRS (Belgium) under 
grant n$^\circ$~CDR-J.0187.16 and CDR-J.0115.18. S. G. {} acknowledges the 
support of Fonds de la Recherche Scientiffique-FNRS  (Belgium). This work was 
also partially supported by the European Cooperation in Science 
and Technology (COST) Actions MP1304 and CA16214.

\appendix

\section{Neutron chemical potential: proof of Eq.~(\lowercase{\ref{mung}})} 
\label{app:mung}

Although the bulk of this paper has been devoted to the inhomogeneities of
the outer and inner crusts, we must recognize that this feature holds only at 
the microscopic level, the isolated nuclei of the former containing no more 
than a few hundred nucleons, and the WS cells of the latter at most a few 
thousand. Thermodynamically the crust, like the core, can be regarded as
homogeneous, in the sense that over an appreciable range of sizes the mass, 
energy and entropy (for $T > 0$) of a piece of crust (or of the core) are 
proportional to its volume. It is then easy to show that the total Gibbs energy
$G$ of the piece can be written as
\beq\label{mung1}
G = \sum_i\,\mu_i\,N_i \quad ,  
\eeq
where $\mu_i$ is the chemical potential of component $i$ and $N_i$ is the 
number of particles of component $i$ in the piece (see, for example, Section
11.3 of \citealt{adkins}). 

Discounting the possible presence of hyperons, we have just four components in
any star: neutrons, protons, electrons and muons (it makes no difference in
the following whether or not the nucleons form clusters or nuclei). Then
given charge neutrality it follows from Eq.~(\ref{mung1}) that the Gibbs energy
per nucleon is 
\beq\label{mung2}
g = \mu_\mathrm{n} + (\mu_\mathrm{p} + \mu_\mu - \mu_\mathrm{n})Y_\mathrm{p} + (\mu_\mathrm{e} - \mu_\mu)Y_\mathrm{e} \quad .
\eeq 
If there is beta equilibrium, as in a neutron star, then Eqs.~(\ref{corea.13a})
and~(\ref{corea.13b}) will hold, and the last two terms of Eq.~(\ref{mung2})
vanish, leaving us with Eq.~(\ref{mung}). The same conclusion follows if there
are no muons, since in that case $Y_\mathrm{e} = Y_\mathrm{p}$.  It is worth noting that 
Eq.~(\ref{mung}) holds for all temperatures, including $T = 0$.

\section{Treatment of lepton gas}
\label{app:lepton}

The electron gas in the inner crust and the core, and the muon gas in the 
latter region, are assumed to be completely uniform and degenerate, 
with special-relativistic effects treated in all generality. Then following, 
for example, Chapter 24 of \citet{cg}, we define
\beq\label{lep.1}
x = \lambda\left(3\pi^2n\right)^{1/3} \quad   ,
\eeq
where $\lambda = \hbar/(mc)$ is the Compton wavelength of the lepton
(electron or muon), $m$ being the corresponding mass, and $n$ is the lepton 
number density. Defining also
\beq\label{lep.2}
g(x) = -8x^3 + 3x(1 + 2x^2)\sqrt{1 + x^2} - 3\sinh^{-1}x  \quad  ,
\eeq
we can write the kinetic-energy density of the leptons as
\beq\label{lep.3}
{\mathcal E}^\mathrm{kin} = \frac{1}{24\pi^2\lambda^3}mc^2g(x)   \quad   , 
\eeq
which reduces in the high-density extreme-relativistic limit to
\beq\label{lep.3A}
{\mathcal E}^\mathrm{kin} = \frac{3}{4}\left(3\pi^2\right)^{1/3}\hbar\,cn^{4/3} 
\quad .
\eeq

Only in the outer crust will deviations from uniformity be 
non-negligible, and in that region we add to the energy density the screening 
correction given by the interpolation formula of \citet{pc2000}, instead of
the Thomas-Fermi expression \citep{sal61} used by \citet{pgc11} (see 
\citealt{chf16-prd} for a recent discussion on this approximation). Then at 
temperatures $T$ much lower than the plasma temperature 
\beq
T_\mathrm{pl}=\frac{\hbar \omega_\mathrm{pl}}{k_\textrm{B}} \quad , 
\eeq
where
\beq
\omega_\mathrm{pl} =\sqrt{\frac{4\pi Z^2 e^2 n_\mathrm{e}}{M^\prime(A,Z) Z}}
\eeq
is the ion-plasma frequency (the ion mass coincides with the nuclear mass since
atoms are fully ionized), the energy density of the screening correction
is given by \citet{pc2000}
\beq
\label{eq:e-pol}
\mathcal{E}_\mathrm{e}^\mathrm{scr} =
 - f_\infty(x_\mathrm{e})
  \left( \frac{4 \pi}{3} \right)^{1/3} e^2 Z^{2/3} n_\mathrm{e}^{4/3} \,
   \left[ 1 + \mathcal{A}(x_\mathrm{e})\,\left(\frac{q}{\Gamma_\mathrm{pl}}\right)^s \right]
\, .
\eeq
Here $x_\mathrm{e}$ is the relativity factor (\ref{lep.1}) for electrons, and
\begin{align}
 &\Gamma_\mathrm{pl} =\frac{Z^2 e^2}{a_\mathrm{n} k_\textrm{B} T_\mathrm{pl}} \, ,\nonumber \\
  & f_\infty(x_\mathrm{e}) = \frac{54}{175}\left(\frac{12}{\pi}\right)^{1/3} \alpha Z^{2/3} b_1\,\sqrt{1+\frac{b_2}{x_\mathrm{e}^2}}\, , \nonumber \\
  & \mathcal{A}(x_\mathrm{e}) = \frac{ b_3+a_3 x_\mathrm{e}^2 }{ 1+b_4 x_\mathrm{e}^2 }\quad ,
\end{align}
in which $a_\mathrm{n} = (4 \pi/3\ n_\mathrm{e}/ Z)^{-1/3}$ is the inter-ion spacing and
$\alpha=e^2/(\hbar c)$ is the fine-structure constant. Also, the parameters $s$
and $b_1$--$b_4$, which depend only on $Z$, are given by \citet{pc2000} as
\begin{align}
  & s = \left( 1+0.01\,(\ln Z)^{3/2} + 0.097\,Z^{-2} \right)^{-1}\, , \nonumber\\
 & b_1 = 1 - a_1 \,Z^{-0.267} + 0.27\,Z^{-1}\, , \nonumber\\ 
 & b_2 = 1 + \frac{2.25}{ Z^{1/3}}\, \frac{1+a_2\,Z^5+0.222\,Z^6 }{ 1+0.222\,Z^6}\, ,\nonumber\\
 & b_3 = \frac{a_4}{1+\ln Z}\, ,
\nonumber\\
 & b_4 = 0.395 \ln Z + 0.347\, Z^{-3/2}\, . \nonumber
\end{align}
where, for a bcc lattice, $a_1=1.1866$, $a_2=0.684$, $a_3=17.9$, $a_4=41.5$, 
and $q=0.205$ \citep{pc2000}. 

Note that the expressions suggested by \citet{pc2000} provide good approximations to 
the energy and pressure at arbitrary temperature, but they produce an unphysically 
slow decrease of the electron-screening contribution to the heat capacity with 
decreasing $T$ below $T_\mathrm{pl}$, i.e., for a quantum crystal. The problem was recognized 
and rectified by \citet{pc10}. However, the expressions of \citet{pc2000} and
\citet{pc10} have the same limit at $T/T_\mathrm{pl}\to0$, Eq.~(\ref{eq:e-pol}), which we use 
here.

In addition to a correction for non-uniformity of the electron gas in the outer
crust we include a correction for Coulomb exchange in all regions of the star.
The general expression that we take for the exchange energy density is 
Eq.~(9) of \citet{sal61}, which can be simplified \citep{engel}, without
approximation, to  
\begin{equation}
\label{lep.5}
\mathcal{E}^\mathrm{ex} = -\alpha \frac{m c^2 x^4}{4\pi^3 \lambda^3}
 \left[1-\frac{3}{2}\left(\frac{\sqrt{1+x^2}}{x}-\frac{\sinh^{-1}x}{x^2}\right)^2\right] 
\, .
\end{equation}
We use this expression as it stands for the electrons in the outer crust and
the muons in the core, while for the electrons of the inner crust and the core
it suffices to take the extreme relativistic limit 
\beq\label{lep.6}
{\mathcal E}_\mathrm{e}^\mathrm{ex} =
 \frac{3e^2}{8}\Big(\frac{3}{\pi}\Big)^{1/3}n_\mathrm{e}^{4/3} 
= \frac{\alpha}{2\pi}{\mathcal E}_\mathrm{e}^\mathrm{kin} \quad  .
\eeq

There is no direct Coulomb term in the core, since strict neutrality prevails
throughout, but there is in the outer and inner crusts. However, in the outer
crust these direct terms constitute the lattice energy $E_L(A, Z)$ of 
Eq.~(\ref{2.1}), while in the inner crust they are represented by the 
${\mathcal E}_\mathrm{ee,ep}^\mathrm{c}(\bm{r})$ term of Eq.~(\ref{3.-1}). Thus we do not 
have to consider any further the direct Coulomb interactions of the electrons,  
either with themselves or with protons.

As in \citet{pgc11}, we also include in the outer crust a correction for 
the electron-correlation energy, using the high-density limit 
\begin{equation}
\label{lep.ce1}
\mathcal{E}_\mathrm{e}^\mathrm{corr} \simeq
 \frac{\alpha^2 k_\mathrm{Fe}^4 \hbar c}{12 \pi^4}
 \left[ -12.51 + \log_{10} (k_\mathrm{Fe} \lambda_\mathrm{e}) \right] \ ,
\end{equation}
where 
\beq\label{lep.ce2}
k_\mathrm{Fe} = (3 \pi^2 n_\mathrm{e})^{1/3} 
\eeq
(see, e.g., \citealt{engel}). This correction is quite 
negligible in the inner crust and core.

Thus, for the total energy density of the electrons,
excluding their rest mass, we write
\beq\label{lep.10}
{\mathcal E}_\mathrm{e} ={\mathcal E_\mathrm{e}}^\mathrm{kin} + {\mathcal E}_\mathrm{e}^\mathrm{scr} +
{\mathcal E}_\mathrm{e}^\mathrm{ex} + {\mathcal E}_\mathrm{e}^\mathrm{corr} \quad ,
\eeq
in which all quantities are as already defined. A similar expression holds for 
muons.

\emph{Pressure.} Defining
\beq\label{lep.7}
f(x) = (2x^3 - 3x)\sqrt{1 + x^2} + 3\sinh^{-1}x  \quad  ,
\eeq
the lepton pressure corresponding to the energy density~(\ref{lep.3}) is
\beq\label{lep.4}
P^\mathrm{kin} = \frac{1}{24\pi^2\lambda^3}m c^2f(x)   \quad .
\eeq
This reduces for the extremely relativistic electrons of the inner crust and
core to
\beq\label{lep.4a}
P^\mathrm{kin} = \frac{\hbar\,c}{12\pi^2}\left(3\pi^2n\right)^{4/3} \quad .
\eeq
The exchange pressure corresponding to the general expression (\ref{lep.5}) is
\begin{multline}
\label{eq:p-exc}
P^\mathrm{ex} = \frac{\mathcal{E}^\mathrm{ex}}{3} - \alpha \frac{m c^2 x^3}{2\pi^3 
\lambda^3} \left(\frac{1}{\sqrt{1+x^2}}-\frac{\sinh^{-1}x}{x}\right)
\\\times 
\left(\frac{\sqrt{1+x^2}}{x} - \frac{\sinh^{-1}x}{x^2} \right)
\quad   .
\end{multline}
For the extremely relativistic electrons this reduces to 
\beq\label{lep.8}
P_\mathrm{e}^\mathrm{ex} = \frac{\mathcal{E}^\mathrm{ex}}{3} 
= \frac{e^2}{8}\Big(\frac{3}{\pi}\Big)^{1/3}n_\mathrm{e}^{4/3} \quad . 
\eeq
In the outer crust there is also an electron-screening contribution to the 
pressure given by
\begin{multline}
\label{eq:p-pol}
P_\mathrm{e}^\mathrm{scr} = \frac{\mathcal{E}_\mathrm{e}^\mathrm{scr}}{3}
 \,\, - \,\,  \left( \frac{4 \pi}{3}\right)^{1/3} e^2 Z^{2/3} n_\mathrm{e}^{4/3}
  \\ \times\!\!
  \Bigg\{ n_\mathrm{e} \frac{\partial f_\infty(x_\mathrm{e})}{\partial x_\mathrm{e}}
   \frac{\partial x_\mathrm{e}}{\partial n_\mathrm{e}} \left[ 1 + \mathcal{A}(x_\mathrm{e})
   \,\left(\frac{q}{\Gamma_\mathrm{pl}}\right)^s \right] 
  \\
+  n_\mathrm{e} f_\infty(x_\mathrm{e})
 \Bigg[ \left( \frac{q}{\Gamma_\mathrm{pl}} \right)^s
 \frac{\partial \mathcal{A}(x_\mathrm{e})}{\partial x_\mathrm{e}} 
 \frac{\partial x_\mathrm{e}}{\partial n_\mathrm{e}}
  \\
  + s q \mathcal{A}(x_\mathrm{e}) 
 \left( \frac{q}{\Gamma_\mathrm{pl}} \right)^{s-1} 
 \frac{\partial (1/\Gamma_\mathrm{pl})}{\partial n_\mathrm{e}} \Bigg] \Bigg\} \quad ,
\end{multline}
with
\begin{align}
& \frac{\partial f_\infty(x_\mathrm{e})}{\partial x_\mathrm{e}} = 
 - \alpha \frac{54}{175} \left( \frac{12}{\pi} \right)^{1/3}
  \frac{b_1 b_2 Z^{2/3}}{x_\mathrm{e}^2 \sqrt{b_2 + x_\mathrm{e}^2}} \, ,
\nonumber\\
&  \frac{\partial \mathcal{A}(x_\mathrm{e})}{\partial x_\mathrm{e}} =
   \frac{2 (a_3 - b_4 b_3) x_\mathrm{e}}{(1+b_4 x_\mathrm{e}^2)^2}  \, ,
\nonumber \\
& \frac{\partial (1/\Gamma_\mathrm{pl})}{\partial n_\mathrm{e}} =
  \frac{1}{\sqrt{6}} \left( \frac{\pi}{6}\right)^{1/6} Z^{-7/6}
   \sqrt{\frac{\hbar c}{\alpha M'(A,Z) c^2}}\ n_\mathrm{e}^{-5/6} \, ,
\nonumber \\
&  \frac{\partial x_\mathrm{e}}{\partial n_\mathrm{e}} = 
  (3 \pi^2)^{1/3} \frac{\lambda_\mathrm{e}}{3} n_\mathrm{e}^{-2/3} \, .
\end{align}

The pressure corresponding to the electron-correlation energy, 
Eq.~(\ref{lep.ce1}), is given by
\begin{equation}
\label{lep.ce3}
P_\mathrm{e}^\mathrm{corr} = \frac{1}{3} \mathcal{E}_\mathrm{e}^\mathrm{c} +
 \frac{\alpha^2 k_\mathrm{Fe}^4 \hbar c}{36 \pi^4 \ln(10)}\ .
\end{equation}

Thus, for the total pressure exerted by the electrons we have
\beq\label{lep.100}
P_\mathrm{e} = P_\mathrm{e}^\mathrm{kin} +  P_\mathrm{e}^\mathrm{scr} + P_\mathrm{e}^\mathrm{ex} + P_\mathrm{e}^\mathrm{corr} \quad ,
\eeq
in which all quantities are as already defined, with a similar expression
for muons.

\emph{Chemical potentials.} For the chemical potential of the electrons we have 
\beq\label{lep.11}
\mu_\mathrm{e} = \frac{\partial {\mathcal E}_\mathrm{e}}{\partial n_\mathrm{e}} + m_\mathrm{e}c^2
= \frac{P_\mathrm{e}}{n_\mathrm{e}} + e_\mathrm{e} + m_\mathrm{e}c^2 \quad ,
\eeq
where $e_\mathrm{e} = {\mathcal E}_\mathrm{e}/n_\mathrm{e}$ is the total 
energy per electron, excluding the rest mass. With ${\mathcal E}_\mathrm{e}$ 
given by Eq.~(\ref{lep.10}), we have 
\beq\label{lep.12}
 \mu_\mathrm{e} - m_\mathrm{e}c^2 = \mu_\mathrm{e}^\mathrm{kin} + \mu_\mathrm{e}^\mathrm{ex} + \mu_\mathrm{e}^\mathrm{scr} +
\mu_\mathrm{e}^\mathrm{corr} \quad .
\eeq
For the first term here Eqs.~(\ref{lep.3}) and (\ref{lep.4}) lead simply to
the familiar Fermi energy of a free fermion gas with the rest mass subtracted,
\begin{align}\label{lep.13}
\mu_\mathrm{e}^\mathrm{kin} &= m_\mathrm{e}c^2\left[(1 + x^2)^{1/2} - 1\right] \nonumber \\
&= (m_\mathrm{e}^2c^4 + \hbar k_\mathrm{Fe}^2 c^2)^{1/2} - m_\mathrm{e}c^2 \quad .
\end{align}
For the second term in Eq.~(\ref{lep.12}) we find from Eqs.~(\ref{lep.5}) and
(\ref{eq:p-exc})
\begin{multline}\label{lep.14}
\mu_\mathrm{e}^\mathrm{ex} = \frac{4}{3}e_\mathrm{e}^\mathrm{ex} - \frac{3\alpha}{2\pi}m_\mathrm{e}c^2
\left(\frac{1}{\sqrt{1+x^2}}-\frac{\sinh^{-1}x}{x}\right)
\\\times
\left(\frac{\sqrt{1+x^2}}{x} - \frac{\sinh^{-1}x}{x^2} \right) \quad ,
\end{multline}
which reduces in the extreme relativistic limit to
\beq\label{lep.15}
\mu_\mathrm{e}^\mathrm{ex} = \frac{4}{3}e_\mathrm{e}^\mathrm{ex}   \quad .
\eeq
For the screening term in Eq.~(\ref{lep.12}) we find from Eqs.~(\ref{eq:e-pol}) 
and~(\ref{eq:p-pol}) 
\begin{multline}\label{lep.16}
\mu_\mathrm{e}^\mathrm{scr} = \frac{4}{3}e_\mathrm{e}^\mathrm{scr}
 - \left(\frac{4\pi}{3}\right)^{1/3} e^2 Z^{2/3} n_\mathrm{e}^{1/3} 
  \\ \times
\Bigg\{n_\mathrm{e} \frac {\partial f_\infty(x_\mathrm{e})}{\partial x_\mathrm{e}} 
\frac{\partial x_\mathrm{e}}{\partial n_\mathrm{e}} 
\left[1 + \mathcal{A}(x_\mathrm{e})\,\left(\frac{q}{\Gamma_\mathrm{pl}}\right)^s \right]
 \\
+ 
n_\mathrm{e} f_\infty(x_\mathrm{e})
 \Bigg[\left(\frac{q}{\Gamma_\mathrm{pl}}\right)^s 
\frac{\partial \mathcal{A}(x_\mathrm{e})}{\partial x_\mathrm{e}} 
\frac{\partial x_\mathrm{e}}{\partial n_\mathrm{e}}
\\
 + s q \mathcal{A}(x_\mathrm{e}) 
\left(\frac{q}{\Gamma_\mathrm{pl}} \right)^{s-1} 
\frac{\partial (1/\Gamma_\mathrm{pl})}{\partial n_\mathrm{e}} \Bigg] \Bigg\}  \quad , 
\end{multline}
while from Eqs.~(\ref{lep.ce1}) and~(\ref{lep.ce3}) we find for the last term in
Eq.~(\ref{lep.12})
\begin{equation}
\label{lep.17}
\mu_\mathrm{e}^\mathrm{corr} = \frac{4}{3} e_\mathrm{e}^\mathrm{corr} + 
\frac{\alpha^2 k_\mathrm{Fe} \hbar c}{12 \pi^2 \ln(10)}\ .
\end{equation}

All these expressions for the electron chemical potential hold in all regions
of the neutron star, and equally well for the muon chemical potential. 
Note, however, that since we neglect the last two terms in Eq.~(\ref{lep.12})
everywhere except in the outer crust they have no relevance for muons.

\section{Analytic representations}
\label{analytic}

For each of our four functionals BSk22, BSk24, BSk25, and BSk26, we construct 
parametrizations in terms of analytic functions for the quantities of
astrophysical interest that we have calculated, following the approach
previously developed by \citet{hp04,pfncpg12}. In the case of the energy 
per nucleon and pressure we were able to find unified fits to a single
continuous analytic function of the number density $\bar{n}$ over a broad 
density range covering the entire neutron star, i.e., the outer and inner 
crusts and the core (in the core we have $\bar{n} \equiv n$). These unified 
fits smear away all density discontinuities 
between layers of different composition, and can be useful for hydrodynamic 
modeling. Other quantities that are required in modeling neutron-star structure
and evolution, such as particle fractions, are given by separate 
parametrizations for the inner crust and the core.
All parametrizations presented in this Appendix have been implemented in Fortran subroutines,
freely available at the Ioffe Institute website.\footnote{http://www.ioffe.ru/astro/NSG/BSk/}

\subsection{Energy as a function of density}
\label{sect:rho_vs_n}

For each of the four functionals, the 
equilibrium energy per baryon that we have calculated is fitted to a single 
continuous analytic function of the number density $\bar{n}$ over all three
regions of the star. We recall that the equilibrium energy $e_\mathrm{eq}$ per nucleon
that we have shown in the figures and tables always has the neutron mass 
subtracted out: see Eqs.~(\ref{2.1}), (\ref{3.-4}) and (\ref{corea.2}).
Our fitting function consists of a constant term and three parts 
corresponding respectively to low, moderate, and high densities, matched 
together by appropriate weight functions $w_{1,2}(\bar{n})$, thus
\begin{multline}\label{Efit}
 e_\mathrm{eq} = e_\mathrm{gr} +  
 \frac{(p_1\bar{n})^{7/6}}{1+\sqrt{p_2\bar{n}}}\,
 \frac{1+\sqrt{p_4\bar{n}}}{(1+\sqrt{p_3\bar{n}})(1+\sqrt{p_5\bar{n}})}
 \,w_1(\bar{n}) \\
 +p_6\bar{n}^{p_7}\,(1+p_8\bar{n})\,\big[1-w_1(\bar{n})\big]
 \,w_2(\bar{n}) \\
 +\frac{(p_{10}\bar{n})^{p_{11}}}{1+p_{12}\bar{n}}
 \,\big[1-w_2(\bar{n})\big],
\end{multline}
where
\begin{align}
  w_1(\bar{n}) &= \frac{1}{1+p_9\bar{n}},
\qquad
  w_2(\bar{n}) = \frac{1}{1+(p_{13}\bar{n})^{p_{14}}}.
\end{align}
The requirement that $^{56}$Fe be fitted for the vanishingly small values      
of $\bar{n}$ in the outer crust fixes the constant term at
$e_\mathrm{gr}=-9.1536$ MeV. The numerical coefficients $p_i$ are given in
Table~\ref{tab:Efit}, with $e_\mathrm{eq}$ measured in MeV and $\bar{n}$ in fm$^{-3}$.
The fit is valid over the range $10^{-9}$ fm$^{-3} \lesssim \bar{n}
\lesssim 3$ fm$^{-3}$, the typical error of Eq.~(\ref{Efit}) over this range
being $\lesssim0.5$\% for all three functionals; the
maximum errors ($<3$\%) occur at the phase boundaries, because the
dependence $e_\mathrm{eq}(\bar{n})$ is fitted by a continuous function across
the discontinuities of the argument $\bar{n}$.

\begin{table}
\centering
\caption[]{Parameters of Eq.~(\ref{Efit}).}
\label{tab:Efit}
\begin{tabular}{r|cccc}
\hline
\multicolumn{1}{c|}{\rule[-1ex]{0pt}{4ex}$i$} & \multicolumn{4}{c||}{$p_i$}\\
  & BSk22 & BSk24 & BSk25 & BSk26 \\
\hline
1  & \rule{0pt}{2.7ex}7.02$\times10^{8}$  & 6.59$\times10^{8}$   & 6.411$\times10^{8}$ & 1.3995$\times10^{8}$
                                                                                             \\
2  & 1.133$\times10^{11}$ & 9.49$\times10^{10}$  & 8.76$\times10^{10}$ & 1.066$\times10^{9}$ \\
3  & 6.19$\times10^{7}$   & 6.95$\times10^{7}$   & 7.40$\times10^{7}$  & 7.472$\times10^{8}$ \\
4  & 4.54$\times10^{6}$   & 5.63$\times10^{6}$   & 6.31$\times10^{6}$  & 3.599$\times10^{7}$ \\
5  & 5.46$\times10^{5}$   & 6.51$\times10^{5}$   & 7.13$\times10^{5}$  & 2.906$\times10^{6}$ \\
6  & 15.24                & 19.37                & 22.11               & 0.3051              \\
7  & 0.0683               & 0.1028               & 0.1217              & $-0.54068$          \\
8  & 8.86  & 4.09   & 2.54  & 388.5  \\
9  & 4611  & 6726   & 8317  & 776.5  \\
10 & 48.07 & 29.57  & 25.63 & 15.435 \\
11 & 2.697 & 2.6728 & 2.507 & 2.2483 \\
12 & 81.7  & 19.51  & 7.92  & 0.3029 \\
13 & 7.05  & 4.39   & 3.92  & 18.66  \\
14 & 1.50  & 1.75   & 2.06  & 0.569  \\
\hline
\end{tabular}
\end{table}

For some purposes it is more convenient to deal with the total mass-energy
density $\rho$ than with the energy per nucleon $e_\mathrm{eq}$. The function
$\rho(\bar{n})$ is then equally well fitted by the function (\ref{Efit})
through the identity~(\ref{2.2}).

The analytic fit (\ref{Efit}) can be also used for the calculation of the 
baryon number density $\bar{n}$ and energy $e_\mathrm{eq}$ as functions of mass 
density $\rho$. For this purpose, we employ an iterative procedure, using
the secant method for the logarithms (see, for example, Section 9.2 of 
\citealt{numrec}),
\begin{equation}
\ln\bar{n}_{i+1} = \ln\bar{n}_{i} +
\ln(\rho/\rho_{i})\ln(\bar{n}_{i-1}/\bar{n}_i)/\ln(\rho_{i-1}/\rho_{i}).
\end{equation}
For each $i$th approximation to the number density, $\bar{n}_i$, 
Eq.~(\ref{Efit}) provides the mass-energy density estimate
$\rho_{i} =\bar{n}_i\,(e_\mathrm{eq,i} + M_\mathrm{n}c^2)/c^2$, which is used to correct 
$\bar{n}$ at the next iteration. Two starting values, $n_0$ and $n_1$, are
required. For the first of these we take $\bar{n}_0=\rho c^2/e_0$, which,
with $e_0 = e_\mathrm{gr} + M_\mathrm{n}c^2$ = 930.4118 MeV, is exact at vanishingly small 
$\rho$, while for the second we take $\bar{n}_1=\bar{n}_0\rho/\rho_0$. 
The procedure rapidly converges: a fractional accuracy better than
$10^{-6}$ is reached in 1\,--\,2 iterations for crustal densities and in
no more than 5 iterations for the core (within the stability and causality 
limits listed in Tables~\ref{table:Mmax} and \ref{table:caus}).

\subsection{Pressure as a function of density}
\label{sect:Pfit}

For all four functionals we fit the calculated pressure, like the energy per 
nucleon, to a single continuous analytic function of the number density 
$\bar{n}$ that covers the entire star. It is convenient to work through the 
mass-energy density $\rho$, expressing it in terms of $\bar{n}$ through
Eqs.~(\ref{Efit}) and (\ref{2.2}). Our fitting function for the pressure then  
has the same form as in \citet{pfncpg12},
\begin{eqnarray}
  \log_{10}P &=& K +
    \frac{p_1+p_2\xi+p_3\xi^3}{1+p_4\,\xi}\,
        \left\{\exp\left[p_5(\xi-p_6)\right]+1\right\}^{-1}
\nonumber\\&&
     + \, (p_7+p_8\xi)\,
        \left\{\exp\left[p_9(p_6-\xi)\right]+1\right\}^{-1}
\nonumber\\&&
     + \, (p_{10}+p_{11}\xi)\,
\left\{\exp\left[p_{12}(p_{13}-\xi)\right]+1\right\}^{-1}
\nonumber\\&&
     + \, (p_{14}+p_{15}\xi)\,
\left\{\exp\left[p_{16}(p_{17}-\xi)\right]+1\right\}^{-1}
\nonumber\\&&\hspace*{-2em}
     + \, \frac{p_{18}}{1+
     [p_{20}\,(\xi-p_{19})]^2}
     + \frac{p_{21}}{1+
     [p_{23}\,(\xi-p_{22})]^2}
\,\, ,
\label{Pfit}
\end{eqnarray}
where $\xi\equiv \log_{10}(\rho/\textrm{g cm}^{-3})$. Setting $K$ = 0 gives the
pressure in units of dyn~cm$^{-2}$, setting $K$ = -33.2047 gives it in units
of MeV$\cdot$fm$^{-3}$, the units of the figures and tables of this paper. The 
parameters $p_i$ are given in Table~\ref{tab:Pfit}. The typical fit error of 
$P$ is $\approx1$\% for $6\lesssim\xi\lesssim16$. The maximum errors of 
$\lesssim 4$\% are reached at
phase boundaries, because the dependence $P(\rho)$ is fitted by a
continuous function across the discontinuities of the argument $\rho$.

\begin{table}
\centering
\caption[]{Parameters of Eq.~(\ref{Pfit}).}
\label{tab:Pfit}
\begin{tabular}{r|cccc}
\hline
\multicolumn{1}{c|}{\rule[-1ex]{0pt}{4ex}$i$} &
\multicolumn{4}{c||}{$p_i$} \\
  & BSk22 & BSk24 & BSk25 & BSk26 \\
\hline\rule{0pt}{2.7ex}
1  & 6.682      & 6.795     & 7.210     & 3.672     \\
2  & 5.651      & 5.552     & 5.196     & 7.844     \\
3  & 0.00459    & 0.00435   & 0.00328   & 0.00876   \\
4  & 0.14359    & 0.13963   & 0.12516   & 0.22604   \\
5  & 2.681      & 3.636     & 4.624     & 3.129     \\
6  & 11.972     & 11.943    & 12.16     & 11.939    \\
7  & 13.993     & 13.848    & 9.348     & 13.738    \\
8  & 1.2904     & 1.3031    & 1.6624    & 1.3389    \\
9  & 2.665      & 3.644     & 4.660     & 3.112     \\
10 & $-$27.787  & $-$30.840 & $-$28.232 & $-$23.031 \\
11 & 2.0140     & 2.2322    & 2.0638    & 1.6264    \\
12 & 4.09       & 4.65      & 5.27      & 4.83      \\
13 & 14.135   & 14.290  & 14.365   & 14.272    \\
14 & 28.03    & 30.08   & 29.10    & 23.28     \\
15 & $-$1.921 & $-$2.080& $-$2.130 & $-$1.542  \\
16 & 1.08     & 1.10    & 0.865    & 2.10      \\
17 & 14.89    & 14.71   & 14.66    & 15.31     \\
18 & 0.098    & 0.099   & 0.069    & 0.083     \\
19 & 11.67    & 11.66   & 11.65    & 11.66     \\
20 & 4.75     & 5.00    & 6.30     & 6.16      \\
21 & $-$0.037 & $-$0.095& $-$0.172 & $-$0.042  \\
22 & 14.10    & 14.15   & 14.18    & 14.18     \\
23 & 11.9     & 9.1     & 8.6      & 14.8      \\
\hline
\end{tabular}
\end{table}

\subsection{Particle numbers}
\label{sect:numfit}

\subsubsection{Particle numbers in the inner crust}
\label{sect:numfit_crust}

As we see from Fig.~\ref{Zic}, the equilibrium values
$Z_\mathrm{eq}$ of the number of protons in a WS cell are
discontinuous at certain densities. At the values of
$\bar{n}$ below the proton drip density, $Z_\mathrm{eq}$ is
a constant integer between the discontinuities, and is
presented by Tables \ref{tabZ22}\,--\,\ref{tabZ26}. 
In the ETF domain beyond 
proton drip, however, $Z_\mathrm{eq}$ is a smooth continuous
function. The values of this 
function at any given $\bar{n}$
are subject to uncertainty, because the minimum of energy as
a function of $Z$ is very shallow (see Fig.~\ref{eZic2}).
Within this uncertainty, $Z_\mathrm{eq}$ can
be roughly expressed as
\begin{equation}
Z_\mathrm{eq}= p_1 + p_2\bar{n} - p_3\,\left[\mathrm{max}(0,\bar{n}-p_4)
\right]^2,
\label{Zcell_fit}
\end{equation}
with the parameters $p_i$ given in Table~\ref{tab:Zcell_fit}.
The differences between this fit and the data are shown in the lower panel of
Figs.~\ref{Zic_zooma}, \ref{Zic_zoomb}.

\begin{table}
\centering
\caption[]{Parameters of Eq.~(\ref{Zcell_fit}).}
\label{tab:Zcell_fit}
\begin{tabular}{c|cccc}
\hline
 EoS & $p_1$ & $p_2$
 [fm$^3$] & $p_3$  &
 $p_4$   [fm$^{-3}$] \\
\hline 
BSk22  &  14.5 &  220 &  $6\times10^{4}$ &  0.066 \\
BSk24  & $-68$ & 1770 &  $8\times10^{5}$ &  0.075 \\
BSk25  &$-445$ & 7380 &  $4\times10^{6}$ &  0.0806\\
BSk26  &$-40.7$& 1237 &         0        &  0     \\
\hline
\end{tabular}
\end{table}

Jumps of $Z$ from one integer to
another throughout the inner crust are accompanied by
simultaneous jumps of the number of neutrons $N$ in such a way
that the proton fraction
$Y_\mathrm{p}=Z/(Z+N)$ is continuous (see Fig.~\ref{Yic}). It can be fitted as
\begin{equation}
  Y_\mathrm{p} =  \left(p_1\bar{n}^{-3/4}-p_2\right)
     (1+p_3\bar{n})\,[1+(p_4\bar{n})^4],
\label{Y_e_crust}
\end{equation}
where the parameters $p_i$ are given in
Table~\ref{tab:Y_e_crust}, assuming that $\bar{n}$ is
measured in fm$^{-3}$. The differences between
the fit and the data are shown in the lower panel of
Fig.~\ref{Yic}.

\begin{table}
\centering
\caption[]{Parameters of Eq.~(\ref{Y_e_crust}).}
\label{tab:Y_e_crust}
\begin{tabular}{c|cccc}
\hline
 EoS & $p_1$ & $p_2$ & $p_3$  & $p_4$ \\
\hline 
BSk22 &$6.08\times10^{-4}$~ & $3.42\times10^{-3}$~ & 133.6~ & 15.89\\
BSk24 &$5.91\times10^{-4}$~ & $2.88\times10^{-3}$~ & 196.3~ & 12.14\\
BSk25 &$5.82\times10^{-4}$~ & $2.67\times10^{-3}$~ & 247.2~ & 10.15\\
BSk26 &$5.99\times10^{-4}$~ & $1.66\times10^{-3}$~ & 137.5~ & ~8.91\\
\hline
\end{tabular}
\end{table}

It is also of interest to parametrize the cluster component $Z_\mathrm{cl}$ of
$Z_\mathrm{eq}$, as given by Eq.~(\ref{3.2Ca}). At all densities
$Z_\mathrm{cl}$ can be approximated analytically by
\begin{equation}
Z_\mathrm{cl} = Z_\mathrm{eq} - \left(\tilde{Z}^{p_3} + Z_\mathrm{eq}^{p_3}
\right)^{1/p_3}   \quad   ,
\label{Zfree2}
\end{equation}
where the parameter $p_3$, which is negative, is given in the last column of
Table~\ref{tab:Zfree}, and
\begin{equation}
\tilde{Z} =
\left(\frac{p_1}{1-x}\right)^2\,
(\sqrt{x}+p_2 x^4) \quad .
 \label{Zfree3}
\end{equation}
In this last equation the parameters $p_1$ and $p_2$ are given in
Table~\ref{tab:Zfree}, while
       \begin{equation}
        x=\frac{\bar{n}-\bar{n}_\mathrm{nd}}{
	 n_\mathrm{cc}-\bar{n}_\mathrm{nd}}  \quad  ,
	 \label{Zfree1}
	 \end{equation}
which runs from 0 to 1 in the inner crust, $\bar{n}_\mathrm{nd}$ representing
the neutron-drip density and $n_\mathrm{cc}$ the density at the
core-crust interface.
Actually, $\tilde{Z}$ itself, as given by Eq.~(\ref{Zfree3}), is a good
approximation to $Z_\mathrm{f}$ for $\tilde{Z} \ll Z_\mathrm{eq}$, but for higher values of
$\tilde{Z}$ the correction (\ref{Zfree2}) is needed to ensure that
$Z_\mathrm{cl}$ does not become negative.

With the number of free protons being defined by Eq.~(\ref{eqpf}), the number
fraction of free protons relative to all nucleons is given by
$Y_\mathrm{pf}=Z_\mathrm{f}/A = Y_\mathrm{p} Z_\mathrm{f}/Z_\mathrm{eq}$ and thus
can be represented analytically using the analytic fits already made. It is
shown in the upper panel of Fig.~\ref{pfic}, and the  differences between the
fit and the data are shown in the lower panel.

\begin{table}
\centering
\caption[]{Parameters of Eqs.~(\ref{Zfree2}) and (\ref{Zfree3}).}
\label{tab:Zfree}
\begin{tabular}{c|ccc}
\hline
 EoS & $p_1$ & $p_2$ & $p_3$ \\
\hline 
BSk22  &0.057 & 18.0 & $-0.97$ \\
BSk24  &0.069 & 15.5 & $-0.80$ \\
BSk25  &0.089 &  7.3 & $-0.59$ \\
BSk26  &0.060 & 28.7 & $-0.88$ \\
\hline
\end{tabular}
\end{table}

The equilibrium values $N_\mathrm{eq}$ of the number of neutrons in a WS cell can be
expressed by the identity
\beq\label{jmp1}
N_\mathrm{eq} = Z_\mathrm{eq}\left(\frac{1}{Y_\mathrm{p}} - 1\right) \quad.
\eeq
Now $Z_\mathrm{eq}$ has been parametrized in the proton-drip region by 
Eq.~(\ref{Zcell_fit}), while at lower densities in the inner crust it takes
well defined integer values. Moreover, $Y_\mathrm{p}$ is parametrized throughout the
inner crust by Eq.~(\ref{Y_e_crust}). Thus Eq.~(\ref{jmp1}) suffices to
parametrize $N_\mathrm{eq}$ at all densities in the inner crust; it defines the
curve showing $N_\mathrm{eq}$ in Fig.~\ref{Nica} and~\ref{Nicb}. The deviations 
between this parametrization and the data are shown in the lower panels of 
Figs.~\ref{Nica} and~\ref{Nicb}. 

The fraction of neutrons in the crust is $Y_\mathrm{n}=1-Y_\mathrm{p}$. 
In the inner crust, $N_f$ neutrons per unit WS cell are considered as free,
where $N_f$ is given by Eqs.~(\ref{3.2Cb}) and (\ref{eqnf}).  
The number fraction of the free neutrons relative to
the total number of baryons can be approximated as
\begin{equation}
   Y_\mathrm{nf} = \frac{p_1 x +p_4 x^{p_5}}{
       1 + p_2 x +p_3 x^2}
     + p_6 \,x \,\exp[ p_7\,(x-1) ],
\label{Ynf}
\end{equation}
where $x$ is the same as in Eq.~(\ref{Zfree1}), and the parameters $p_i$ are
given in Table~\ref{tab:Ynf}. The differences between
the fit and the data are shown in the lower panel of
Fig.~\ref{nfic}.

\begin{table}
\centering
\caption[]{Parameters of Eq.~(\ref{Ynf}).}
\label{tab:Ynf}
\begin{tabular}{ccccc}
\hline
        &  BSk22   & BSk24    &  BSk25   & BSk26        \\
\hline
  $p_1$ & 173.2 &  195.2  & 203.7  &  216.4             \\
  $p_2$ & 204.8 &  233.8  & 251.1  &  323.0             \\
  $p_3$ &  57.3 &   96.8  & 132.2  &  $1.482\times10^4$ \\
  $p_4$ &  61.4 &   86.8  & 107.7  &  $1.241\times10^4$ \\
  $p_5$ & 1.86  &  2.41   & 3.37   &  1.99              \\
  $p_6$ & 0.080 &  0.116  & 0.154  &  0.062             \\
  $p_7$ & 18    &  29     & 56     &  10.4              \\
\hline
\end{tabular}
\end{table}

With $Z_\mathrm{eq}, Y_\mathrm{p}$ and $Y_\mathrm{nf}$ already parametrized, $N_\mathrm{cl}$ can now
be parametrized through the identity
\beq\label{jmp3}
N_\mathrm{cl} =Z_\mathrm{eq}\left(\frac{1 - Y_\mathrm{p} - Y_\mathrm{nf}}{Y_\mathrm{p}}\right) \quad ;
\eeq
it defines the curves showing $N_\mathrm{cl}$ in Figs.~\ref{Nica} and~\ref{Nicb}. The 
deviations between this parametrization and the data are shown in the lower 
panels of Figs.~\ref{Nica} and~\ref{Nicb}.
 
\subsubsection{Particle numbers in the core}

At $n> n_\mathrm{cc}$, the nuclear clusters
disappear, so that all baryons become essentially free:
$Y_\mathrm{nf}=Y_\mathrm{n}$, $Y_\mathrm{pf}=Y_\mathrm{p}$. When the density increases still
further, the chemical potential of electrons continues to
increase. Eventually it exceeds the muon rest energy $m_\mu c^2 =
105.6584$~MeV. Then free muons are at equilibrium with the
electrons, whence their chemical potentials are the same,
\beq\label{sasha.1}
\mu_\mathrm{e} = \mu_\mu    \quad  ,
\eeq
as follows from Eqs.~(\ref{corea.13a}) and ~(\ref{corea.13b}). 
Thus, neglecting exchange, it follows that the Fermi energies~(\ref{lep.13}) of
the electrons and muons are equal (with the rest masses included), whence for a
given number density of 
electrons $n_\mathrm{e}$, the number density of muons is given by the relation
\begin{equation}
   m_\mu c^2\,\sqrt{1+x_{\mu}^2} =
   m_\mathrm{e} c^2\,\sqrt{1+x_\mathrm{e}^2},
\label{mu_e_equil}
\end{equation}
where
\begin{equation}
  x_\mathrm{e\,(\mu)} = \frac{\hbar}{m_\mathrm{e\,(\mu)} c}\,(3\pi^2
  n_\mathrm{e\,(\mu)})^{1/3}
\label{x_e_mu}
\end{equation}
is the relativity factor at the Fermi surface of the electrons (muons). 
(Actually, the validity of Eq.~(\ref{mu_e_equil}) depends on the muon gas
being degenerate, but since the temperature of the star is not identically
zero it follows that at the threshold where the muons just start to appear  
this condition will not be satisfied. However, the density range over which
we do not have complete degeneracy for the muons is insignificant.)

It follows from Eqs.~(\ref{mu_e_equil}) and (\ref{x_e_mu}) that 
\begin{equation}
   n_\mu = \frac{1}{3\pi^2}\left(
         \frac{m_\mathrm{e} c}{\hbar}\,
         \sqrt{1+x_\mathrm{e}^2-(m_\mu/m_\mathrm{e})^2} \right)^3,
\end{equation}
provided that $1+x_\mathrm{e}^2>(m_\mu/m_\mathrm{e})^2$ (otherwise $n_\mu=0$).
Therefore, it
is sufficient to have a fit to $Y_\mathrm{e}=n_\mathrm{e}/n$ in order
to evaluate $Y_\mu=n_\mu/n$ and $Y_\mathrm{p}=Y_\mathrm{e}+Y_\mu$. We
represent $Y_\mathrm{e}$ by
\begin{equation}     
Y_\mathrm{e}=\frac{p_1+p_2\,n+p_6\,n^{3/2}+p_3\,n^{p_7}}{
         1+p_4\,n^{3/2}+p_5\,n^{p_7}},
\label{Y_e_core}
\end{equation}
with parameters listed in Table~\ref{tab:Y_e_core} (for
$n$ measured in fm$^{-3}$). Fig.~\ref{ycore}
shows electron and muon number fractions as functions of
$n$ in the core of a neutron star and the differences
between the fit (\ref{Y_e_core}) and the numerical data.

\begin{figure*}
\begin{center}
\includegraphics[width=.32\textwidth]{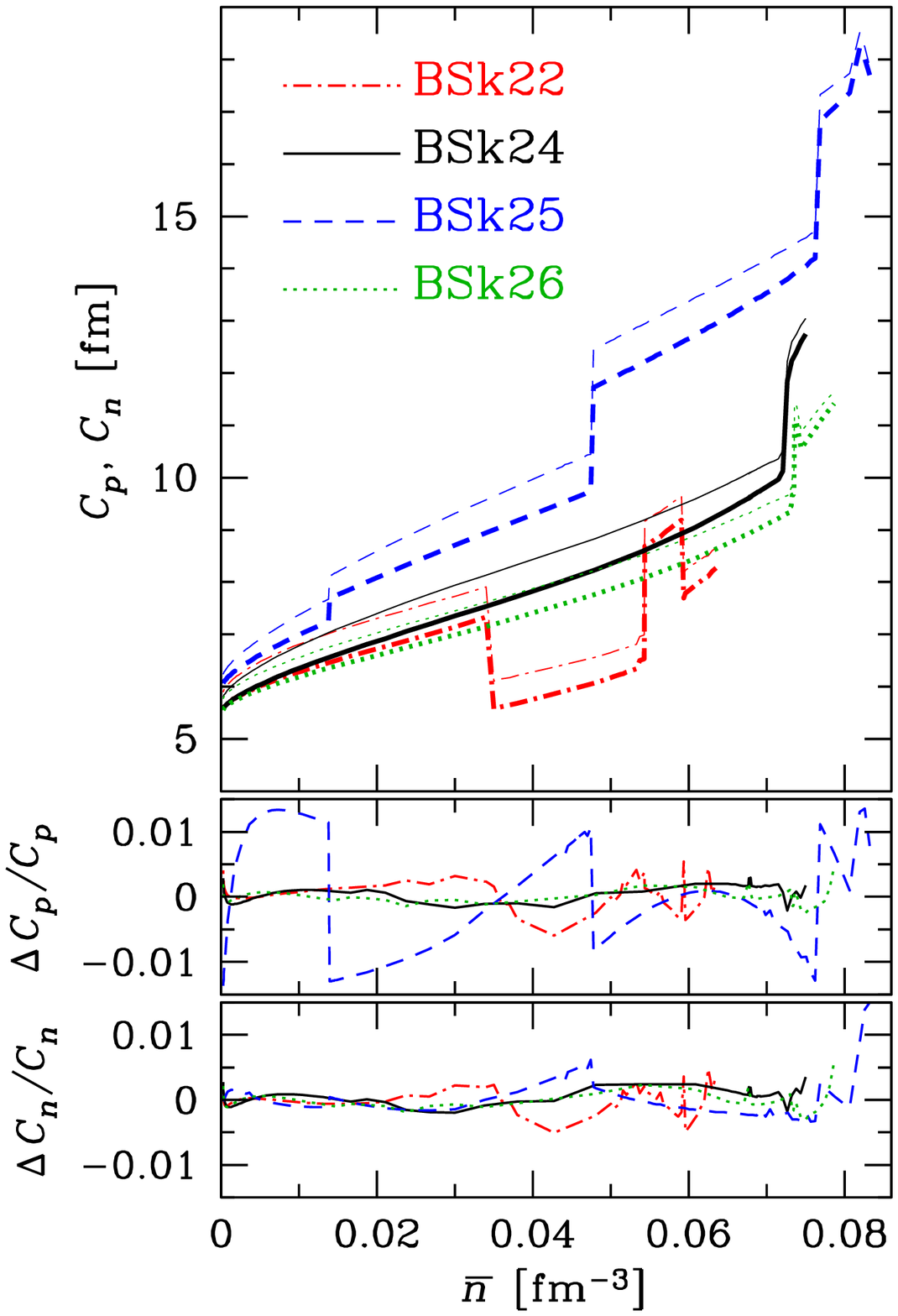}
\includegraphics[width=.32\textwidth]{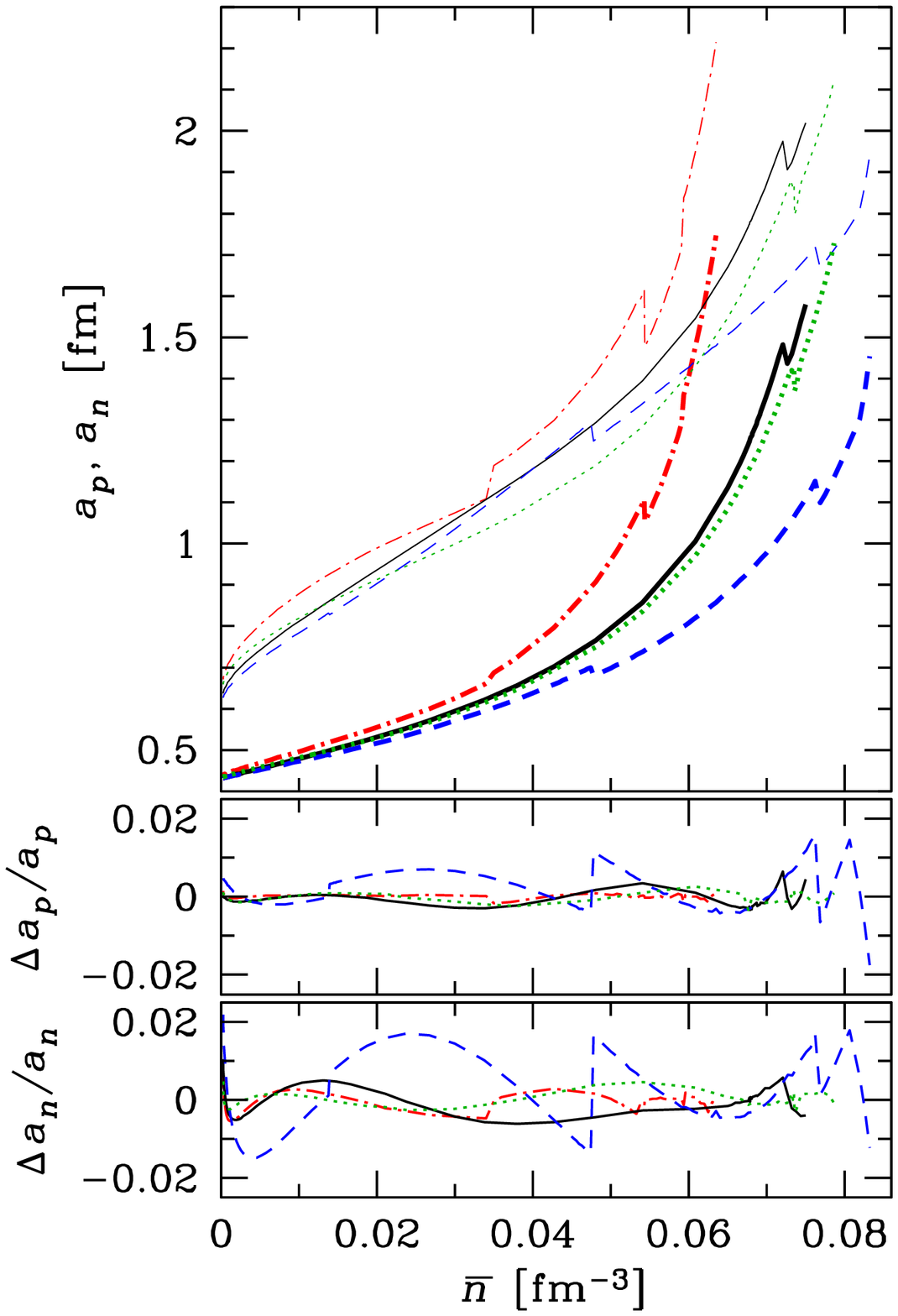}
\includegraphics[width=.32\textwidth]{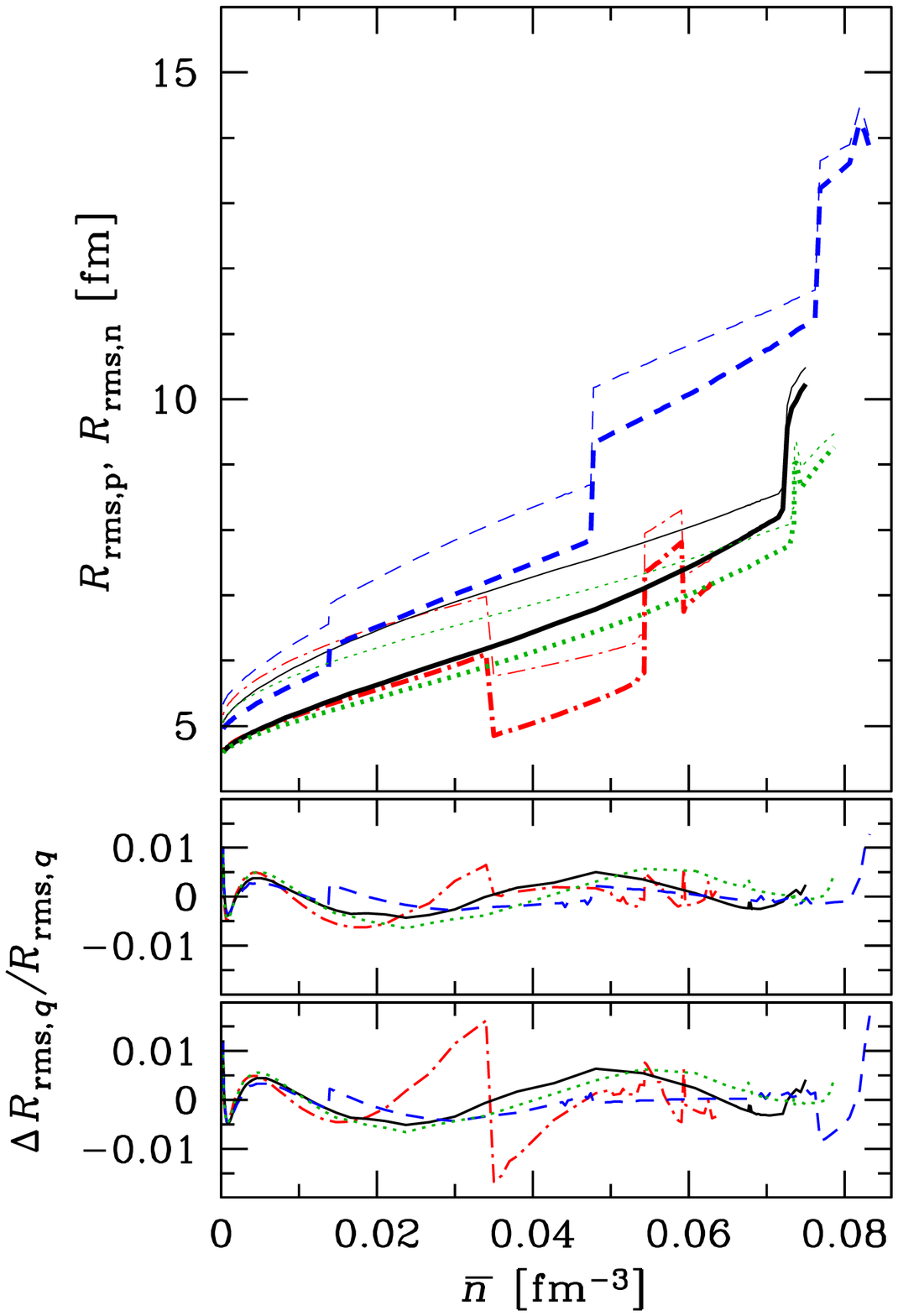}
\end{center}
\caption{Geometrical parameters of density distributions of nucleons in the 
inner crust. Top panels: size parameters $C_q$ (left), diffuseness parameters
$a_q$ (centre), and rms sizes $R_{\mathrm{rms},q}$ (right) of the
distributions of protons ($q=p$, lower thick lines) and
neutrons ($q=n$, upper thin lines) as functions of mean
baryon density in the inner crust of a neutron star for models BSk22,
BSk24, BSk25, and BSk26. 
Middle panels: fractional differences between the fits and
the data for  $C_\mathrm{p}$ (left), $a_\mathrm{p}$ (centre), and $R_\mathrm{rms,p}$
(right).
Bottom panels: fractional differences between the fits and
the data for  $C_\mathrm{n}$ (left), $a_\mathrm{n}$ (centre), and $R_\mathrm{rms, n}$
(right).
(Corrected version.)
}
\label{fig:sizes}
\end{figure*}

\begin{table}
\centering
\caption[]{Parameters of Eq.~(\ref{Y_e_core}).}
\label{tab:Y_e_core}
\begin{tabular}{ccccc}
\hline
        &  BSk22   & BSk24    &  BSk25   & BSk26    \\
\hline
  $p_1$ &$-0.0024$ &  0.0021  & 0.0164   & $-0.0064$ \\
  $p_2$ & 0.456    & 0.581    & 0.410    &  0.690    \\
  $p_3$ & 12.349   & 9.874    & 4.471    &  0.6074   \\
  $p_4$ & 16.15    & 16.20    & 3.89     &  28.11    \\
  $p_5$ & 48.72    & 39.20    & 17.64    &  2.455    \\
  $p_6$ &  0.311   & $-0.148$ & $-0.517$ &  0.344    \\
  $p_7$ & 3.674    & 4        & 4        &  4.6      \\
\hline
\end{tabular}
\end{table}

\subsection{Chemical potentials of nucleons}

\begin{table}
\centering
\caption[]{Parameters of Eq.~(\ref{munsfit1}).}
\label{tab:munsfit}
\begin{tabular}{c|cccc}
\hline
 EoS & $p_1$ & $p_2$ & $p_3$\\
\hline 
BSk22  & 9.41 &  7.62 & 259 \\
BSk24  & 4.027& 18.31 & 106 \\
BSk25  & 1.172& 26.97 &61.9 \\
BSk26  & 8.042&$-6.86$&11.1 \\
\hline
\end{tabular}
\end{table}

In the outer crust, the nucleons are bound in the nuclei. In the inner
crust and in the core, there are free nucleons, and their chemical
potentials can be of interest for some astrophysical problems. In this
section we present fits to these chemical potentials.

\subsubsection{Chemical potential of nucleons in the inner crust}
\label{sect:mufitic}

In the inner crust, the chemical potential of free neutrons without the
rest mass, $\mu_\mathrm{n}^* = \mu_\mathrm{n}-M_\mathrm{n} c^2$,
is fitted by
\beq\label{munsfit1}
  \mu_\mathrm{n}^* = \frac{\mu_\mathrm{id}^*}{
   1+p_1\sqrt{\bar n}+p_2\bar n+p_3\bar n^2},
\eeq
\beq
   \mu_\mathrm{id}^* = \frac{\hbar^2(3\pi^2
   n_\mathrm{nf})^{2/3}}{2M_\mathrm{n}}.
\eeq
Here, $n_\mathrm{nf}=Y_\mathrm{nf}\,\bar n$ is the number density of free neutrons,
$Y_\mathrm{nf}$ is represented by Eq.~(\ref{Ynf}),
$\mu_\mathrm{id}^*$ is the Fermi energy of free neutrons
(without the rest energy) in
the model of a gas of nonrelativistic fermions, and
the denominator in Eq.~(\ref{munsfit1}) for
$\mu_\mathrm{n}^*$ is a correction factor to this model.
Parameters $p_i$ are given in Table~\ref{tab:munsfit}, assuming that
$\bar n$ is measured in fm$^{-3}$. 
Deviations of this parametrization from the computed data are shown
in the lower panel of Fig.~\ref{munic}. 

The proton chemical potential is then parametrized by using the 
beta-equilibrium condition (\ref{mupg}), with $\mu_\mathrm{n}$ given by the
parametrization (\ref{munsfit1}), while we approximate $\mu_\mathrm{e}$ by
the Fermi energy $\mu_\mathrm{e}^\mathrm{kin} +m_\mathrm{e}c^2$, where $\mu_\mathrm{e}^\mathrm{kin}$ is given by
Eq.~(\ref{lep.13}) and Eq.~(\ref{Y_e_crust}) for $Y_\mathrm{p}$; the neglected terms in 
$\mu_\mathrm{e}$ have a negligible impact.
Deviations of this parametrization from the computed data are shown
in the lower panel of Fig.~\ref{mupic}. 

\subsubsection{Chemical potential of nucleons in core}
\label{sect:mufitcore}

In the core, the model of free fermion gas
($\mu_\mathrm{id}^*$) would be an inadequate starting
approximation, because of the great role of strong
interactions. We parametrize the neutron chemical potential
without the rest mass in the core by
\begin{equation}
\mu_\mathrm{n}^* =  
\,\frac{p_1n^{p_2}\,[1+(p_3n)^6]^{p_4}}{
                          [1+
(p_5n)^7]^{p_6}\,[1+1.5(n/n_\mathrm{cc}-1)]}.
\label{mu_core_fit}
\end{equation}
Parameters $p_i$ are given in Table~\ref{tab:mu_core_fit}, assuming that
$n$ is measured in fm$^{-3}$ and $\mu_\mathrm{n}^*$ in MeV.
Deviations of this parametrization from the computed data are shown
in the lower panel of Fig.~\ref{muncore}. 

\begin{table}
\centering
\caption[]{Parameters of Eq.~(\ref{mu_core_fit}).}
\label{tab:mu_core_fit}
\begin{tabular}{c|cccccc}
\hline
 EoS & $p_1$ & $p_2$ & $p_3$ & $p_4$ & $p_5$ & $p_6$ \\
\hline 
BSk22  & 6824 & 2.4322 & 9.250  & 0.10698  & 2.248  & 0.0440 \\
BSk24  & 2807 & 2.1785 & 8.055  & 0.18142  & 1.915  & 0.0716 \\
BSk25  & 1434 & 1.9741 & 7.932  & 0.24071  & 2.009  & 0.1132 \\
BSk26  & 3302 & 2.2416 & 6.268  & 0.14566  & 0.770  & 0.0400 \\
\hline
\end{tabular}
\end{table}

\begin{table*}
\centering
\caption[]{Parameters of Eq.~(\ref{fitshape}).}
\label{tab:fitshape}
\begin{tabular}{c|ccccccc}
\hline
 EoS & $p_1$ & $p_2$ & $p_3$  & $p_4$ & $p_5$ & $p_6$  & $p_7$ \\
\hline 
& \multicolumn{6}{c}{for $C_\mathrm{p}$:} \\
BSk22  &5.50 &$3.04\times10^5$& 4.22 & 2.75  & 0.5 & 0.462 & 0.486 \\
BSk24  &5.527&$5.29\times10^4$& 4.00 & 1.574 &  1  & 0.631 & 0.720 \\
BSk25  &5.71 &  455           & 1.93 & 202   &  3  & 0.197 & 0.438 \\
BSk26  &5.555& $-3.90$        & 0.315&$-60.4$&  2  & 0.294 & 0.444 \\
[-1ex]\multicolumn{8}{c}{\dotfill}\\
& \multicolumn{7}{c}{for $C_\mathrm{n}$:} \\
BSk22  & 5.73 &$4.16\times10^5$& 4.43 &  2.0   & 0.5 &  0.459 & 0.478 \\
BSk24  & 5.706&$1.18\times10^4$& 3.60 & 0.063  &  1  &  0.516 & 0.640 \\
BSk25  & 3.77 & 45.1           &0.855 & $-139$ &  3  &  0.0626& 0.031 \\
BSk26  & 5.775& $-2.51$        &0.250 & $-48.4$&  2  &  0.283 & 0.431 \\
[-1ex]\multicolumn{8}{c}{\dotfill}\\
& \multicolumn{7}{c}{for $a_\mathrm{p}$:} \\
BSk22  & 0.4376& 3.23 & 0.877 & $-731.55$ & 2.568 & $  -29.8       $    & 3.153\\
BSk24  & 0.431 & 4.89 & 1.005 & $-3729 $  & 3.356 & $-6.7\times10^8$    & 9.70 \\
BSk25  & 0.434 & 7.86 & 1.159 & $-2199 $  & 3.355 & $-18.5$             & 3.52 \\
BSk26  & 0.434 & 4.09 & 0.970 & $-2523 $  & 3.244 & $-1.07\times10^{12}$& 12.7 \\
[-1ex]\multicolumn{8}{c}{\dotfill}\\
& \multicolumn{7}{c}{for $a_\mathrm{n}$:} \\
BSk22  & 0.656 & 5.50 & 0.671 & $-1.39\times10^5$ & 4.643 & $  -0.463$ & 1.43 \\
BSk24  & 0.631 & 5.27 & 0.729 & $-2701$           & 3.377 & $-335$     & 4.13 \\
BSk25  & 0.636 & 12.6 & 0.972 & $-1.70\times10^6$ & 6.34  & $-42.6$    & 3.86 \\
BSk26  & 0.633 & 1.94 & 0.511 & $-136.6$          & 2.175 & $-3.65$    & 2.442\\
[-1ex]\multicolumn{8}{c}{\dotfill}\\
& \multicolumn{7}{c}{for $x_\mathrm{p}$:} \\
BSk22  & 0.1023 & 1.292 & 0.5114 & 343 & 2.605 &    1.12           & 2.59  \\
BSk24  & 0.1035 & 1.944 & 0.5717 & 608 & 3.143 &  0.0225           & 1.26  \\
BSk25  & 0.09975& 2.747 & 0.6316 & 1748& 3.906 &$5.94\times10^{-4}$& 0.192 \\
BSk26  & 0.1025 & 1.383 & 0.526  & 6.69& 1.238 &    45.8           & 4.25  \\
[-1ex]\multicolumn{8}{c}{\dotfill}\\
& \multicolumn{7}{c}{for $x_\mathrm{n}$:} \\
BSk22  & 0.1134 & 1.620 & 0.5163 &  132   & 2.41 & 229           & 4.89 \\
BSk24  & 0.1137 & 2.562 & 0.5932 & 3333   & 4.63 & 160           & 4.66 \\
BSk25  & 0.1116 & 3.132 & 0.6207 & $-2.2$ & 2.78 &$9.8\times10^4$& 7.96 \\
BSk26  & 0.1113 & 1.571 & 0.5196 &  2.49  & 0.966&  71           & 4.47 \\
\hline
\end{tabular}
\end{table*}

The proton chemical potential $\mu_\mathrm{p}$ is now parametrized in the core the same
way as it was in the inner crust.
Deviations of this parametrization from the computed data are shown
in the lower panel of Fig.~\ref{mupcore}.

\subsection{Geometrical parameters and sizes of nuclei in the inner crust}
\label{sect:geom}

In applications one sometimes needs more detailed information on microscopic 
distributions of nucleons in the inner crust than is given simply by the 
numbers of free and bound nucleons considered in Section~\ref{sect:numfit_crust}.
For example, cross sections of scattering of electrons on nuclei in the crust 
depend on the charge distribution in a nucleus (e.g., 
\citealt*{kppty99,gyp01}). For the use in such applications, we construct 
analytic approximations for the geometrical parameters $C_q$ and $a_q$, which 
enter Eq.~(\ref{3.2}) and are tabulated as 
functions of $\bar n$. The general form of these approximations is
\begin{equation}
  y
            =
       \frac{p_1+p_2\,\bar n^{p_3}}{
         1+p_4\,\bar n^{p_5}} + p_6 Z_\mathrm{eq} \bar n^{p_7},
\label{fitshape}
\end{equation}
where $y=C_\mathrm{p}$, $a_\mathrm{p}$, $C_\mathrm{n}$, or
$a_\mathrm{n}$. Equation~(\ref{fitshape}) is similar to Eq.~(17) of
\citet{pfncpg12}, but more complicated, mainly because of the last term, 
which appears since $Z_\mathrm{eq}$ varies with increasing density in the crust,
unlike constant $Z_\mathrm{eq}=40$ in \citet{pfncpg12}. As in \citet{pfncpg12}, the
same form of parametrization (\ref{fitshape}) also provides approximations to 
the dimensionless nuclear-size parameters $x_\mathrm{p}$ and $x_\mathrm{n}$, which
enter the expressions for electrical and thermal
conductivities \citep{gyp01} and are related to the
root-mean-square radii of the proton and neutron
distributions in the clusters, $R_\mathrm{rms,p}$ and $R_\mathrm{rms,n}$, via
$x_\mathrm{p,n}=\sqrt{5/3}\,R_\mathrm{rms,p,n}/R$, where $R$
denotes the cell radius.
The parameters $p_i$ of approximation (\ref{fitshape}) for the
parameters $a_q$, $C_q$, and $x_q$
($q=p,n$) are listed  in Table~\ref{tab:fitshape} (for
$\bar n$ measured in fm$^{-3}$). Figure~\ref{fig:sizes}
demonstrates the behavior of $C_q$, $a_q$, and $R_{\mathrm{rms},q}$ as
functions of density (the top panels) and the fractional
errors of the corresponding fits (\ref{fitshape}) (middle and
bottom panels).  The errors of the fits are within
(1\,--\,2)\%.

\bsp	
\label{lastpage}
\end{document}